\documentclass[preprint]{aastex}
\usepackage{emulateapj5}
\usepackage{apjfonts}
\usepackage{natbib}
\usepackage{epsfig}
\input psfig.tex

\def\aa{{A\&A}}

\def\aj{{AJ}}

\def\annrev{{ARA\&A}}
\def\apj{{ApJ}}
\def\apjs{{ApJS}}

\def\deg{$^{\circ}$}

\def\farcs{\hbox{$.\mkern-4mu^{\prime\prime}$}}

\def\hst{{\it HST}}
\def\kms{km s$^{-1}$}

\def\lax{{$\mathrel{\hbox{\rlap{\hbox{\lower4pt\hbox{$\sim$}}}\hbox{$<$}}}$}}
\def\gax{{$\mathrel{\hbox{\rlap{\hbox{\lower4pt\hbox{$\sim$}}}\hbox{$>$}}}$}}
\def\simlt{\lower.5ex\hbox{$\; \buildrel < \over \sim \;$}}
\def\simgt{\lower.5ex\hbox{$\; \buildrel > \over \sim \;$}}
\def\lum{ergs s$^{-1}$}
\def\mbh{{$M_{\rm BH}$}}

\def\mnras{{MNRAS}}

\def\percm2{cm$^{-2}$}

\def\solmass{$M_\odot$}

\def\oiii{[\ion{O}{3}]}

\def\mlb{$M_{\rm BH}-L_{\rm{bul}}$}
\def\mlt{$M_{\rm BH}-L_{\rm{host}}$}
\def\msig{$M_{\rm BH}-\sigma_\star$}
\def\edd{$L_{{\rm bol}}$/{$L_{{\rm Edd}}$}}
\def\lbul{$L_{\rm bul}$}
\def\ser{S\'{e}rsic}
\def\hnr{$L_{\rm host}/L_{\rm nuc}$}
\def\bnr{$L_{\rm bul}/L_{\rm nuc}$}

\def\rbulge{$M_{R,{\rm bul}}$}

\slugcomment{To Appear in {\it
The Astrophysical Journal}.}
\shorttitle{Black Hole Mass vs. Bulge Luminosity Relation}
\shortauthors{KIM et al.}

\begin{document}

\title{The Origin of the Intrinsic Scatter in the Relation 
Between Black Hole Mass and Bulge Luminosity for Nearby Active Galaxies
\altaffilmark{1}}
\author{Minjin Kim\altaffilmark{2,3}, Luis C. Ho\altaffilmark{2}, Chien Y.
Peng\altaffilmark{4}, Aaron J. Barth\altaffilmark{5},  
Myungshin Im\altaffilmark{3}, Paul Martini\altaffilmark{6}, and
Charles H. Nelson\altaffilmark{7}}

\altaffiltext{1}{Based on observations made with the NASA/ESA  {\it Hubble 
Space Telescope}, obtained from the Data Archive at the Space Telescope 
Science Institute, which is operated by the Association of Universities for 
Research in Astronomy (AURA), Inc., under NASA contract NAS5-26555. These 
observations are associated with program AR-10969 and GO-9763.}

\altaffiltext{2}{The Observatories of the Carnegie Institution of Washington, 
813 Santa Barbara Street, Pasadena, CA 91101; mjkim@ociw.edu, lho@ociw.edu.}

\altaffiltext{3}{Department of Physics and Astronomy, Frontier Physics 
Research Division (FPRD), Seoul National
University, Seoul, Korea; mim@astro.snu.ac.kr.}

\altaffiltext{4}{NRC Herzberg Institute of Astrophysics, 5071 West Saanich 
Road, Victoria, British Columbia, Canada V9E2E7; cyp@nrc-cnrc.gc.ca.}

\altaffiltext{5}{Department of Physics and Astronomy, University of 
California at Irvine, 4129 Frederick Reines Hall, Irvine, CA 92697-4575; 
barth@uci.edu.}

\altaffiltext{6}{Center for Cosmology and AstroParticle Physics,
The Ohio State University, 
191 West Woodruff Avenue, OH 43210; martini@astronomy.ohio-state.edu}

\altaffiltext{7}{Physics and Astronomy Department, Drake University, 2507 
University Avenue, Des Moines, IA 50311; charles.nelson@drake.edu}

\begin{abstract}
We investigate the origin of the intrinsic scatter in the correlation between 
black hole mass (\mbh) and bulge luminosity (\lbul) in a sample of 45 massive, 
local ($z$ \lax\ 0.35) type~1 active galactic nuclei (AGNs).  We derive \mbh\ 
from published optical spectra assuming a spherical broad-line region, and 
\lbul\ from detailed two-dimensional decomposition of archival optical 
{\it Hubble Space Telescope}\ images.  AGNs follow the \mlb\ relation of 
inactive galaxies, but the zero point is shifted by an average of $\Delta \log 
M_{\rm BH} \approx -0.3$ dex.  We show that the magnitude of the zero point 
offset, which is responsible for the intrinsic scatter in the \mlb\ relation, 
is correlated with several AGN and host galaxy properties, all of which are 
ultimately related to, or directly impact, the BH mass accretion rate.  At a 
given bulge luminosity, sources with higher Eddington ratios have lower \mbh.  
The zero point offset can be explained by a change in the normalization of the 
virial product used to estimate \mbh, in conjunction with modest BH growth 
($\sim$10\%--40\%) during the AGN phase.  Galaxy mergers and tidal 
interactions appear to play an important role in regulating AGN fueling in 
low-redshift AGNs.
\end{abstract}


\section{Introduction}

Early-type galaxies commonly contain a central black hole (BH) whose mass
strongly correlates with the bulge luminosity (Kormendy \& Richstone 1995;
Magorrian et al. 1998) and stellar velocity dispersion (Gebhardt et al. 2000; 
Ferrarese \& Merritt 2000).  Lower-mass BHs found in late-type spirals and 
spheroidal galaxies follow a similar \msig\ relation (Barth et al. 2005; 
Greene \& Ho 2006a) but apparently a different \mlb\ relation 
(Greene et al. 2008).  The BH-host scaling relations suggest that BHs play  
an important role in galaxy formation and evolution (e.g., Granato et al. 2004; 
Di~Matteo et al. 2005; Robertson et al. 2006).  Understanding the mechanism by 
which BHs coevolve with their hosts impacts current models of cosmological 
structure formation (e.g., Bower et al. 2006; Croton et al. 2006). 

A key, unanswered question is how the BH-host scaling relations originated.  
This issue can be addressed by extending the BH-host scaling 
relations to {\it active}\ galaxies---wherein, the BH, by selection, is 
currently still growing---and by tracking the scaling relations to higher 
redshift to see when and possibly how they were established.  BH masses in 
type~1 (broad-line, unobscured) active galactic nuclei (AGNs) now can be 
routinely estimated to reasonable accuracy ($\sim 0.3-0.5$ dex), from the 
``virial method'' using single-epoch ultraviolet or optical spectra (e.g., 
Kaspi et al. 2000; McLure \& Dunlop 2001; Vestergaard 2002; Greene \& Ho 
2005).  More challenging to obtain are reliable measurements of the underlying 
host galaxy, particularly of the bulge component, which is maximally affected 
by the bright AGN core (e.g., McLure et al. 1999; Floyd et al. 2004; Nelson 
et al. 2004; Greene \& Ho 2006b; Kim et al. 2007), although 
substantial progress has been made.

Recent studies present tantalizing evidence that the BH-host scaling relations 
for active galaxies evolve with redshift, even by $z \approx 0.4$ (Woo et al. 
2006; Treu et al. 2007) and as far back as $z \approx 4$ (Peng et al. 2006a, 
2006b; Shields et al. 2006; Ho 2007).  Compared to local, inactive systems, 
the sense of the evolution is that for a given host galaxy mass or 
gravitational potential, higher redshift AGNs have a larger BH mass than 
similar systems at low redshift.  Taken at 
face value, this suggests that the growth of the BH precedes, or at least 
outpaces, the growth of the galaxy at higher redshifts.  On the other hand, 
local ($z \approx 0$) AGNs seem to behave quite differently. McLure \& Dunlop 
(2002) studied a sample of 72 nearby AGNs and find that they roughly follow 
the same \mlb\ relation defined by inactive galaxies, albeit with a somewhat 
greater scatter.  In the \msig\ relation of AGNs, highly accreting AGNs 
seem to have a different normalization in the sense that they tend to have a 
lower \mbh\ for a given $\sigma_\star$ (Greene \& Ho 2006b; Shen et al. 2008).
A similar trend is seen by Ho et al. (2008), using H~I line widths to constrain 
the gravitational potential of the underlying host galaxy. 
%
%At the moment, it is unclear whether the difference is due to an actual 
%difference in \mbh\ or a difference in the normalization factor.

As a concrete step toward establishing a robust $z=0$ baseline for comparison 
with high-$z$ studies, this series of papers (Kim et al. 2007, 2008; M. Kim 
et al., in preparation) focuses on quantifying the local \mlb\ 
%%%%%%%%%%%%%%%%%%%%%%%%%%%%%%%%%%%%%%%%%%%%%%%%%
\psfig{file=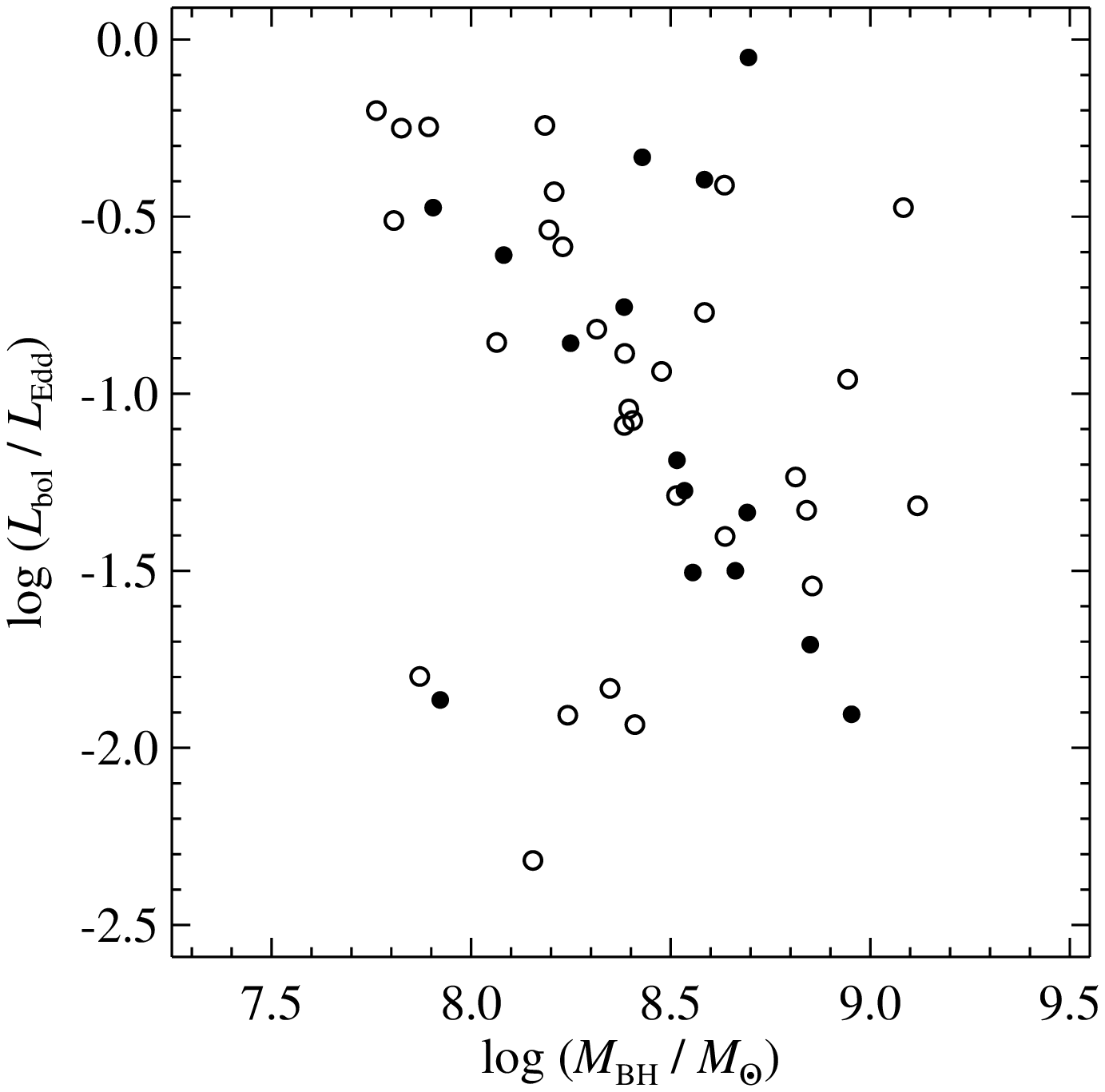,width=8.5cm,angle=0}
\figcaption[fig1.ps]
{Distribution of \mbh\ vs. Eddington ratio. Radio-loud and radio-quiet objects 
are denoted by filled and open symbols, respectively.
\label{fig1}}
\vskip 0.3cm
%%%%%%%%%%%%%%%%%%%%%%%%%%%%%%%%%%%%%%%%%%%%%%%%%%
\noindent
relation for 
active galaxies.  Among nearby 
inactive galaxies, BH mass 
correlates only marginally less tightly with bulge luminosity or mass than with bulge stellar 
velocity dispersion (Marconi \& Hunt 2003; H\"aring \& Rix 2004; Novak et al. 2006).  
Moreover, the \mlb\ relation for active galaxies shows no large systematic 
differences from that of inactive galaxies (McLure \& Dunlop 2002).  This 
suggests 
that the \mlb\ relation can be used as a useful substitute for the 
\msig\ relation, being an especially effective observational tool 
to track the cosmological evolution of the BH-galaxy connection 
(Peng et al. 2006a, 2006b; Treu et al. 2007).  Whereas stellar velocity 
dispersions are difficult, if not impossible, to measure for distant quasars, 
for example, photometric measurements of quasar hosts continue to be feasible 
even out to high redshifts, either through direct imaging (e.g., Kukula et al. 
2001; Ridgway et al. 2001) or through strong lensing magnification (Peng et 
al. 2006b).  In a companion paper, Kim et al. (2008) demonstrate that the 
bulge luminosity of type~1 AGNs can be measured to a reasonable accuracy
($\sim 0.5$ mag) in {\it Hubble Space Telescope (HST)}\ images, even in the 
regime when the active nucleus far outshines the galaxy.

Instead of characterizing the full \mlb\ relation for AGNs, this paper 
restricts itself to only one important aspect: the origin of the intrinsic 
scatter. By choosing a sample  for which we can estimate reliable BH masses 
and for which we can derive robust measurements of bulge luminosity from \hst\ 
images, our objective is to quantify the true intrinsic scatter of the 
relation and to characterize possible variations of the scatter with physical 
properties of the AGN, host galaxy, or environment.  By elucidating the 
physical drivers that influence the scatter of the \mlb\ relation, we hope to 
gain insights on how the BH-host galaxy relations were established.

This paper is structured as follows. We describe the sample selection in 
\S{2}. We present the image-fitting procedure for measuring bulge
luminosities and our image decomposition results in \S{3}.
We investigate the \mlb\ relation for type~1 AGNs in \S 4.
Finally, Section 5 discusses the origin of
the intrinsic 
%%%%%%%%%%%%%%%%%%%%%%%%%%%%%%%%%%%%%%%%%%%%%%%%%
\psfig{file=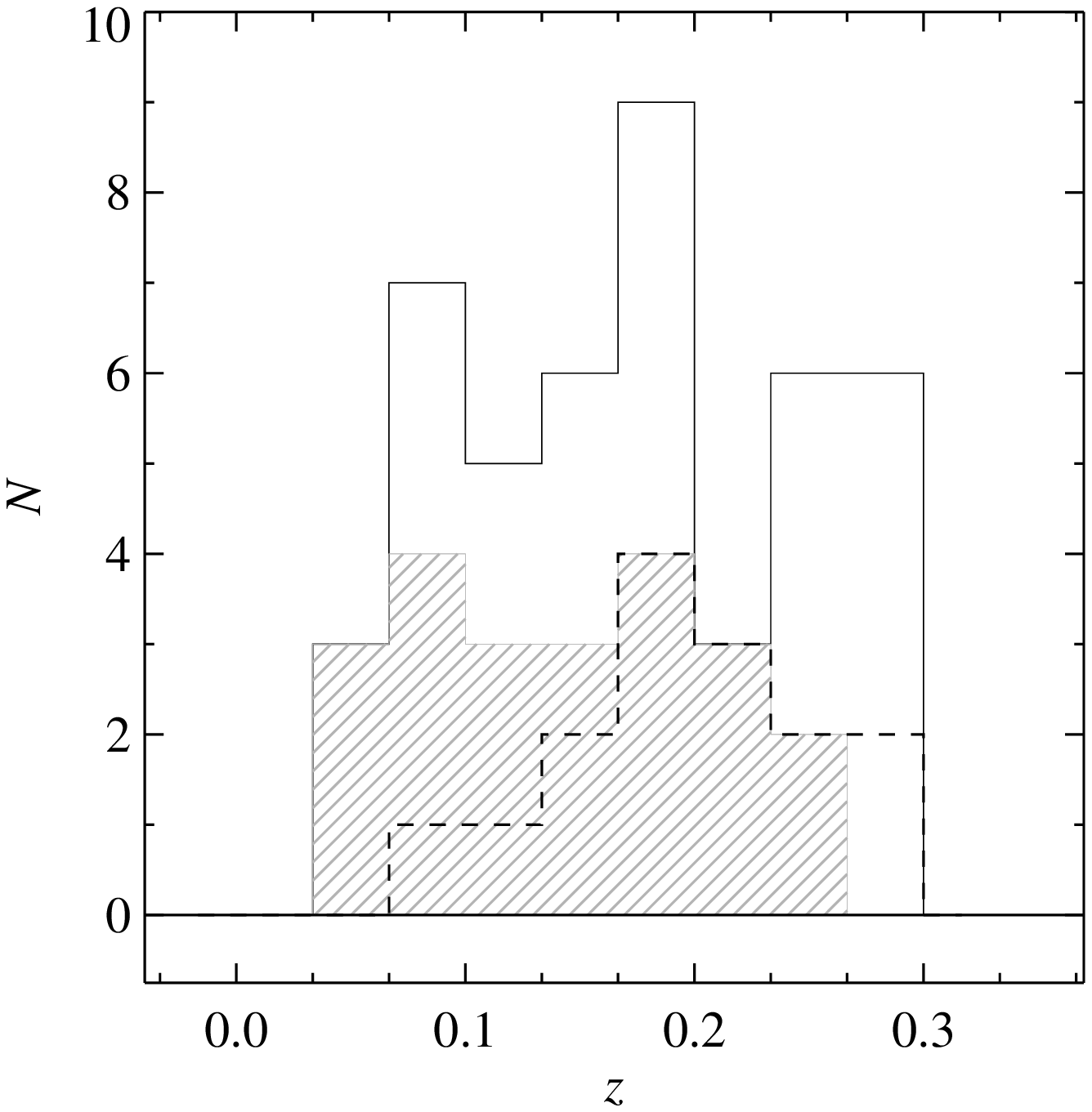,width=8.5cm,angle=0}
\figcaption[fig2.ps]
{Distribution of redshifts for our sample.  The open histogram shows the total
sample; the hatched histogram shows the objects with low Eddington ratio
(\edd $\leq 0.1$); the dashed histogram shows the radio-loud objects.
\label{fig2}}
\vskip 0.3cm
\noindent
%%%%%%%%%%%%%%%%%%%%%%%%%%%%%%%%%%%%%%%%%%%%%%%%%%
scatter in the \mlb\ relation, ending with a summary in \S6.  
Throughout we adopt the following cosmological 
parameters: $H_0 = 100\,h = 71 \, $\kms~Mpc$^{-1}$,
$\Omega_m = 0.27$, and $\Omega_{\Lambda} = 0.75$ (Spergel et al. 2003).

\section{Sample Selection}

Our initial selection begins with all AGNs known to possess broad emission
lines (type~1 objects), and have reasonably deep, and non-highly
saturated optical images in the \hst\ public archive. 
Since we are interested in establishing the local 
\mlb\ relation, we only consider sources with $z$ \lax\ 0.35.
Next, we carefully search the literature for published measurements 
of broad emission-line widths (either H$\alpha$ or H$\beta$), which are needed 
for calculating \mbh.  Although this step is necessarily somewhat subjective, 
we try to be consistent in selecting only objects that have line widths with 
published error bars \lax 10\%. The availability of spectrophotometric 
measurements is not essential for us because our BH masses ultimately make use 
of the nuclear luminosities from our photometric decomposition of the nucleus 
(\S 3.3).  The above screening process yields approximately 200 objects.  

Since the principal aim of this work is to study the intrinsic scatter of the 
\mlb\ relation, it is imperative that we choose objects for which we can 
obtain the most reliable estimates of the two primary quantities of interest, 
\mbh\ and \lbul.  Guided by this overriding goal, we purposely restrict our
sample to the upper end of the \mbh\ distribution.  All else being equal, this 
selection criterion biases the sample toward more luminous, more massive, 
earlier-type hosts for which we can derive more reliable bulge parameters 
because the structural decomposition will be less complicated than in 
later-type systems.  An added benefit of this mass selection is that our 
sample will consist of close analogs of 
higher-redshift quasars.  For concreteness, we choose sources
with \mbh\ $> 10^{7.8}$ \solmass.  This limit\footnote{Our BH masses assume a 
smaller geometrical factor than that used in Onken et al. (2004), by a factor 
of 1.8. Thus, \mbh\ = $10^{7.8}$ \solmass\ on our scale is equivalent to \mbh\ 
= $10^8$ \solmass\ on the scale of Onken et al.  We note that that our main 
conclusions do not rely on the geometrical factor. This issue is discussed in 
\S 5.2.} is admittedly somewhat 
%%%%%%%%%%%%%%%%%%%%%%%%%%%%%%%%%%%%%%%%%%%%%%%%%%%%%%%%%%%%%%%%%%%%%%%%%
%BoundingBox: 10 10 610 790
\begin{figure*}[t]
\centerline{\psfig{file=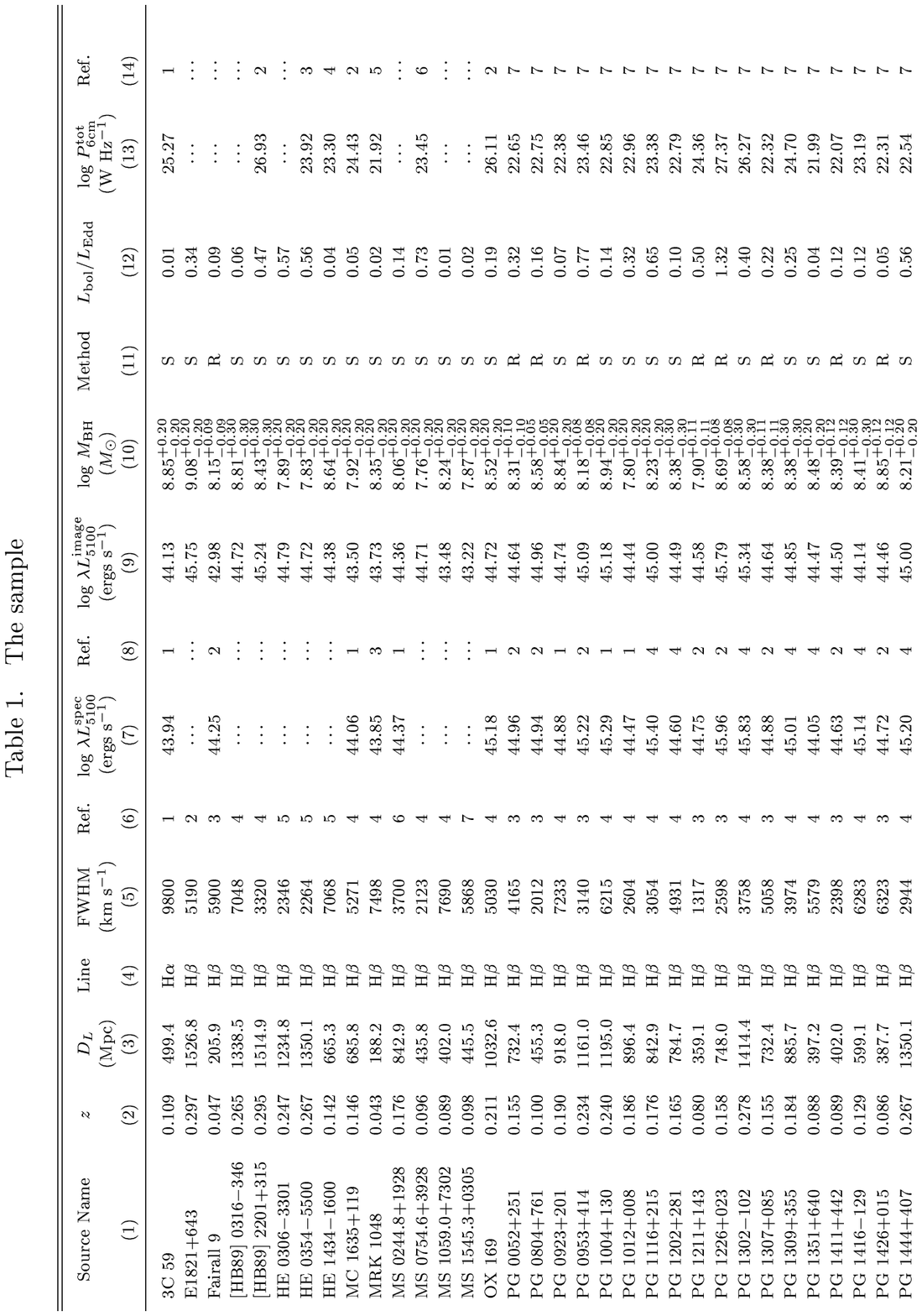,height=0.99\textheight,angle=0}}
\end{figure*}
%%%%%%%%%%%%%%%%%%%%%%%%%%%%%%%%%%%%%%%%%%%%%%%%%%%%%%%%%%%%%%%%%%%%%%%%%%
\clearpage
%%%%%%%%%%%%%%%%%%%%%%%%%%%%%%%%%%%%%%%%%%%%%%%%%%%%%%%%%%%%%%%%%%%%%%%%%
%BoundingBox: 10 10 610 790
\begin{figure*}[t]
\centerline{\psfig{file=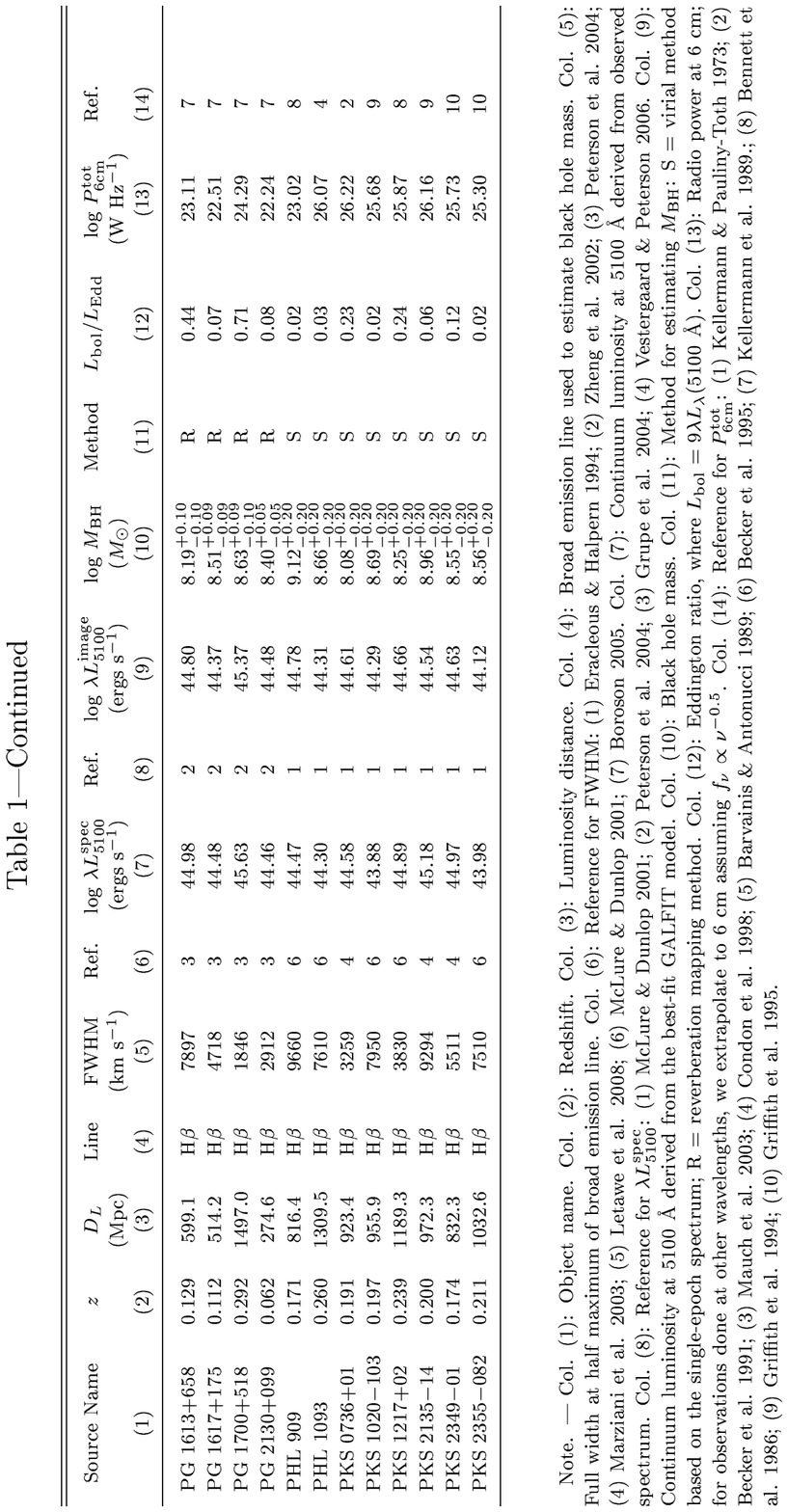,height=0.99\textheight,angle=0}}
\end{figure*}
%%%%%%%%%%%%%%%%%%%%%%%%%%%%%%%%%%%%%%%%%%%%%%%%%%%%%%%%%%%%%%%%%%%%%%%%%%
\clearpage
%%%%%%%%%%%%%%%%%%%%%%%%%%%%%%%%%%%%%%%%%%%%%%%%%%%%%%%%%%%%%%%%%%%%%%%%%
%BoundingBox: 10 10 610 790
\begin{figure*}[t]
\centerline{\psfig{file=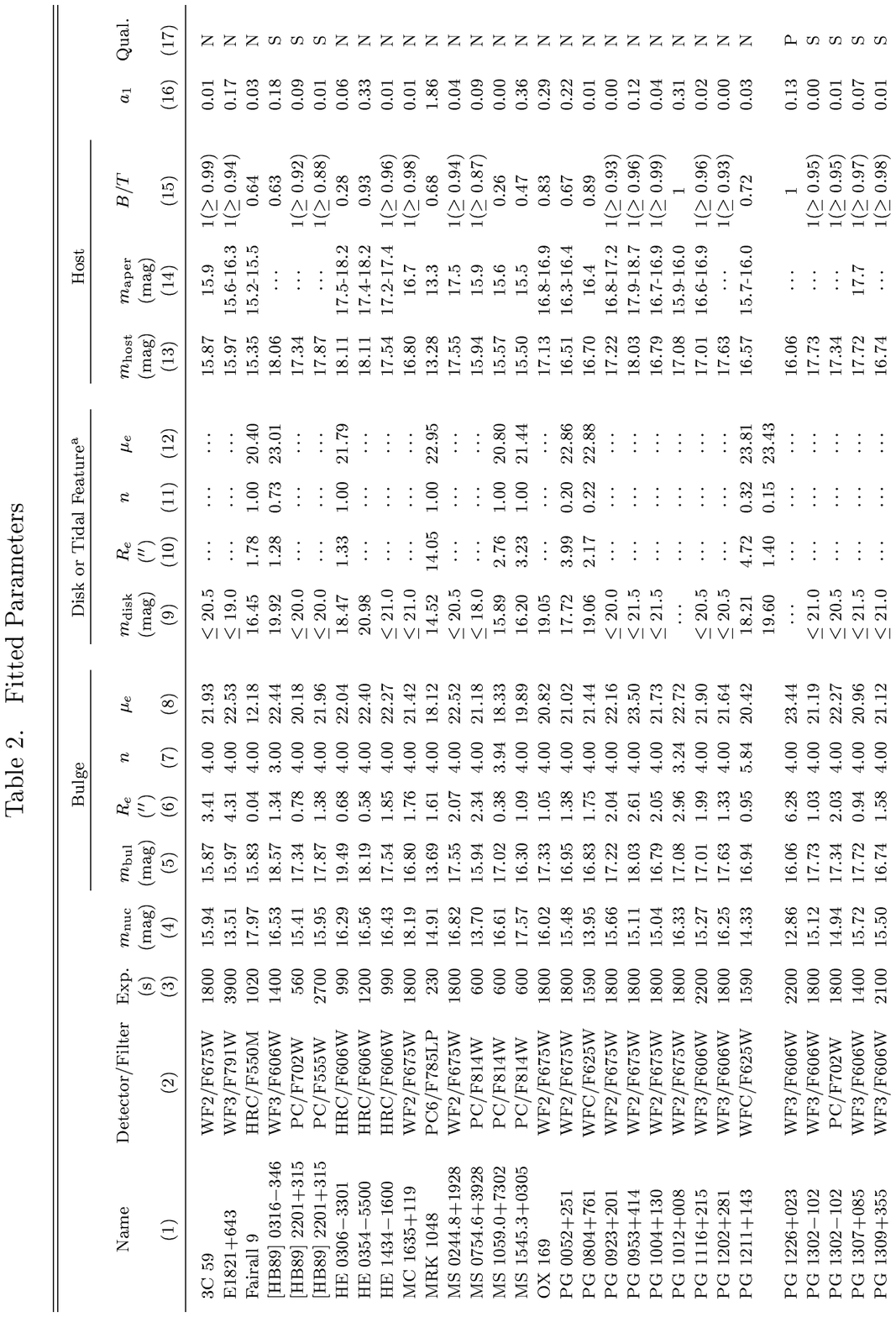,height=0.99\textheight,angle=0}}
\end{figure*}
%%%%%%%%%%%%%%%%%%%%%%%%%%%%%%%%%%%%%%%%%%%%%%%%%%%%%%%%%%%%%%%%%%%%%%%%%%
\clearpage
%%%%%%%%%%%%%%%%%%%%%%%%%%%%%%%%%%%%%%%%%%%%%%%%%%%%%%%%%%%%%%%%%%%%%%%%%
%BoundingBox: 10 10 610 790
\begin{figure*}[t]
\centerline{\psfig{file=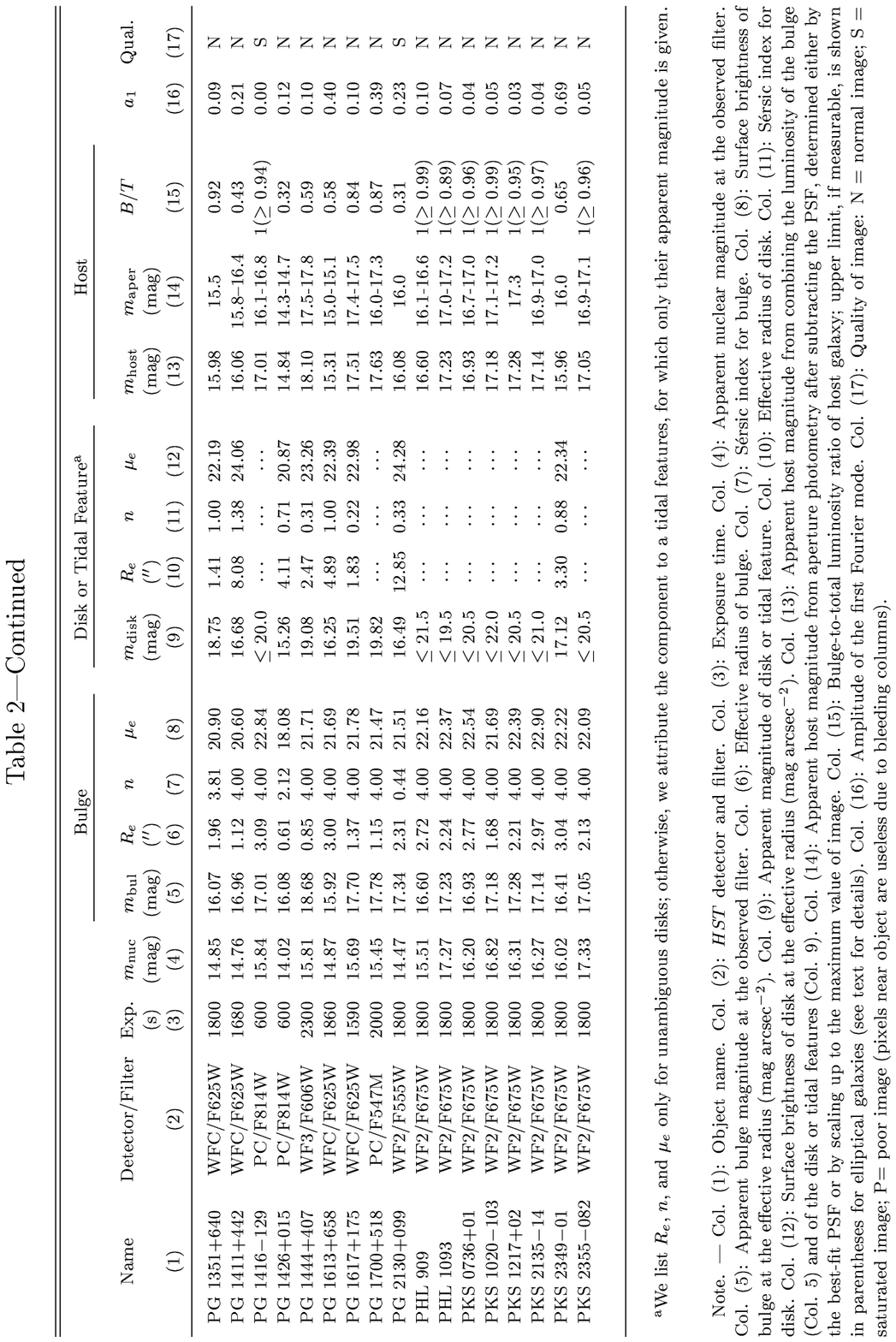,height=0.99\textheight,angle=0}}
\end{figure*}
%%%%%%%%%%%%%%%%%%%%%%%%%%%%%%%%%%%%%%%%%%%%%%%%%%%%%%%%%%%%%%%%%%%%%%%%%%
\clearpage
%%%%%%%%%%%%%%%%%%%%%%%%%%%%%%%%%%%%%%%%%%%%%%%%%
\psfig{file=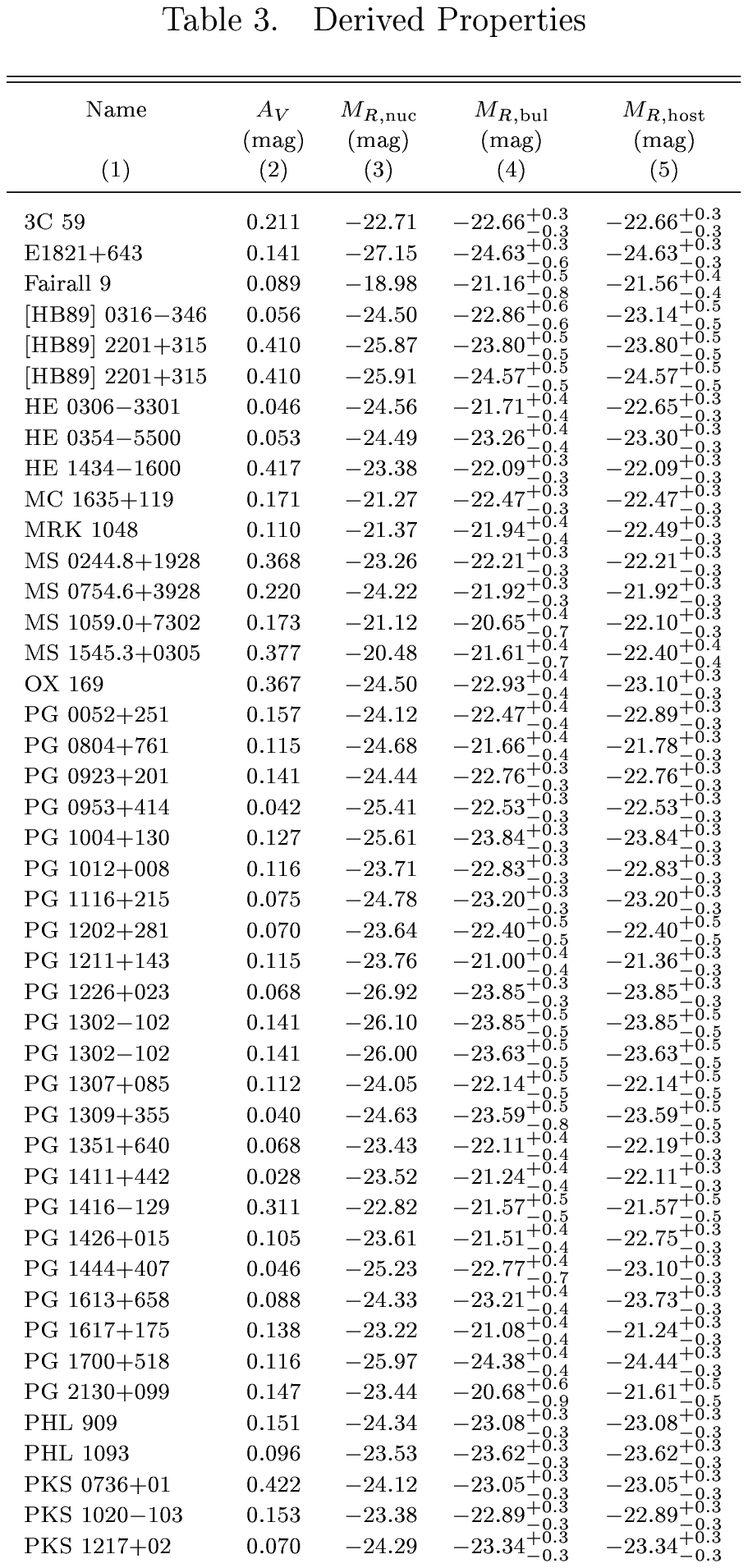,height=9.5in,angle=0}
%%%%%%%%%%%%%%%%%%%%%%%%%%%%%%%%%%%%%%%%%%%%%%%%%%
\clearpage
%%%%%%%%%%%%%%%%%%%%%%%%%%%%%%%%%%%%%%%%%%%%%%%%%
%\hskip 0.0truein
\psfig{file=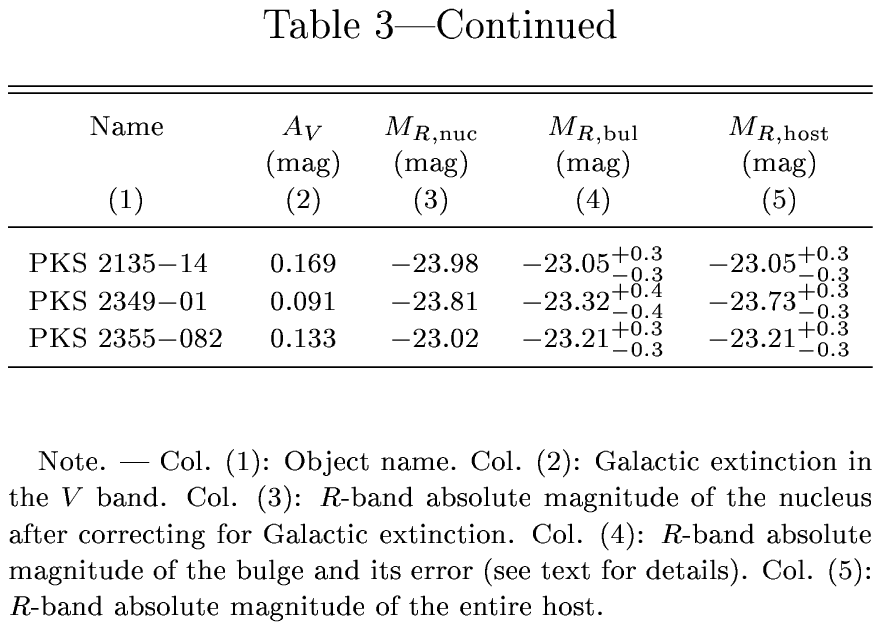,height=9.5in,angle=0}
%%%%%%%%%%%%%%%%%%%%%%%%%%%%%%%%%%%%%%%%%%%%%%%%%%
\clearpage
%%%%%%%%%%%%%%%%%%%%%%%%%%%%%%%%%%%%%%%%%%%%%%%%%
\psfig{file=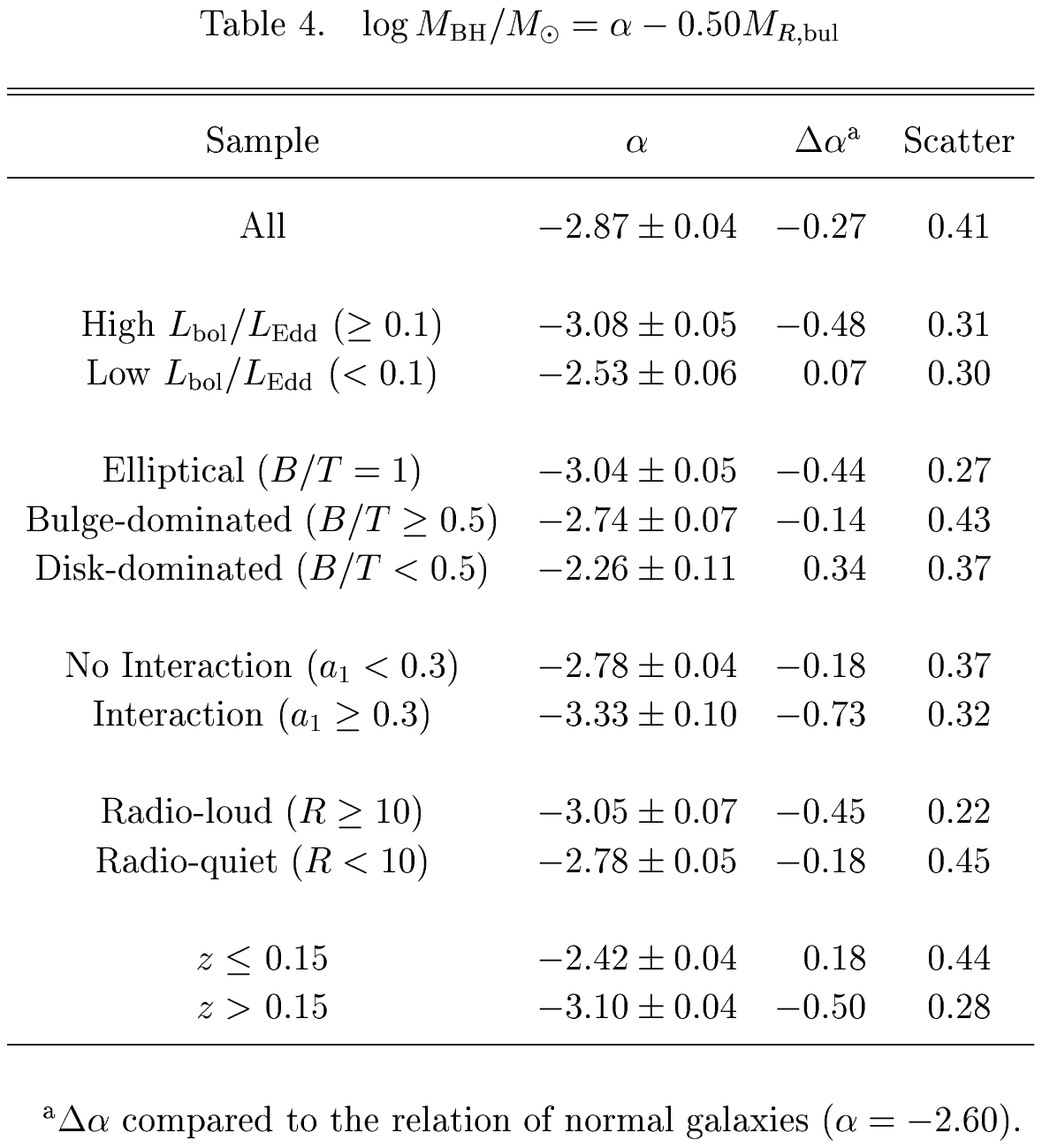,height=9.5in,angle=0}
%%%%%%%%%%%%%%%%%%%%%%%%%%%%%%%%%%%%%%%%%%%%%%%%%%
\clearpage
%%%%%%%%%%%%%%%%%%%%%%%%%%%%%%%%%%%%%%%%%%%%%%%%%
\psfig{file=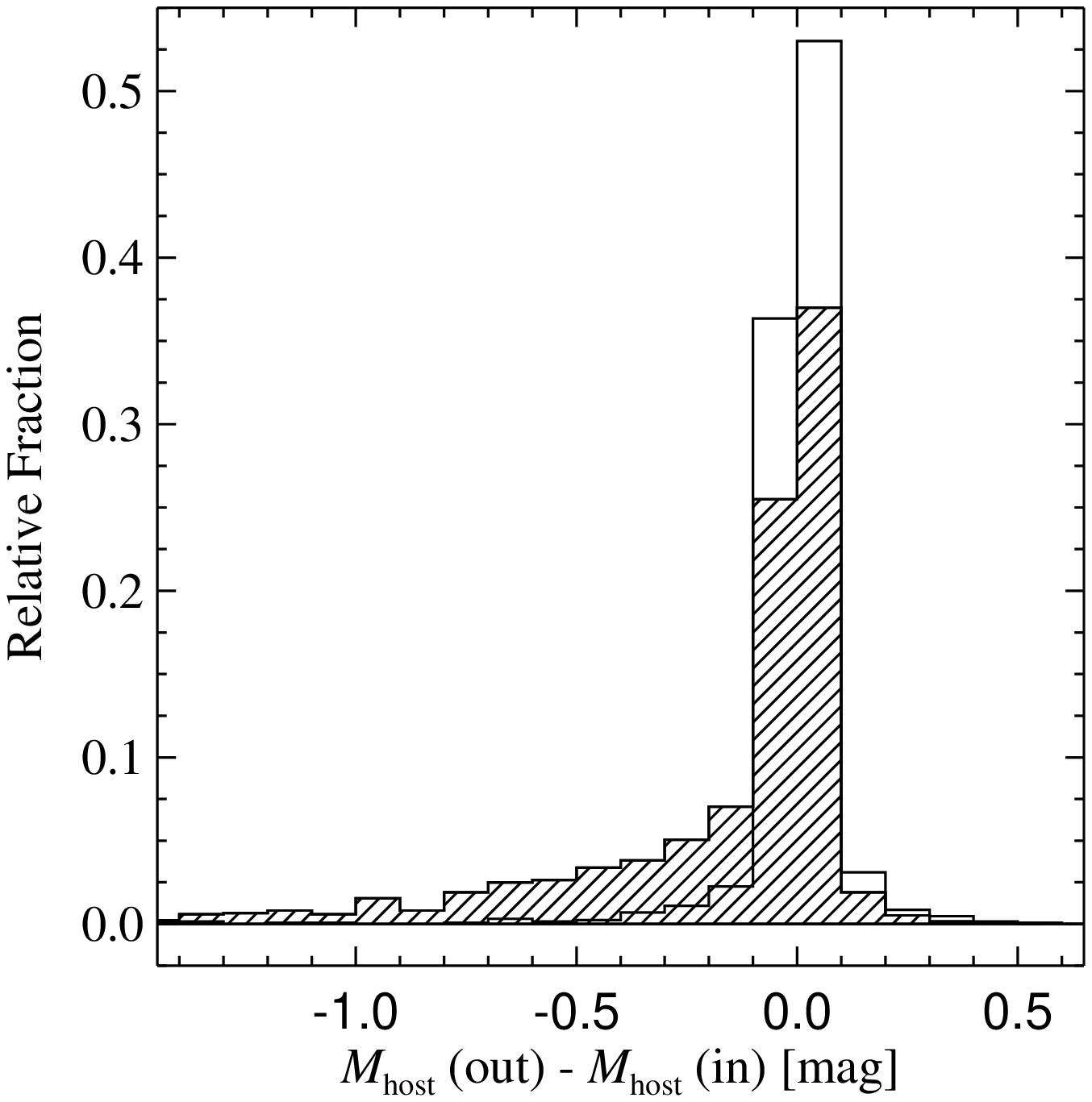,width=8.5cm,angle=0}
\figcaption[fig3.ps]
{Distribution of the differences in host galaxy magnitudes from the 2-D
imaging-fitting simulations of Kim et al. (2008).  Artificial images of AGN
host galaxies with a range of input parameters were generated, and GALFIT was
used to recover the input parameters.  The open histograms denote the errors
for idealized conditions in which the fitting was done with the same PSF as
used for making the input images.  The hatched histograms give the errors
for the more realistic situation that accounts for PSF mismatch (see Kim et
al. 2008 for details).  PSF mismatch causes significant systematic errors.
\label{fig3}}
\vskip 0.3cm
%%%%%%%%%%%%%%%%%%%%%%%%%%%%%%%%%%%%%%%%%%%%%%%%%%
%
\noindent
arbitrary, but it yields a sizable sample of 45 objects. Our objects
(Table 1) span $\sim 2$ orders of magnitude in accretion rate (Eddington 
ratio) over a relatively narrow range ($\sim 1$ dex) in \mbh\ (Fig.~1) and 
redshift (Fig.~2).  From radio data assembled from the literature, 
33\% (15/45) of the sample 
is radio-loud, defined by $R \geq 10$, where
$R \equiv f_\nu({\rm 6~cm})/f_\nu(4400$~\AA) (Kellermann et al. 1989).

%%%%%%%%%%%%%%%%%%%%%%%%%%%%%%%%%%%%%%%%%%%%%%%%%
%%BoundingBox: 0 27 595 347
\begin{figure*}
\psfig{file=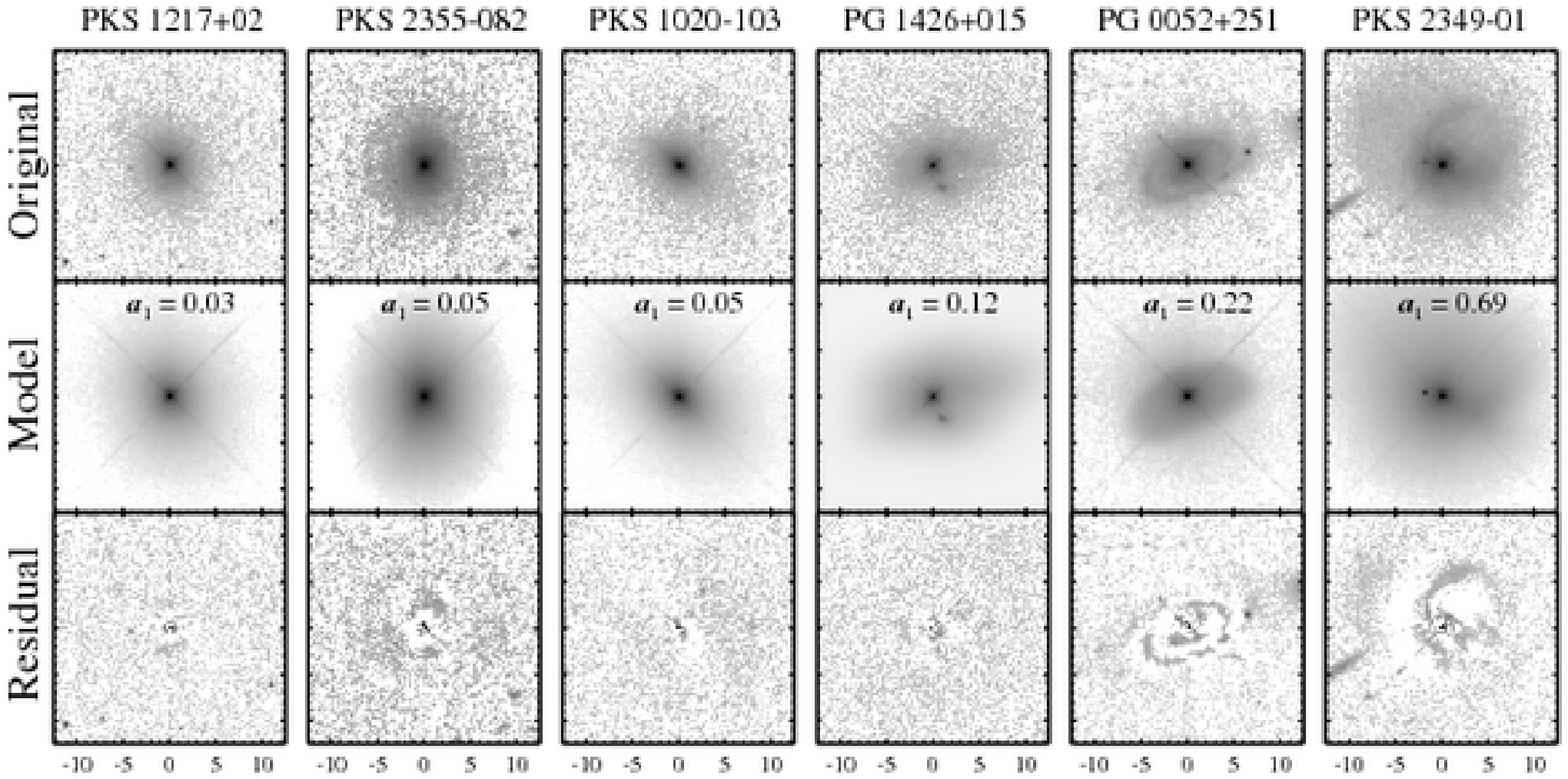,width=18.5cm,angle=0}
\figurenum{4}
\figcaption[fig4.ps]
{Examples of non-interacting and interacting objects, based on the strength
of the first Fourier mode ($a_1$) measured with GALFIT.  From left to right, the
six objects are arranged in the order of increasing $a_1$.  The first three, with
$a_1$ \lax\ 0.1, show essentially no signs of perturbation.  The last three, with
$a_1$ \gax\ 0.1, are increasingly disturbed morphologically.  In each column, we
show, from top to bottom, the original image, model, and residual. The units
of the images are in arcseconds.
\label{fig4}}
\end{figure*}
%%%%%%%%%%%%%%%%%%%%%%%%%%%%%%%%%%%%%%%%%%%%%%%%%%
%\clearpage

\section{Analysis}

\subsection{Structural Decomposition}

Our companion paper (Kim et al. 2008) discusses in detail our technique for 
decomposing the \hst\ images of AGN host galaxies.  We performed extensive 
simulations to quantify the performance of the two-dimensional (2-D) 
image-fitting code GALFIT (Peng et al. 2002) under conditions typically 
encountered in AGN host galaxy images contained in the \hst\ archive, similar 
to those analyzed in this study.  We paid particular attention to quantifying 
systematic uncertainties in estimating the photometric parameters of the 
bulge component and devised strategies for mitigating them.

In the regime where the bright, unresolved active nucleus dominates over the 
light of the host galaxy, our ability to extract the bulge luminosity 
depends sensitively on the properties of the point-spread function (PSF).
PSF mismatch systematically biases the derived bulge luminosities to high 
values, by as much as 0.5--1 mag (Fig.~3).  PSF mismatch occurs as a result of 
variations in time, location on the detector, and, to a lesser extent, 
differences in the spectral energy distribution between the PSF star and the
science target.  The dominant effect, however, comes from the fact that 
{\it HST}\ PSFs are undersampled.  Because the PSFs are not Nyquist-sampled, 
it is impossible to shift by a sub-pixel unit to perfectly align the PSF star 
with the AGN core.  Kim et al. (2008) show that this fundamental problem 
outweighs most other concerns, including the choice of actually observed 
(stellar) or synthetic (TinyTim; Krist 1995) PSFs.  They demonstrate 
that the undersampling problem can be significantly alleviated by broadening 
both the science image and the PSF image to critical sampling [full width at 
half maximum (FWHM) $\sim 2$ pixels].  This is the strategy we adopt here. 
Although the choice of real versus synthetic PSFs is secondary, 
we use observed PSF stars whenever possible.  When these are 
not available, we use TinyTim PSFs.

As in our companion paper, we use an updated version of GALFIT 
(C.~Y. Peng et al., in preparation)\footnote{{\tt http://www.ociw.edu/$\sim$peng/work/galfit/newfeatures.html}}.  The code simultaneously fits multiple 
components to model the host galaxy, with the freedom to use Fourier modes to 
accommodate complex, nonaxisymmetric features such as tidal distortions or even 
spiral arms.  These improvements allow us to obtain a more accurate 
decomposition of the structural components of the host, an important 
consideration for our aim of deriving robust bulge luminosities.

We model the active nucleus with a synthetic PSF and the host galaxy with 
ellipses represented by a \ser\ (1968) function:

\begin{equation}
I(r) = I_e~{\rm exp} \left[ -b_n \left(\frac{r}{r_e}\right)^{1/n}-1 \right],
\end{equation}
\noindent
where $r_e$ is the effective radius, $I_e$ is the
intensity at $r_e$, $n$ is the \ser\ index, and $b_n$ satisfies

\begin{equation}
\int_0^{\infty}I(r) 2 \pi r dr = 2 \int_0^{r_e} I(r) 2 \pi r dr.
\end{equation}

\noindent
The \ser\ function reduces to an exponential profile for $n=1$ and a 
de~Vaucouleurs (1948) profile for $n=4$.

Several different fits are done for each object.  We determine the sky value 
from a growth curve analysis of the image.  We begin with the simplest 
possible option of modeling the host with a single \ser\ component, allowing 
the index $n$ to be free as well as fixing it to $n=1$ and $n=4$. Our 
simulations show that when the AGN is much brighter than the bulge allowing 
$n$ to be free can sometimes lead to spurious results.  In these situations, 
it is better to fix $n$ to specific values in the fit and then empirically 
bracket the resulting host luminosity from the range of acceptable models, 
as judged from the relative change of $\chi^2$ and visual examination of 
the model residuals.  With the \ser\ index fixed, the program solves for the 
following free parameters: for the host galaxy, these are the position, 
effective radius $R_e$, surface brightness $\mu_e$ at $R_e$, axis ratio, and 
position angle; for the AGN, these are the position and nuclear magnitude, 
$m_{\rm nuc}$.  Figure~56 gives an example of a single-component fit for 
PKS 2355$-$082.

We perform two-component (bulge+disk) fits to objects that clearly have 
a disk component in the original image or that show significant extended 
structure in 
the residual image after subtraction of a single-component model.  We try 
three cases: (1) $n=4$ for bulge and $n=1$ for disk; (2) $n$ = free for bulge 
and $n=1$ for disk; and (3) $n$ = free for bulge and $n$ = free for disk. 
We permit the bulge to have $n < 4$ to allow for the possibility of 
pseudobulges (Kormendy \& Kennicutt 2004), and the disk can deviate from a 
pure exponential profile if it is severely distorted, if it has bar or 
ring-like structures, or if it represents tidal features and arcs.  
In most cases when a disk is present, it tends to have a smaller \ser\ $n$, 
a larger $R_e$, and often a smaller axial ratio than the bulge.
The disk component adds four additional free 
parameters, namely position, $R_e$, $\mu_e$, axis ratio, and  position angle.
Figure~24 gives an example of a bulge+disk fit for MS 1059.0+7302.

The choice between a single-component and a double-component fit for the host 
galaxy is not always clear.  Adding 
extra free parameters obviously yields a 
better fit.  The question is whether the extra component is clearly required 
and physically meaningful.  For example, when PSF mismatch is 
particularly 
severe, the extra component might simply be attempting to account for the 
large residuals from the poor PSF model.  In such situations the 
``extra'' component tends to have 
unusual properties such as suspiciously 
tiny $R_e$ or extreme values of $n$.  In practice, complicated fits often 
require some degree of judgment call, but we have tried to err on the side of 
caution and generally invoke extra components only when they are absolutely 
needed.  In most cases, we admit an additional component only if it is 
clearly visible in the original image or in the residual image.

A disk component may be present but undetectable in shallow images (e.g., 
Bennert et al. 2008). It is thus very useful to place an upper limit on the 
disk component even if no disk is required by the best-fit model of the \hst\ 
image.  For objects with no directly detected disks, we derive upper limits 
for the disk component by placing artificial (face-on) disks covering a wide 
range of luminosity on the science image, assuming that the host galaxy 
follows the relation between bulge-to-total light ratio ($B/T$) and the ratio 
of disk size to bulge effective radius derived from nearby early-type galaxies 
(de~Jong et al. 2004).  We then fit the simulated images with two-component 
(bulge+disk) models.  The disk luminosity at which the program fails to 
recover the input value gives an estimate of the upper limit for the disk 
component.  We did not attempt to derive disk upper limits for systems that 
are exceptionally complicated (e.g., highly distorted, close companions, etc.).

The updated version of GALFIT also has the ability to model spiral
arms in the disk.  The spiral structure is created by a hyperbolic tangent 
rotation function with the following parameters: bar length, outer spiral 
radius, rotation rate, sky inclination, and position angle.  The details of 
the spiral structures are created by high-order Fourier modes.  
Figure~41 shows the fit for PG 1411+442, whose disk component shows two 
prominent spiral arms.

A unique aspect of our analysis is that we attempt to quantitatively 
estimate the degree to which the host galaxy exhibits nonaxisymmetric 
distortions.  Morphological disturbances may be signatures of recent mergers 
or tidal interactions, which might trigger or enhance AGN fueling.
While a variety of techniques have been devised to characterize morphological 
asymmetry in inactive galaxies (e.g., Conselice et al. 2000; Lotz et al.
2004), they cannot be readily extended to galaxies containing bright AGNs
because the central point source can dominate the asymmetry signal.  The 
latest version of GALFIT implements 
asymmetry parameters as an integral part of the image-fitting process.  This 
is accomplished by introducing higher-order Fourier modes to change the shape 
of the host galaxy from axisymmetric ellipses into more complicated shapes.  
All the while the light profile of the host galaxy model would still decline 
as a \ser\ profile in every direction from the peak.  In this scheme, the 
strength of an external perturbation on the host galaxy would sensitively 
register as high-amplitude Fourier modes, with phase angles that reflect the 
direction of the perturbation.

If the residual image shows significant nonaxisymmetry, we adopted a Fourier 
component to fit it.  The Fourier mode has the following form:

\begin{equation}
r(x,y)=r_0(x,y) \left[1+\sum^{N}_{m=1}a_m {\rm cos}(m(\theta+\pi_m))\right].
\end{equation}

\noindent
In this expression, $\theta= {\rm arctan}[(y-y_c)/(x-x_c)q]$, where 
($x_c,y_c$) is the centroid of the ellipse, $q$ is the axis ratio, $r_0(x,y)$ 
is the generalized ellipse, $a_m$ is the amplitude for mode $m$, and $\pi_m$ 
is the phase angle for mode $m$.  The Fourier mode is always coupled with a 
general single (e.g., \ser) component and shows how much a component is 
perturbed from the perfect ellipsoid. Thus, the Fourier mode allows us to
{\it quantify}\ the degree of asymmetry.  In principle, we can use an infinite 
number of Fourier modes, but in practice we find that four modes ($m=1,3,4,5$) 
are enough to fit the asymmetrical structures encountered in our sample. 
Figure~4 illustrates a series of objects with increasing strengths in 
$a_1$.  Sources with $a_1$ \lax\ 0.1 show little to no obvious signs of 
morphological perturbation, whereas those with $a_1$ \gax\ 0.1 appear 
increasingly disturbed.

Despite the significant new features of the updated version of GALFIT, we 
note that the derived values of many standard parameters (e.g., size and 
luminosity) are not substantially different between the original and new 
versions of the code.  In situations where there are differences,
the new features of the code allow for better convergence, especially when it 
comes to multi-component decompositions.  The updated version of GALFIT 
has been tested extensively by us, but for the sake of brevity we defer 
a full discussion of the technical details to an upcoming paper 
(C. Y. Peng et al., in preparation).

Table 2 summarizes the results of the structural decomposition.  For each 
object, we list the best-fit nuclear and photometric parameters for the 
bulge, disk or tidal feature, and for the overall host galaxy; the parameter 
$a_1$ is also tabulated.

\subsection{Bulge Luminosities}

The error bars on the bulge measurements are influenced by a number of 
systematic uncertainties.  Using the simulations in Kim et al. (2008) 
as a guide, the final error budget on the bulge luminosity was estimated 
as follows.  For sources with \bnr\ $\geq 0.2$, $\sigma \approx 
\pm 0.3\,{\rm mag}$, whereas $\sigma \approx \pm 0.4\,{\rm mag}$ if \bnr\ 
$< 0.2$.  On top of these values, additional uncertainties are introduced if 
bulge-to-disk decomposition is required ($\sim 0.1\,{\rm mag}$), if saturation 
occurs ($\sim 0.2\,{\rm mag}$), or if the image contains substantial inner 
fine structure ($\sim 0.3\,{\rm mag}$). 
We also determine the error of the host luminosity ($m_{\rm host}$) 
from the simulations at a given \hnr. According to Kim et al., the error on 
the nucleus magnitude is $\sim \pm 0.1$ mag.

Because of the complexity of the GALFIT decomposition, it is worthwhile to 
cross-check our 2-D parametric fits with a nonparametric estimate of the 
host luminosity (see, e.g., Greene et al. 2008).  To perform this test, we 
remove the nucleus from each source simply by subtracting a shifted, scaled 
PSF model from the peak of the AGN core.  After masking out obvious companions 
and foreground stars, we sum up the remaining flux to estimate the total host 
galaxy magnitude.  As a separate test, we compute the host galaxy flux after 
subtracting the PSF component derived from the best-fitting GALFIT model 
for the entire image.  These two tests give the range of values tabulated 
as $m_{\rm aper}$ in Table~2.  Comparison of these estimates with the host 
magnitudes obtained from the parametric fits ($m_{\rm host}$) shows reasonably 
good agreement for the majority of the sources.  The few cases in 
which $m_{\rm aper}$ is substantially brighter than $m_{\rm host}$ can be 
attributed to large residuals from PSF mismatch and contamination from 
neighboring sources.
 
A significant number of the sources in our sample overlap with those studied 
by Dunlop et al. (2003), affording an independent, external check of our 
analysis.  Dunlop et al. also performed 2-D decomposition of their sample, but 
they fitted the host galaxies with only a single component, modeled as either 
a classical de~Vaucouleurs ($n=4$) bulge or an exponential ($n=1$) disk.  
After accounting for differences in the adopted cosmological parameters, we 
find, not surprisingly, that for bulge-dominated sources our bulge magnitudes 
generally agree well (to within 0.1 mag) with those given by Dunlop et al. 
The exceptions are objects with large, nearby neighbors and sources with 
multiple components.  Whereas we perform 
a simultaneous 2-D fit of all nearby sources that could potentially affect 
our target of interest, Dunlop et al. simply masked them out.  This 
could lead to systematic errors in the derived properties of the primary host.

A particularly striking example is PG~1012+008, 
which is an obviously interacting system consisting of three galaxies.  
Simultaneously accounting for the subcomponents, including an off-centered 
disk, our best-fit model yields a 
bulge with $m_{\rm bul}$ = 17.1 (F675W) and an effective radius of 
$R_e$ = 2\farcs 96, or 9.3 kpc.  By contrast, Dunlop et al. obtain 
$m_{\rm bul}$ = 16.4 mag and $R_e$ = 5\farcs75, which corresponds to 
18 kpc using our assumed distance of 896 Mpc.  For sources that clearly 
contain both a bulge and a disk (e.g., PKS~2349$-$01), our two-component fits 
yield more robust bulge luminosities.  Lastly, for completeness, 
we note that our final nuclear magnitudes (Table~3), converted to the $R$ 
band, are systematically brighter by 0.4 mag compared to those given in Dunlop 
et al.  This difference can be traced to the different assumptions used 
for calculating the $k$-correction.  Dunlop et al. assumed that the AGN 
spectrum can be represented by a single power law ($f_\nu \propto \nu^{-2}$), 
whereas we use the empirical quasar composite spectrum of Vanden Berk et al. 
(2001).

Five of our objects (MS 0754.6+3928, MS 1059.0+7302, MS 1545.3+0305, 
PG 1416$-$129, and PG 1426+015) overlap with the sample studied by Schade et 
al. (2000), who also performed 2-D fits to derive photometric parameters for 
the host galaxies.  The two studies show significant difference in the sense 
that Schade et al. tend to underestimate the nuclear magnitudes on average 
by 0.2 mag and to overestimate the bulge magnitudes by $\sim$0.7 mag. Some
objects show particularly striking disagreement.  In our analysis, the host 
galaxy of MS 1059.0+7302 is well described by a 
bulge+disk model.  Our best fit yields the $m_{\rm bul} = 17.02$ mag and 
$R_e$ = 0\farcs38 for the bulge and $m_{\rm disk} = 15.89$ mag and 
$R_e$ = 2\farcs76 for the disk.  By contrast, Schade et al. find 
$m_{\rm bul} = 15.53$ mag and $R_e$ = 3\farcs47 for the bulge and 
$m_{\rm disk} = 16.68$ mag and $R_e$ = 1\farcs28 for the disk.
We attribute the discrepancy between our results and those of Schade et al. 
to a difference in methodology.  Schade et al. fitted the \hst\ images 
simultaneously with ground-based images.  Although the ground-based images 
are deeper, they have a much broader and less stable PSF than the \hst\ 
images; it is difficult to know how this effect impacts the fitting results.
Other differences stem from the model adopted in the fit.  In our work,  
PG 1426+015 is best fit 
%%%%%%%%%%%%%%%%%%%%%%%%%%%%%%%%%%%%%%%%%%%%%%%%%
%%BoundingBox: 0 375 595 652
\psfig{file=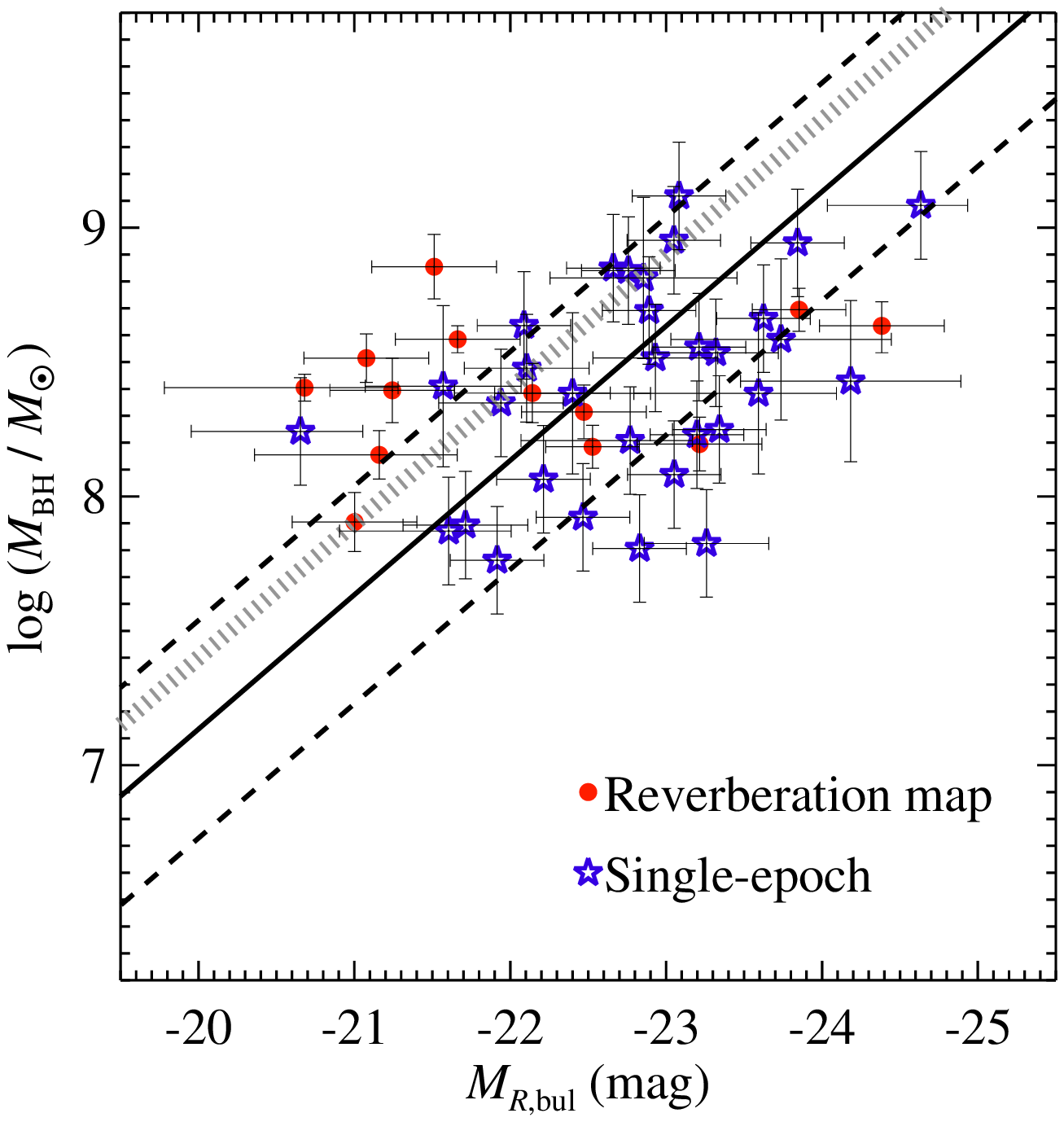,width=8.5cm,angle=0}
\figurenum{5}
\figcaption[fig5.ps]
{Correlation between BH mass and absolute $R$-band magnitude for the bulge.  
The BH masses are derived from reverberation mapping ({\it red circles}) or 
from single-epoch spectra 
({\it blue stars}), and the bulge luminosities are based on
the GALFIT decomposition.  The values of \mbh\ estimated
from single-epoch spectra are assumed to have a systematic uncertainty of
0.2 dex.  The best fit is plotted as a solid line, and the intrinsic scatter is 
denoted by the dashed lines. 
The \mlb\ relation for inactive galaxies is shown by the thick hatched line.
\label{fig5}}
\vskip 0.3cm
%%%%%%%%%%%%%%%%%%%%%%%%%%%%%%%%%%%%%%%%%%%%%%%%%%
\noindent
with a two-component bulge+disk model with a 
significant $a_1$ Fourier mode, whereas Schade et al. employed only a 
single-component bulge for the host. If we adopt a single-component model, 
our results agree well with those of Schade et al.

The bulge magnitudes listed in Table~2 were derived from images taken in 
different filters.  For our subsequent analysis, we need to convert the 
magnitudes to a single standard bandpass at $z$ = 0.  For ease of comparison
with the \mlb\ relation of inactive galaxies (Bettoni et al. 2003), we
choose the $R$ band as the reference.  
We perform the color conversion of the observed magnitude in the various
\hst\ filters to the $R$ band and 
apply $k$-correction using galaxy template spectra
from Calzetti et al. (1994) and Kinney et al. (1996). We assume that the bulge 
component has the spectrum of an elliptical galaxy and 
that the disk 
component is approximated by a late-type (Sc) galaxy. For the images taken in 
the F814W filter, we employ the template spectrum of a starburst galaxy for the 
disk component because the template spectrum of a late-type galaxy is 
unavailable in this wavelength regime.
The final $R$-band absolute magnitudes are given in Table~3.

\subsection{Black Hole Masses}

The BH masses for type~1 AGNs can be estimated from the virial product \mbh\
$\approx f R v^2/G$, where $R$ is the radius of the broad-line region (BLR),
$v$ is the line width of the BLR gas represented by FWHM$_{\rm H\beta}$, the 
FWHM of the broad H$\beta$ line, and $f$ is a factor of order unity that 
depends on the structure, dynamics, and inclination angle of the BLR.  
Direct measurements of $R$ through reverberation mapping are available only 
for a small number of sources (Peterson et al. 2004), but fortunately this 
quantity can be estimated from the correlation between $R$ and luminosity 
(Kaspi et al. 2000, 2005).  The virial product is, however, uncertain by the 
normalizing factor $f$.  Assuming that the BLR is spherical and has an 
isotropic velocity field, $f = 0.75$, and the latest radius-luminosity 
relation from Bentz et al. (2006) yields,

\begin{equation}
M_{\rm BH} = 5.5 \times 10^6 \, M_{\odot}\,
\left(\frac{\lambda L_{5100}} {10^{44} \, {\rm ergs\,s^{-1}}}\right)^{0.52} 
\left(\frac{{\rm FWHM}_{\rm H\beta}}{10^3 \, {\rm km \, s^{-1}}}\right)^{2.0},
\end{equation}

\noindent
where $\lambda L_{5100}$ is the continuum luminosity at 5100 \AA\ 
(see Greene \& Ho 2007b for details).  

The continuum luminosity can be estimated either through spectrophotometry or
through our image analysis.  Spectrophotometry has the advantage that the
specific continuum flux at the desired wavelength can be directly measured
without making assumptions about the spectral shape.  On the other hand,
accurate absolute spectrophotometry is nontrivial to achieve and is rarely
available for most objects in the literature.  Moreover, ground-based
apertures invariably blend the nucleus with at least part of the host.  By
contrast, our careful image decomposition yields a clean, unambiguous
measurement of the nuclear continuum.  Our nuclear magnitudes have a typical
uncertainty (dominated by systematic effects from PSF mismatch) of $\sim$0.1
mag.  We need to assume a spectrum (we choose the quasar template from
Vanden~Berk et al. 2001) in order to estimate the continuum luminosity at 5100
\AA, but the amount of extrapolation for our filters is minimal.  A more
significant uncertainty comes from temporal variability between our
photometric measurements and the literature-based spectral observations used
to obtain FWHM$_{\rm H\beta}$.  Nevertheless, AGNs of the luminosity
considered here usually vary by only $\sim 0.13$ mag on long timescales
(e.g., Giveon et al.  1999).  The widths of broad emission lines in type~1 AGNs
typically have an uncertainty of $\sim10\%$ (e.g., Marziani et al. 2003). 
Taking all of these factors into consideration, we estimate that they 
introduce a measurement uncertainty of only $\sim$0.2 dex in \mbh.  The 
largest source of uncertainty for the single-epoch masses, however, probably 
comes from the intrinsic scatter of the radius-luminosity relation, which is 
estimated to be $\sim 0.4$ dex (Bentz et al. 2006).
According to Peterson et al. (2004), the values of \mbh\ derived
from reverberation mapping are accurate to $\sim 30$\%, or 0.1 dex.
We note that there is an additional uncertainty on the geometrical factor
($f$). For instance, Collin et al. (2006) argued that $f$ might be dependent
on the accretion rate. We visit this issue in \S{5.2}.

\section{The \mlb\ Relation for Type 1 AGNs}

Figure 5 shows the \mlb\ relation for our sample of type~1 AGNs. Objects 
with \mbh\ derived from reverberation mapping are encoded separately from 
those based on single-epoch spectra.  We assume that the \mlb\ relation 
follows a log-log relation 

\begin{equation}
\log (M_{\rm BH}/M_\odot) = \alpha + \beta M_{R,{\rm bul}}. 
\end{equation}

\noindent
The thick hatched line represents the \mlb\ relation for inactive galaxies 
from Bettoni et al. (2003). Converted to our cosmology (see Peng et al. 
2006a), the best-fitting relation for inactive galaxies is described 
by $\alpha = -2.6$ and $\beta = -0.5$, with a scatter of 0.4 dex.

We estimate $\alpha$ and $\beta$ for the active sample by minimizing $\chi^2$, 
defined as

\begin{equation}
\chi^2 \equiv \sum^{N}_{i=1} \frac
{(y_i-\alpha-\beta x_i)^2}{\epsilon^2_{yi}+\beta^2\epsilon^2_{xi}},
\end{equation}

\noindent
%%%%%%%%%%%%%%%%%%%%%%%%%%%%%%%%%%%%%%%%%%%%%%%%%
\psfig{file=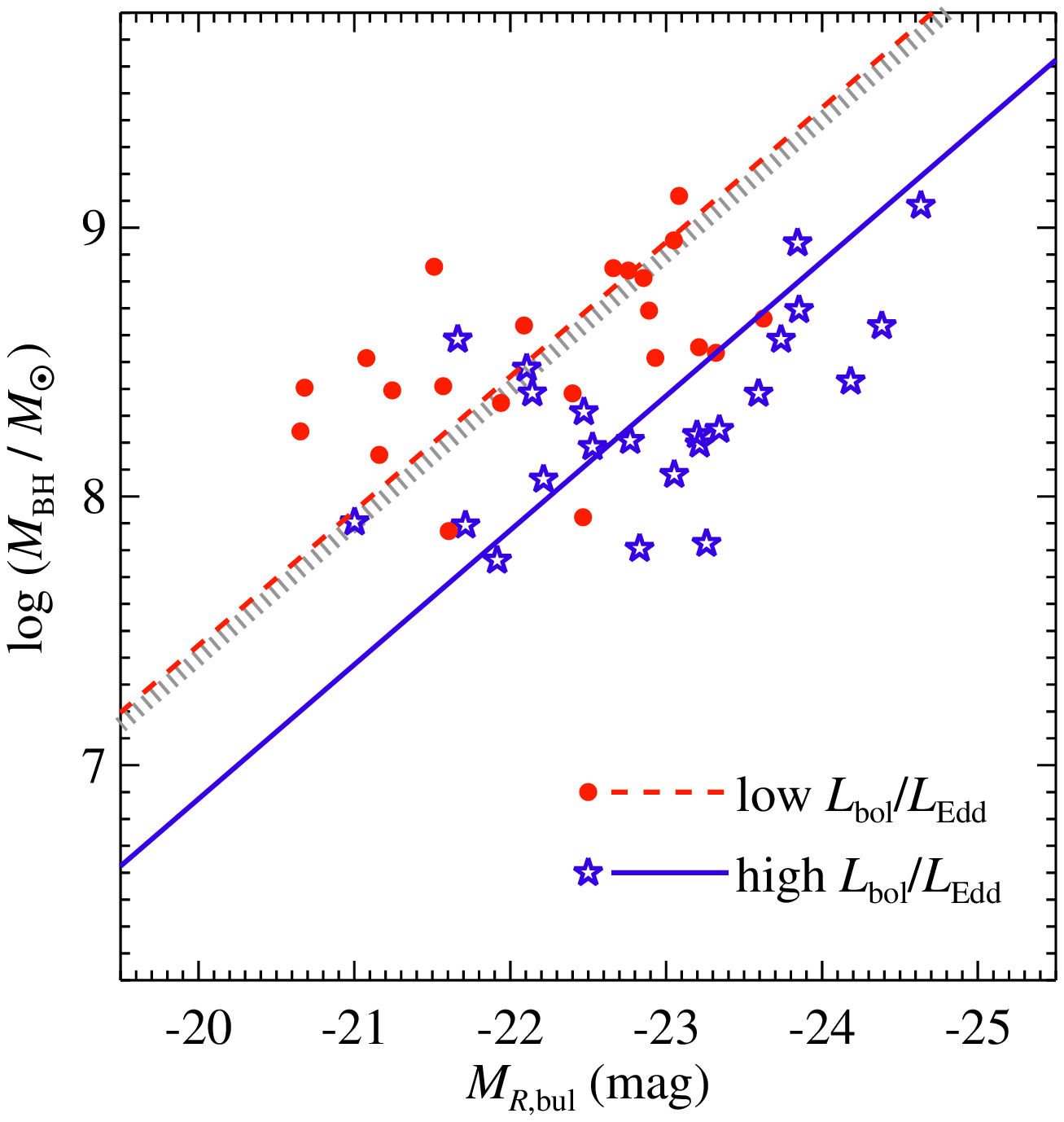,width=8.5cm,angle=0}
\figurenum{6}
\figcaption[fig6.ps]
{Dependence of the \mlb\ relation on Eddington ratio.  Blue stars and
solid line represent the correlation and the best fit for high-Eddington ratio
objects (\edd $\geq 0.1$). Red circles and dashed line represent
low-Eddington ratio objects (\edd $< 0.1$).  We fixed the slope to $-0.5$,
as derived for the inactive galaxies, whose relation is denoted by the thick
hatched line.  Measurement uncertainties of 0.1 and 0.2 dex are
adopted for \mbh\ estimated from reverberation mapping and single-epoch 
spectra, respectively.
\label{fig6}}
\vskip 0.3cm
%%%%%%%%%%%%%%%%%%%%%%%%%%%%%%%%%%%%%%%%%%%%%%%%%%
\noindent
where $y= \log (M_{\rm BH}/M_\odot)$, $x=M_{R,{\rm bul}}$, and $\epsilon_{yi}$ 
and $\epsilon_{xi}$ are the measurement errors of $y$ and $x$, respectively 
(Tremaine et al. 2002).  
This method treats both $x$ and $y$ as independent 
variables and accounts for asymmetric uncertainties for each.

The estimation of the \mlb\ relation depends on $\epsilon_{yi}$, 
the choice of errors for \mbh.  If we adopt an uncertainty of 0.4 dex for the 
masses based on single-epoch spectra, the $\chi^2$ value is practically 
dominated by the 
reverberation-mapped objects because their uncertainties are a factor of 4 
smaller, resulting in a \mlb\ relation strongly biased toward the 
reverberation-mapped subsample.  
For concreteness, we assume that uncertainties on the single-epoch masses
are 0.2 dex, which is a typical measurement error (\S{3.3}).
As Figure 5 shows, our sample 
of AGNs cluster around the fiducial \mlb\ relation of inactive galaxies 
with significant scatter.  The formal fit for the AGNs has 
a slope of $\beta = -0.26\pm0.05$, flatter than for inactive galaxies 
($\beta = -0.5$), 
but because of the limited 
dynamic range in \mbh, we do not regard the AGN fit to be robust.  A more 
meaningful exercise is to fix the slope of the relation to the value 
for inactive galaxies and then examine the offset and scatter of the AGN 
sample.  Fixing $\beta$ to $-0.5$, the AGN sample has 
$\Delta \alpha = -0.3$ and an rms scatter of 0.4 dex.

\subsection{Dependence on Eddington Ratio}

To understand the physical origin of the intrinsic scatter in the \mlb\ 
relation, we divide the sample into two bins in Eddington ratio, at \edd\ = 
0.1.  The Eddington luminosity is defined as $L_{\rm Edd} = 1.26\times 10^{38} 
(M_{\rm BH}/M_\odot)$ \lum, and the bolometric luminosity is estimated assuming
$L_{\rm bol} = 9 \, \lambda L_{5100}$ (Kaspi et al. 2000). 
Figure 6 shows a clear offset between the two subsamples.
At a given \mbh, objects with high Eddington ratios tend to be hosted by more 
luminous bulges, or, alternatively, at a given bulge luminosity they tend to 
have less massive BHs.  In order to quantify the offset between the two 
subsamples, we fix the slope to that of the \mlb\ relation for inactive 
%%%%%%%%%%%%%%%%%%%%%%%%%%%%%%%%%%%%%%%%%%%%%%%%%
\psfig{file=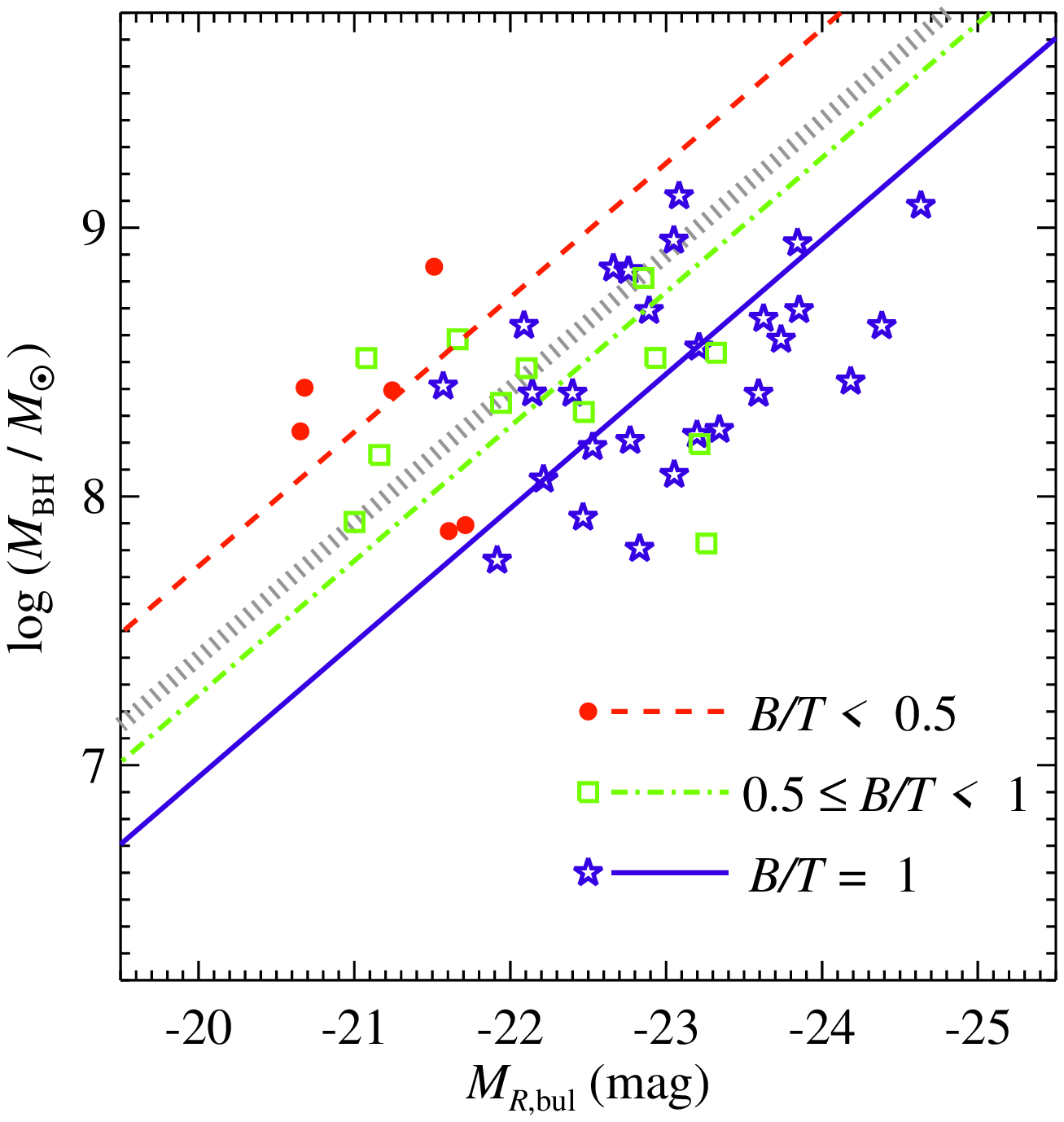,width=8.5cm,angle=0}
\figurenum{7}
\figcaption[fig7.ps]
{Similar to Figure 6, except that here we show the dependence on galaxy
morphology: elliptical galaxies ($B/T = 1$; {\it blue stars and solid line}),
bulge-dominated systems ($0.5 \leq B/T < 1$; {\it green squares and dash-dotted
line}), and disk-dominated systems ($B/T < 0.5$; {\it red circles and 
dashed line}).
\label{fig7}}
\vskip 0.3cm
\noindent
%%%%%%%%%%%%%%%%%%%%%%%%%%%%%%%%%%%%%%%%%%%%%%%%%%
galaxies.  At a fixed \rbulge, the offset in \mbh\ is $\sim
-0.6$ dex; at a 
fixed \mbh, the offset in \rbulge\ is $\sim 1.2$ mag.  We note that these 
offsets are much larger than the measurement errors.  Performing a 
Kolmogorov-Smirnov test to evaluate how \rbulge\ in the two subsamples is 
distributed, we find that the null hypothesis that the two subsamples are 
drawn from the same parent population can be rejected with a probability of 
97.3\%.  As discussed in \S 5.2, the segregation between the two subsamples
really do reflect intrinsic differences in Eddington ratios rather than 
uncertainties in the determination of BH mass.

\subsection{Dependence on Morphological Type}

The availability of robust structural decomposition gives us an opportunity to 
examine possible trends with morphological type.  Using the  measured values 
of bulge-to-total luminosity ratio ($B/T$; Table 2) and the correlation 
between morphological type and $B/T$ in normal, inactive galaxies (Simien \& 
de~Vaucouleurs 1986), we divide the 
sample into three subgroups: $B/T = 1$ (ellipticals), $0.5 \leq B/T < 1$ 
(bulge-dominated), and $B/T < 0.5$ (disk-dominated).  Figure 7 (see also Table 
4) shows that the zero point of the \mlb\ relation, and possibly scatter, may 
depend on $B/T$, although given the limited statistics we regard the evidence 
as tentative.  Ellipticals and early-type, bulge-dominated galaxies appear 
virtually indistinguishable, but later-type, disk-dominated systems 
($B/T < 0.5$) appear distinctly offset to {\it larger}\ \mbh\ (by $\sim 
0.4-0.6$ dex) at a fixed \rbulge.  The magnitude of the offset is much larger 
than possible systematic biases in bulge luminosities resulting from 
uncertainties in bulge-to-disk decomposition ($\sim 0.2$ mag; see Fig. 14 in 
Kim et al. 2008).  As discussed in \S5.1, in many instances 
our two-component fits may not correspond strictly to a bulge+disk 
decomposition but rather to a bulge+tidal feature decomposition. 
%%%%%%%%%%%%%%%%%%%%%%%%%%%%%%%%%%%%%%%%%%%%%%%%%
\psfig{file=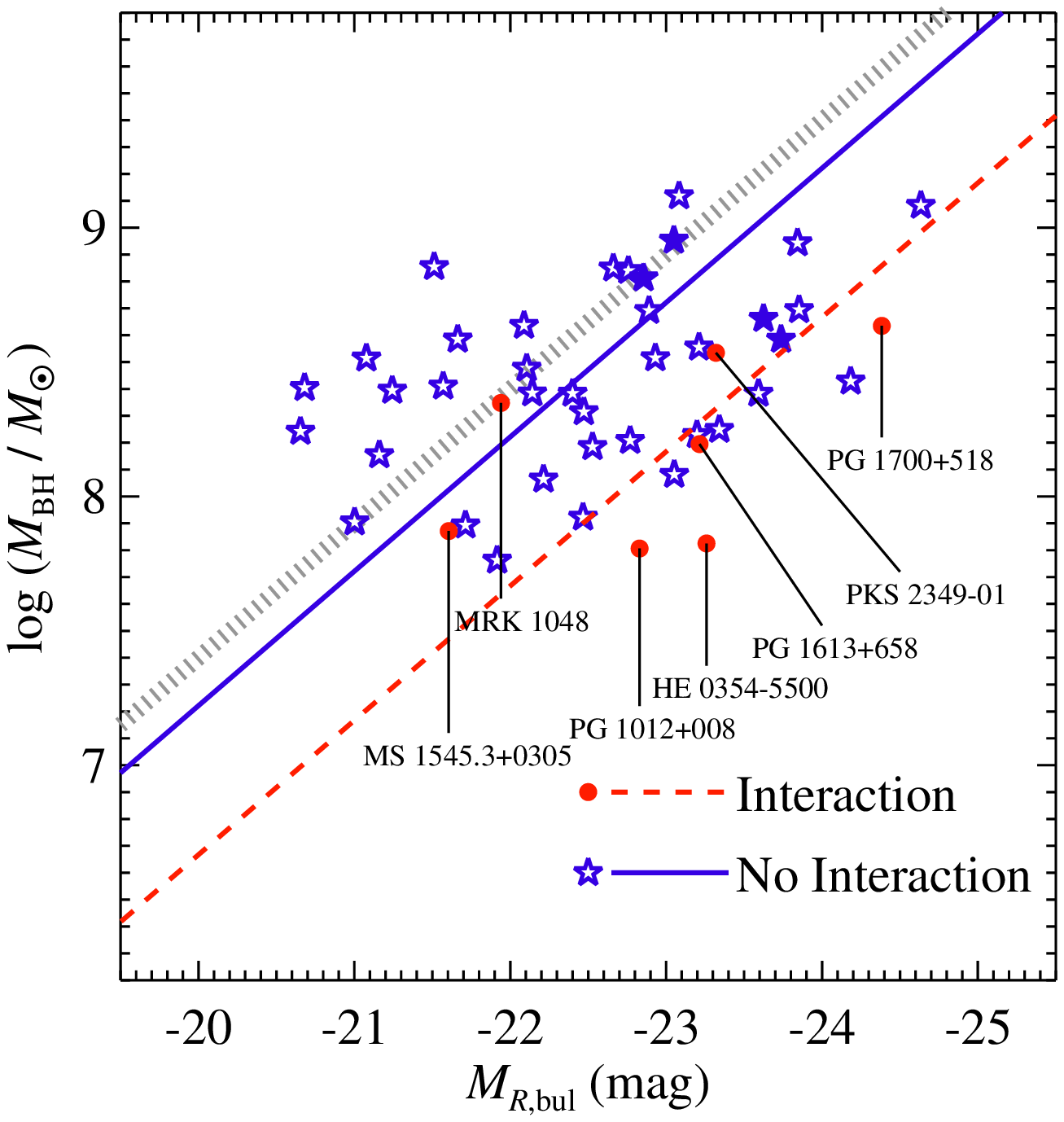,width=8.5cm,angle=0}
\figurenum{8}
\figcaption[fig8.ps]
{Similar to Figure 6, except that here we show the dependence on the degree of
interaction.  Non-interacting or mildly interacting objects ($a_1 < 0.3$) are 
denoted by blue stars and the solid line, while strongly interacting systems 
($a_1 \geq 0.3$), with their names labeled, are denoted by red circles and the 
dashed line.  Four sources with probable minor companions are marked
as filled blue stars.  The \mlb\ relation for inactive galaxies is shown 
by the thick hatched line.
\label{fig8}}
\vskip 0.3cm
%%%%%%%%%%%%%%%%%%%%%%%%%%%%%%%%%%%%%%%%%%%%%%%%%%

\subsection{Dependence on Tidal Interaction}

We make use of the quantitative measure of galaxy asymmetry, $a_1$, to study 
the possible effect of tidal interaction.  Given our small sample size, we 
simply divide it into two bins according to the value of $a_1$.  As shown
in Figure 4, $a_1 \approx 0.1$ seems to provide a useful empirical boundary 
between objects that are disturbed morphologically ($a_1 \geq 0.1$) 
from those that are not ($a_1 < 0.1$).  With this threshold for $a_1$, 
however, the two populations show no obvious segregation in the \mlb\ plane.
But with the boundary set at a higher threshold of $a_1 = 0.3$, Figure 8 
illustrates that four out of the 
six objects in our sample with the clearest signs of morphological disturbance 
do preferentially seem to lie among the most extreme negative outliers in the
\mlb\ relation.

\subsection{Dependence on Radio Properties}

The physical drivers responsible for the generation of jets and radio emission 
in AGNs are still unclear.  Suggestions have included BH mass (e.g., Laor 
2000), accretion rate (e.g., Ho 2002), and host galaxy morphology, which might 
ultimately be linked to BH spin (Sikora et al. 2007).  Figure 9 shows that a 
clear separation exists between radio-loud objects and radio-quiet objects.
Radio-loud sources lie preferentially below the \mlb\ relation of inactive 
galaxies and radio-quiet sources.

\subsection{Dependence on Redshift}

The redshift range of our objects is small ($0 < z < 0.35$), and our sample 
was not designed to test for evolutionary effects.  Nevertheless, even by 
$z = 0.36$ Woo et al. (2006) and Treu et al. (2007) have claimed that AGNs 
already show evidence of evolution in the BH-host galaxy scaling relations.  
Dividing the sources into two bins in redshift (Fig. 10), it appears that the 
two subsamples are offset from each other in 
%%%%%%%%%%%%%%%%%%%%%%%%%%%%%%%%%%%%%%%%%%%%%%%%%
\psfig{file=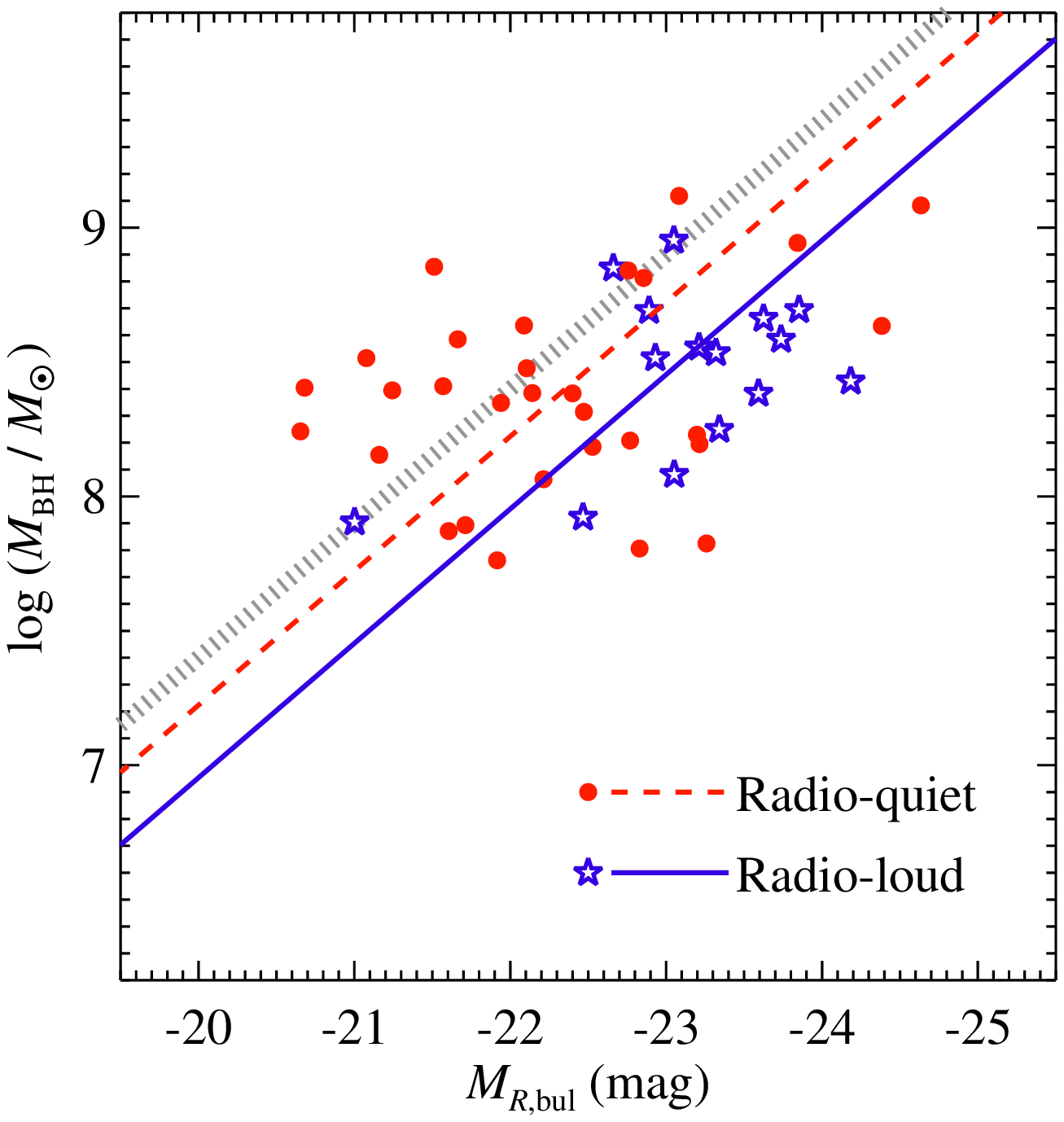,width=8.5cm,angle=0}
\figurenum{9}
\figcaption[fig9.ps]
{Similar to Figure 6, except that here we show the dependence on radio
emission: radio-loud sources are marked with blue stars and solid line, and 
radio-quiet sources are marked with red circles and dotted line.  The 
\mlb\ relation for inactive galaxies is shown by the thick hatched line.
\label{fig9}}
\vskip 0.3cm
%%%%%%%%%%%%%%%%%%%%%%%%%%%%%%%%%%%%%%%%%%%%%%%%%%
\noindent 
the sense that lower-redshift 
sources have a higher \mbh\ at a given \rbulge.   This trend, however, is 
probably a selection effect because low-Eddington ratio, later-type galaxies 
tend to be closer.  Indeed, the $z \leq 0.15$ subsample 
has $\langle L_{\rm bol}/L_{\rm Edd} \rangle = 0.11$ and $\langle B/T \rangle 
= 0.71$, to be compared with $\langle L_{\rm bol}/L_{\rm Edd} \rangle = 0.24$ 
and $\langle B/T \rangle = 0.94$ for the $z > 0.15$ subsample.

\section{Discussion}

\subsection{Which is the Primary Variable?}

We have assembled a sample of local massive type~1 AGNs with reliable 
spectroscopic data and host galaxy photometric measurements to investigate the 
origin of 
%%%%%%%%%%%%%%%%%%%%%%%%%%%%%%%%%%%%%%%%%%%%%%%%%
\psfig{file=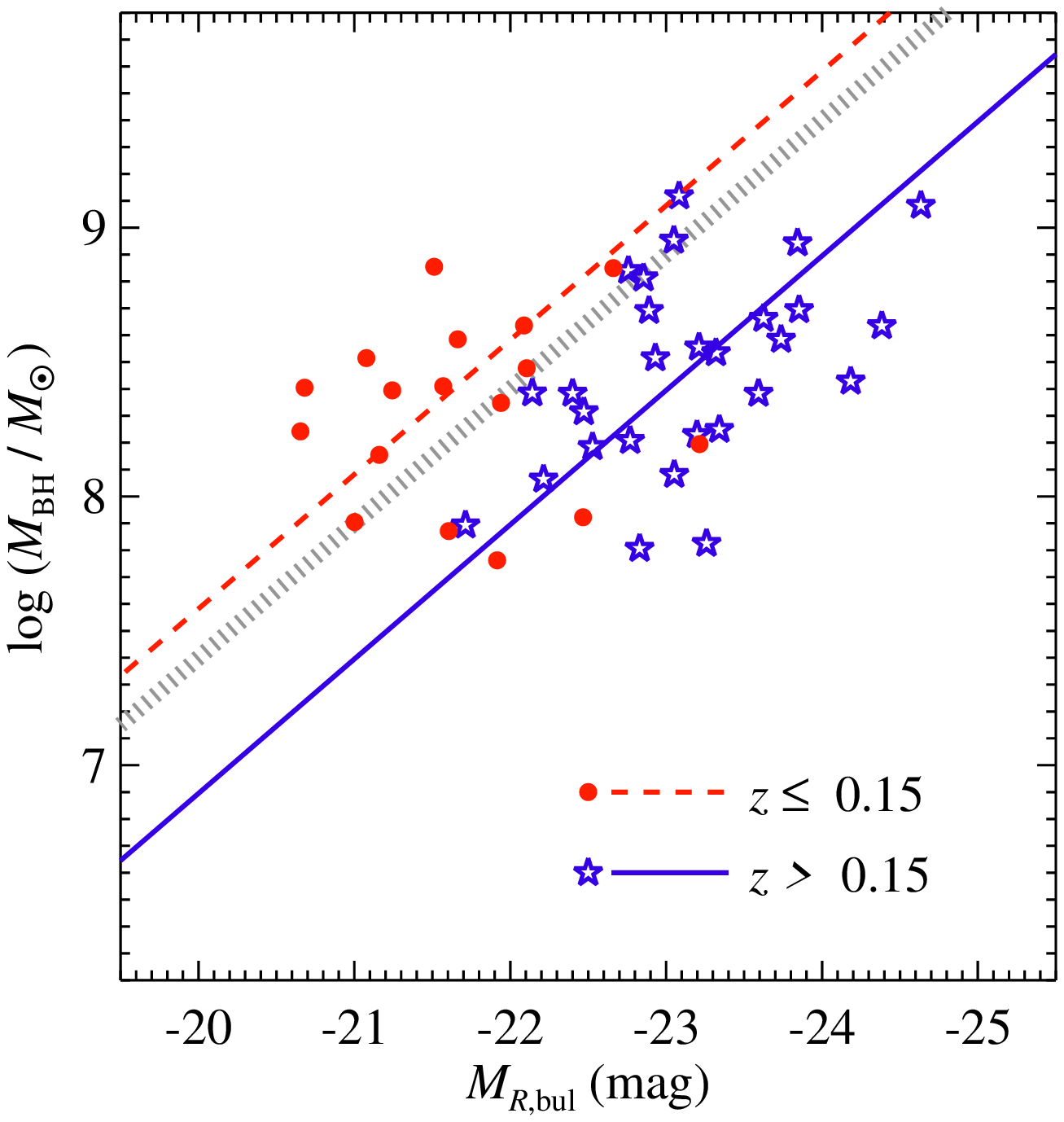,width=8.5cm,angle=0}
\figurenum{10}
\figcaption[fig10.ps]
{Similar to Figure 6, except that here we show the dependence on redshift:
low-redshift sources are marked with red circles and dashed line, and 
high-redshift sources are marked with blue stars and solid line.  The 
\mlb\ relation for inactive galaxies is shown by the thick hatched line.
\label{fig10}}
\vskip 0.3cm
%%%%%%%%%%%%%%%%%%%%%%%%%%%%%%%%%%%%%%%%%%%%%%%%%%
\noindent
the intrinsic scatter in 
the correlation between BH mass and bulge 
luminosity.  Assuming a geometrical factor of $f = 0.75$ for the BLR, we find 
that the AGNs in our sample lie systematically below the \mlb\ relation of 
inactive galaxies by an average offset of $\Delta \alpha \approx -0.3$ dex.  
Moreover, we have shown that the magnitude of the offset correlates with 
secondary parameters connected with the AGN (Eddington ratio and degree of 
radio-loudness) and host galaxy (redshift, bulge-to-disk ratio, and 
signs of morphological disturbance).

Among the several variables that correlate with the offset in the \mlb\
relation, the only one that appears unphysical is that related to redshift.
As we noted in \S4.5, the apparent dependence on redshift most likely reflects 
the selection effect that higher redshift sources tend to be biased toward 
higher Eddington ratios (e.g., Boyle et al.  2000) and more luminous, earlier 
Hubble types.  If we divide the sample into two at $z = 0.15$, the \mlb\ 
relation for both the nearby and distant halves continue to exhibit the 
dependence on \edd\ and $B/T$ that we see for the full sample.

Still, among the rest of the variables that correlate with the zero point 
offset in the 
\mlb\ relation, which is the primary one?  Given that many galaxy and AGN 
parameters are mutually correlated, this is not a trivial question to answer.  
We propose that the primary physical driver is the mass accretion rate, as 
reflected in the Eddington ratio.  We argue that the host morphology and 
degree of tidal disturbance directly affect the AGN accretion rate, and that 
the accretion rate, in turn, is linked to the radio-loudness parameter.

Although our 2-D fits indicate that the hosts of many of our AGNs have a 
non-zero disk component apart from a bulge, it is important to recognize that, 
with few exceptions (Fairall 9,  HE 0306$-$3301, MS 1059.0+7302, PG 2130+99), 
most of the sources in our sample do not have regular, normal disks.  The 
vast majority of the sample---by construction when we imposed the \mbh\ 
cut---is decisively bulge-dominated.  Many of the features that we attribute 
to a ``disk,'' in fact, are simply diffuse, extended features above and beyond 
the dominant bulge component, which we have parameterized using a single 
S\'ersic function.  There is no a priori reason why the bulge should be 
defined in such a manner, that it cannot have a more complex light 
distribution, especially at large radii.  In other instances, the extra-bulge 
component is highly disturbed and almost certainly of tidal origin.  With 
few exceptions (Bennert et al. 2008), these
features have generally never been measured before quantitatively in AGN host 
galaxies.  However, it is entirely debatable whether any of these structures 
truly belongs to or will ever settle into a normal disk component.   Instead, 
we surmise that many of the tidal tails and extended, distorted features, in 
fact, should be considered as part of the {\it bulge in formation}.  They are 
reminiscent of morphological signatures attributed to the late, advanced 
stages of gas-rich mergers (e.g., Barnes \& Hernquist 1996; Lotz et al. 2008) 
or possibly even gas-poor (``dry'') mergers (e.g., van~Dokkum 2005; Naab et 
al. 2006).  Plausible examples of this phenomenon in our sample include [HB89] 
0316$-$346 (Fig. 15), HE 1434$-$1600 (Fig. 19), and PG 1012+008 (Fig. 32).  In 
support of this hypothesis, Figure 11{\it b}\ illustrates that the 
morphological segregation seen in the \mlb\ relation (Fig. 7) essentially 
disappears when the bulge luminosity is replaced with the {\it total}\ 
luminosity of the host.  The scatter also goes down slightly, from 0.40 dex 
to 0.36 dex in the \mlt\ relation.  The dependence on Eddington 
ratio, however, remains (Fig. 11{\it a}).

The above interpretation offers a plausible explanation for the apparent link 
between morphological type and accretion rate, which otherwise is somewhat 
perplexing.  Within our sample it is the apparently earlier-type, more 
bulge-dominated systems that actually have {\it higher}\ accretion rates.  
This runs counter to the trend normally seen in nearby AGNs (e.g., Heckman et 
al.  2004; Greene \& Ho 2007a) and the general tendency for present-day 
early-type galaxies to be more gas-poor than late-type galaxies.  However, if 
high-mass, luminous AGNs 
result from the aftermath of gas-rich galaxy-galaxy mergers (e.g., Sanders et 
al. 1988; Hopkins et al. 2006), our results imply that it is during the most 
advanced stages of the merger that accretion on the central BH attains its
maximum rate.  Some of the most highly accreting objects (those with large 
\edd) in our sample, which coincide among those with the largest offset in the 
\mlb\ relation, also happen to be among the ones that show the most conspicuous 
morphological signatures of tidal perturbations, as measured by the  Fourier 
parameter $a_1$ (see Fig. 8).  These include HE 0354$-$5500 (\edd = 0.57; 
$a_1 = 0.33$), PG 1012+008 (\edd = 0.32; $a_1 = 0.32$), PG 1613+658 (\edd = 
0.44; $a_1 = 0.40$), and PG 1700+518 (\edd = 0.71; $a_1 = 0.39$).  However, 
not every object with a high \edd\ has a large value of $a_1$.  This may imply
that not all quasar episodes 
are triggered by major galaxy interactions, or that 
accretion can proceed at a substantial rate even after the tidal features have 
disappeared.   To the extent that the peak star formation rate in the merger 
has already subsided during this phase, our scenario offers an additional 
explanation for why type~1 AGNs contain intermediate-age stars (Kauffmann et 
al.  2003) but generally not much concurrent star formation (Ho 2005; Kim 
et al. 2006).
%%%%%%%%%%%%%%%%%%%%%%%%%%%%%%%%%%%%%%%%%%%%%%%%%
%%BoundingBox: 0 375 595 652
\begin{figure*}
\psfig{file=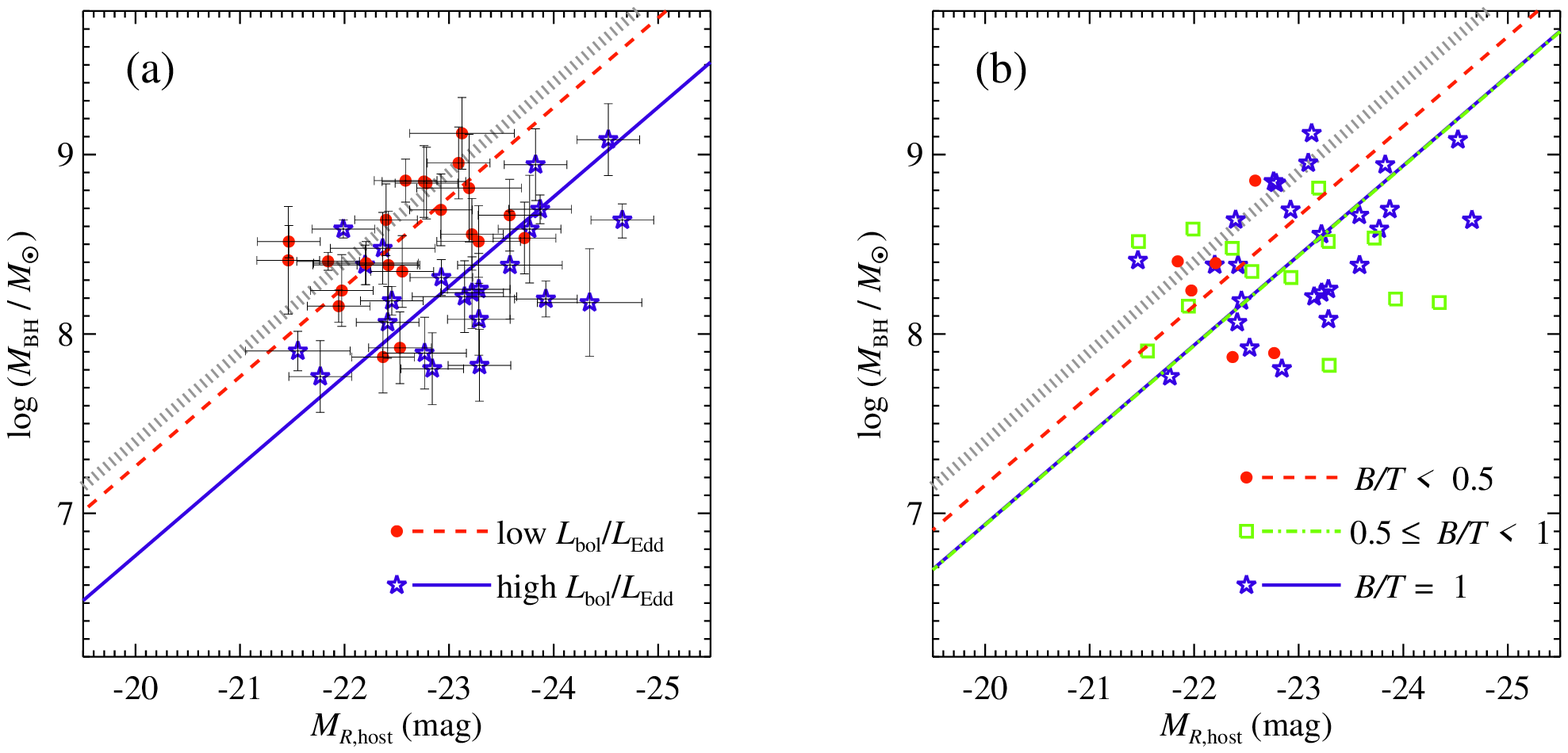,width=18.5cm,angle=0}
\figurenum{11}
\figcaption[fig11.ps]
{Correlation between BH mass and absolute $R$-band magnitude for the
{\it total}\ emission from the host galaxy.  ({\it a}) The symbols are the 
same as in Figure 6; the dependence on Eddington ratio still remains.  
({\it b}) The symbols are the same as in Figure 7; the dependence on galaxy 
morphology is much weaker. The relation between \mbh\ and bulge luminosity for 
inactive galaxies is denoted by the thick hatched line.
\label{fig11}}
\end{figure*}
%%%%%%%%%%%%%%%%%%%%%%%%%%%%%%%%%%%%%%%%%%%%%%%%%%
%\noindent

Within our sample, four AGNs have compact sources---plausibly small 
accreted companions---projected close to the primary host galaxy.  These may 
be examples of minor mergers.  None has a large value of $a_1$. Two of the 
four lie exactly on the \mlb\ relation of inactive galaxies (Fig. 8) and have 
relatively low Eddington ratios (\edd\ = 0.05 for PG 1426+015 and \edd\ = 0.06 
for PKS 2135$-$14).  The other two lie offset below the relation, but only 
PG 1302$-$102 has a large \edd\ (0.4); PHL 1093 has \edd\ = 0.03.  Thus, within
our limited statistics, we have no evidence that minor mergers play a 
significant role in elevating the accretion rate in AGNs.

Within this backdrop, we can offer a tentative explanation for the zero point 
difference between radio-loud and radio-quiet sources (Fig. 9), one that 
ultimately links the generation of powerful radio jets to the BH accretion 
rate and/or host galaxy morphology.  But first we 
should clarify some terminology.  There are two popular definitions of 
``radio-loud'' AGNs in the literature, and it is important not to confuse them.
One common usage of this term refers to sources that are classified solely by 
their radio-to-optical flux ratio ($R$) as defined by Kellermann et al. 
(1989)\footnote{Terashima \& Wilson (2003) advocated a closely related 
radio-loudness parameter $R_{\rm X}$ based on the radio-to-X-ray flux ratio.}, 
regardless of their radio power.  On this basis, the vast majority of AGNs in 
the local Universe (Ho 2008), most with extremely low luminosities, qualify as 
being radio-loud, with the degree of radio-loudness increasing with 
{\it decreasing}\ \edd\ (Ho 2002; Terashima \& Wilson 2003; Greene et al. 
2006).  The radio emission in most of these low-power sources is largely 
dominated by a compact core, and any jet-like features are confined to 
sub-galactic scales.  The host galaxies encompass all morphological types, 
including spiral galaxies (Ho \& Peng 2001).  The Milky Way's Sgr~A$^\star$ is 
a familiar example.  The second commonly used definition of radio-loudness 
is less clear-cut, but it involves some combination of relative ($R \gg 10$)
and absolute ($P_{\rm rad}$ \gax\ $10^{23-24}$ W Hz$^{-1}$) measures of radio 
power.  When detected, the radio jets have super-galactic dimensions and are 
highly collimated.  The radio-loud sources in this study, with median $R = 
820$ and $P_{\rm 6cm}^{\rm tot} = 7\times10^{25}$ W Hz$^{-1}$, belong to this 
second category.  Strong radio sources of this variety invariably reside in 
early-type galaxies (e.g., McLure et al. 1999) and are associated with 
{\it high}\ accretion rates (e.g., Maccarone et al. 2003; K\"{o}rding et al. 
2006).  Nevertheless, only a minority of highly accreting AGNs in massive, 
early-type host galaxies are radio-loud.  There is no universally accepted 
explanation for this longstanding quandary.  One possibility is that a 
necessary ingredient for the generation of powerful, collimated jets is the 
existence of a BH with a large spin (e.g., Sikora et al. 2007), which is more 
easily attained in merger-driven accretion events during the formation of 
giant elliptical galaxies than in disk or spiral galaxies (Volonteri et al. 
2007).  

\subsection{Physical Interpretation}

The principal conclusions of our study are that the zero point of the \mlb\ 
relation for AGNs is offset from that of inactive galaxies and that the 
magnitude of the offset correlates with several physical properties of the AGN 
and host ultimately connected to the accretion rate.  These effects account 
for the bulk of the observed intrinsic scatter in the AGN \mlb\ relation.

Interestingly, other studies of nearby AGNs using different probes of the host 
galaxy have independently arrived at very similar conclusions.  Onken et al. 
(2004; see also Nelson et al. 2004), analyzing a small sample of type~1 AGNs
with available stellar velocity dispersions and BH masses determined through
reverberation mapping, concluded that the active systems lie offset below
the \msig\ relation of inactive galaxies by $\sim 0.2$ dex 
if one assumes a spherical BLR.   Greene \& Ho
(2006b) obtained an almost identical result using a much larger sample
of type~1 AGNs with measured $\sigma_\star$ and \mbh\ estimated from
single-epoch spectra.  These conclusions have been reaffirmed by Shen et al.
(2008) in their analysis of composite AGN spectra derived from the
Sloan Digital Sky Survey.  A different approach was taken by Ho et al. (2008)
through 21~cm H~I observations. Using the maximum rotational velocity and
total dynamical mass of the galaxy as new variables to represent the
gravitational potential of the host, they find that both quantities strongly
correlate with \mbh.  In qualitative agreement with the results from this
study, Ho et al. find that the zero point in the scaling relations depends
primarily on the accretion rate in the sense that, at a given galaxy rotation
velocity and especially dynamical mass, AGNs with higher Eddington ratios
have systematically lower \mbh\ than those with lower Eddington ratios.

These trends can be interpreted in one of two ways.  In the first instance, 
we note that AGNs, by selection, have actively growing BHs.  We can envision 
that, at a fixed bulge potential (velocity dispersion, luminosity, mass), 
active galaxies have less massive BHs than inactive galaxies {\it if}\ the 
bulk of the star formation precedes, and is not well synchronized with, a major 
accretion event.  This particular time sequence, which is required in order to 
imprint a net negative offset in the \mbh\ versus host galaxy relations, seems 
to be supported by the prevalence of post-starburst signatures 
(Canalizo \& Stockton 2001; Kauffmann et al. 2003) 
as well as the low levels of ongoing star formation (Ho 2005; Kim 
et al. 2006) found in type~1 AGNs.  If the AGN phase lasts, say, for 
$\sim 10^8$ yrs, which is near the upper end of the currently estimated 
lifetimes (Martini 2004), then a $10^8$ \solmass\ BH would increase its mass 
by a factor of 2 ($\sim 0.3$ dex) if it is radiating at \edd\ = 0.5 with 
a radiative efficiency of 0.1.  This example is merely illustrative.  In 
reality, the AGNs in our sample span a wider range of \edd\ and the AGN 
lifetimes may be shorter.  Nevertheless, on average we expect luminous AGNs to 
lie systematically below inactive galaxies on the \mlb\ relation.  Moreover, 
our study, as do those of Ho et al. (2008) and Shen et al. (2008), further 
shows that the magnitude of the zero point offset depends on the Eddington 
ratio: the BH masses of high-Eddington ratio AGNs have more catching up to
do than the BH masses of lower-Eddington ratio AGNs.  Such a systematic trend 
can only come about if the accretion rate directly relates to the evolutionary
phase of the accretion event.  This seems plausible, in light of the apparent 
association between accretion rate and the degree of tidal perturbation.

Instead of the BH being undermassive, perhaps it is actually the bulge that is 
overluminous, by  $\Delta M_{R,{\rm bul}} \approx 0.5-0.6$ mag.  First, we 
dismiss the possibility that the luminosity enhancement could be due to 
contamination from nebular emission from the narrow-line region.  Although the 
spatial extent of the narrow-line region in quasars can reach several kpc 
(Bennert et al. 2002), substantially overlapping with the bulge, the typical 
\oiii\ luminosities in our sample ($\langle L_{\rm [O III]} \rangle = 10^{42}$ 
ergs s$^{-1}$) contribute less than 3\% to the luminosity of the bulge 
($\langle$\rbulge$\rangle=-22.75$ mag).  Given the evidence outlined in \S 5.1 
that the most extreme outliers in the \mlb\ seem to have undergone a recent 
merger or tidal interaction, a more likely possibility is that
the bulge luminosity may be moderately enhanced by the latest episode of 
central star formation.  Indeed, for a small sample of reverberation-mapped 
Seyfert 1 galaxies with stellar velocity dispersion and bulge luminosity 
measurements, Nelson et al. (2004) have shown that these objects are somewhat 
brighter ($\sim 0.4$ mag) than inactive galaxies at a given velocity 
dispersion.  They attributed this offset in the Faber-Jackson (1976) relation 
to younger stellar populations in AGNs.  Boosting the $R$-band luminosity by 
0.5 mag requires $\sim 15$\% of the stellar mass to come from a 1 Gyr 
population with solar metallicity (Nelson et al. 2004).  While this offers a 
plausible explanation for the offset in the \mlb\ relation seen in our sample, 
it cannot account for the fact that the most recent and largest samples of 
nearby type 1 AGNs, statistically at least, show negative offsets (with 
respect to inactive galaxies) when \mbh\ is compared to {\it all}\ bulge or 
host galaxy parameters (stellar velocity dispersion, bulge luminosity, total 
galaxy dynamical mass).  The direction of the offset is the same (at a given 
galaxy parameter, \mbh\ is lower), and the magnitude of the offset is also 
similar ($\sim 0.2-0.3$ dex).  Although this clearly needs to be verified with 
a large sample that has reliable measurements of both velocity dispersion and 
bulge luminosities, the most recent studies suggest, contrary to Nelson et al. 
(2004), that local type 1 AGNs actually do not depart from the standard 
Faber-Jackson (1976) relation.  Furthermore, for a large sample of sources with 
measurements of rotational velocity and total galaxy luminosity, Ho et al. 
(2008) show that type 1 AGNs show no obvious deviations from the Tully-Fisher 
(1977) relation of inactive galaxies.  In light of these considerations, we 
favor the view that the negative offset in the \mlb\ relation represents 
a deficit in \mbh\ rather than an excess in \lbul.

Alternatively, we can assert that both active and inactive galaxies 
intrinsically should obey the {\it same}\ BH mass versus host scaling 
relations. From this standpoint, the zero point offset between AGNs and inactive
galaxies, as well as the variations of the offset with \edd, can be viewed as a 
systematic {\it underestimate}\ of the true value of \mbh.  Recall that our BH
masses are based on a virial product assuming a spherical distribution of BLR 
clouds with isotropic velocities, for which the geometric factor is $f = 0.75$.
If, for example, the BLR (or at least the portion of it that predominantly 
emits the Balmer lines) has a flattened, disk-like geometry with kinematics 
dominated by rotation, and on the scale of the BLR type~1 sources happen to be 
preferentially more face-on to our line of sight, then we systematically 
underestimate the deprojected rotation velocity and hence \mbh.  Wu \& Han 
(2001) invoked this line of reasoning to interpret the offset of type~1 AGNs 
on the \msig\ relation and concluded that on average their BLRs are inclined by 
$\langle i \rangle \approx 36$\deg.  The same argument can, in principle, be 
applied to the observed offset in the \mlb\ relation.  While the effect of 
inclination probably enters at some level (see also Collin et al. 2006), it 
cannot account for the fact that the magnitude of the offset depends on \edd.  
The latter is not a trivial consequence of the mass being underestimated 
because sources with high \edd\ truly {\it do}\ exhibit characteristically 
distinct X-ray, optical, and radio properties (Boller et al. 1996; Boroson 
2002; Greene et al. 2006).  Whatever the physical origin of the offset (Onken 
et al. 2004; Collin et al. 2006; Marconi et al. 2008), we can empirically 
adjust the normalization factor of the virial product by forcing the AGN 
sample to agree with the fiducial reference of inactive galaxies.  To remove 
the zero point offset of $\Delta \alpha \approx -0.2$ to $-0.3$ dex, then, the 
normalization should be increased by a factor of $\sim 1.6-2$, from 
$f = 0.75$ to $f \approx 1.2-1.5$.  For the most extreme offsets of 
$\Delta \alpha \approx -0.6$, $f$ increases to $\sim 3$.

Without additional information, unfortunately, the above two alternative
explanations---an undermassive \mbh\ versus an underestimated $f$ factor---are
degenerate. It is easy to imagine that both effects must operate jointly.
On the one hand, the BHs in AGNs are, after all, gaining mass.  On the 
other hand, as discussed in Collin et al. (2006), there are multiple 
reasons to believe that the BLR has a nonspherical geometry and that the
Eddington ratio may influence its structure and dynamics.   The only way
to resolve this degeneracy is to obtain independent estimates of \mbh\ 
for AGNs that do not rely on the BLR virial technique.  To date, efforts 
to apply resolved stellar dynamical techniques to reverberation-mapped AGNs 
have yielded very rough estimates of \mbh\ for only a couple of sources 
(NGC 3227: Davies et al. 2006; Hicks \& Malkan 2008; NGC 4151: Onken et al. 
2007), and thus attempting to cross-calibrate the two techniques is still 
far too premature.  BH mass estimators based on X-ray variability seem 
more promising.  Gierli\'nski et al. (2008; see also Hayashida et al. 1998) 
find that, for accreting BHs in their hard spectral state, the amplitude of 
their high-frequency X-ray variability scales inversely with \mbh\ over a very 
wide range of masses.  For a small subset of nearby type~1 AGNs, the 
X-ray-derived masses show rough agreement with \mbh\ obtained through 
reverberation mapping assuming $f \approx 1.2$.  A very similar conclusion was 
reached by Niko\l ajuk et al. (2006).  From comparison of \mbh\ for 
reverberation-mapped sources with masses obtained using the X-ray excess 
variance method, these authors estimate $f = 1.06\pm 0.26$, which, 
interestingly, lies in between the values of the $f$ factor for the two 
extreme alternatives discussed above.  

Assuming that the X-ray-derived normalization factor truly does represent the 
correct normalization factor for the virial masses, then the inferred growth 
rates for \mbh\ in luminous AGNs are much more modest, from typically as 
little as 10\%--40\% (0.05--0.15 dex) to at most 280\% (0.45 dex).  

Still, we note that the tendency for BHs in local AGNs to be less massive than 
the BHs in inactive galaxies of similar type runs counter to the trend 
observed at higher redshift.  Already by $z=0.36$, type~1 AGNs begin to depart 
from the local \msig\ and \mlb\ relations (Woo et al. 2006; Treu 
et al. 2007), but in the {\it opposite}\ direction as that seen at lower 
redshifts: at a given $\sigma_\star$ or \lbul, AGNs are offset compared to 
local inactive systems by $\Delta \log M_{\rm BH} \approx +0.5$ dex.
This trend has now been extended by 
Woo et al. (2008) out to $z = 0.57$ using stellar velocity dispersion 
measurements.  At even higher redshifts, direct $\sigma_\star$ measurements 
are no longer feasible, but surrogate estimates of $\sigma_\star$ using narrow 
emission lines (Salviander et al. 2007) as well as probes of the host galaxy 
using imaging (Peng et al. 2006a, b) and CO emission lines (Shields et al. 
2006; Ho 2007) support the notion that the growth of the BHs in AGNs have 
been decoupled from, and outpaced, the underlying host.  

\section{Summary}

We performed two-dimensional structural decomposition of a sample of 45 
nearby ($z$ \lax\ 0.35) type~1 AGNs with available archival optical \hst\ 
images and published spectroscopic data.  We calculated virial BH masses 
assuming a spherical BLR with isotropic velocities.  Using a new 
version of the versatile code GALFIT, we derived detailed fits to the 
structural components of the host galaxies, yielding not only robust
measurements of the bulge luminosities with realistic error bars but also, 
for the first time, quantitative estimates of nonaxisymmetric features such 
as extended disks and tidal arms.  

Our principal aim is to understand the origin of the intrinsic scatter in the 
\mlb\ of active galaxies over a restricted range of BH masses (\mbh\ 
$\approx\,10^{8.5\pm0.5}$ \solmass).  While AGNs closely follow the \mlb\ 
relation of inactive galaxies, we find that the intrinsic scatter is 
substantial (0.40 dex) and that the zero point of the relation is shifted by 
an average of $\Delta \log M_{\rm BH} \approx -0.3$ dex.  The magnitude of the 
zero point offset in the \mlb\ relation depends on properties of the AGN 
(Eddington ratio and radio-loudness parameter) and the host galaxy 
(morphological type and degree of tidal perturbation).  We argue that the 
principal physical parameter responsible for the variation in zero point is 
the BH accretion rate, as reflected in the Eddington ratio.  We suggest that 
galaxy mergers and tidal interactions play a substantial role in boosting the 
accretion rate in this sample of AGNs.  A significant fraction of the zero 
point offset in the \mlb\ relation can be explained if the virial BH masses 
have been underestimated, as indicated from comparison with independent masses 
derived from X-ray variability techniques.  After accounting for this change 
in the normalization of the virial BH mass scale, we estimate that BHs during 
the AGN phase experience a modest growth of $\sim$10\%--40\% in mass.

\acknowledgements

We are grateful to the referee for providing a timely and helpful review.
We thank James Dunlop and Ross McLure for useful discussions.  This work was
supported by the Carnegie Institution of Washington and by NASA through grants
HST-AR-10969 and HST-GO-9763 from the Space Telescope Science Institute, which 
is operated by the Association of Universities for Research in Astronomy, 
Inc., under NASA contract NAS5-26555).  M.K. and M.I. acknowledge the support 
of the Korea Science and Engineering Foundation (KOSEF) through the 
Astrophysical Research Center for the Structure and Evolution of the Cosmos 
(ARCSEC).  C.Y.P. is grateful for support through the Plaskett Fellowship 
program of Herzberg Institute of Astrophysics and the STScI Institute 
Fellowship program.  Research by A.J.B. was also supported by by NSF grant 
AST-0548198.

\appendix

\section{Notes on Individual Objects}

Comments on the fitting results for individual objects are given here.

\bigskip

{\it 3C 59} (Fig. 12) --- \ \ 
The host can be fit with classical bulge represented by
a de~Vaucouleurs ($n=4$) profile.

{\it E 1821+643} (Fig. 13) --- \ \
The best fit shows that the bulge is slightly disturbed ($a_1=0.17$). 
There is no evidence of a disk.

{\it Fairall 9} (Fig. 14) --- \ \
The host requires a bulge and a disk.  A ring-like structure, which we do not 
model, is also seen in the residuals.  The bulge appears be quite compact, 
with an effective radius of 1.6 pixels, although it is not well decomposed 
from the nucleus.  Thus, the bulge luminosity might be highly uncertain. 

{\it [HB89] 0316$-$346} (Fig. 15) --- \ \
Highly disturbed object with prominent tidal features. We fit the host
with an $n=4$ bulge and two disk components with Fourier 
modes.  One of the disk components is not centered on the nucleus.

{\it [HB89] 2201$+$315} (Fig. 16) --- \ \
This object was observed in two different filters (F555W and F702W), but both 
images are saturated in the core. The bulge luminosities derived from the two 
different images (corrected to the $R$ band) differ by 0.7 mag. We fit the 
host with a single $n=4$ bulge in both images. The residual image from the 
long (2700~s) exposure shows possible signs of an extended disk. 

{\it HE 0306$-$3301} (Fig. 17) --- \ \
The best-fit result shows an elongated bulge component ($b/a \approx 0.45$) 
and a disk with a spiral arm. 
The signal-to-noise ratio (S/N) of the observed PSF star is lower than 
that of the science image.

{\it HE 0354$-$5500} (Fig. 18) --- \ \
This appears to be a merging system. We fit the host galaxy with a bulge
($n=4$) and an off-centered tidal-like feature with a high-amplitude Fourier 
mode ($a_1=0.33$).
The S/N of the observed PSF star is lower than that of the science image.

{\it HE 1434$-$1600} (Fig. 19) --- \ \
We fit the host with only a single bulge component ($m_{\rm bul} \approx 
17.5$ mag), although the residual image shows evidence for arcs and ripples, 
which may be evidence of a recent collision (Letawe et al. 2004). 
Alternatively, if the host is fit with two components, the bulge magnitude
becomes 18.6 mag. 
The S/N of the observed PSF star is lower than that of the science image.

{\it MC 1635+119} (Fig. 20) --- \ \
We fit the host with a single bulge component ($n=4$).

{\it MRK 1048} (Fig. 21) --- \ \
This object has a close companion and an extremely large but faint, off-center 
tidal feature.  However, neither of these has a large effect on the fitting 
result.  We fit the host with a classical bulge ($n=4$) and a tidal tail.

{\it MS 0244.8+1928} (Fig. 22) --- \ \
The host is reasonably well represented by an $n$ = 4 bulge, although the 
residuals indicate that there might be an additional faint, outer envelope.

{\it MS 0754.6+3928} (Fig. 23) --- \ \
The host is reasonably well represented by an $n$ = 4 bulge, although the 
residuals indicate that there might be an additional faint, outer envelope.

{\it MS 1059.0+7302} (Fig. 24) --- \ \
The best-fitting model requires a bulge and a disk.  
There is a ring-like structure in the residual image.

{\it MS 1545.3+0305} (Fig. 25) --- \ \
The best fit requires a bulge ($n=4$) and a disturbed exponential disk.  A 
ring-like structure is prominent in the original and residual images.

{\it OX 169} (Fig. 26) --- \ \
There are significant residuals after subtracting the best-fit observed PSF, 
which might be due to PSF mismatch. The fit with the TinyTim PSF results in 
a brighter AGN by 0.1 mag. The host contains a bulge ($n=3.9$) and a 
prominent, extended, asymmetric structure that is fit with a S\'ersic 
component with a small $n$ ($\sim 0.2$).  

{\it PG 0052+251} (Fig. 27) --- \ \
It appears to have a tidally disturbed spiral arm, which may be related to 
the two small galaxies in its vicinity. We fit this object with a classical 
bulge ($n=4$) and a truncated disk with a small $n$ ($\sim 0.2$). 

{\it PG 0804+761} (Fig. 28) --- \ \
There appears to be a faint central feature that resembles a bar or highly 
inclined disk-like structure. 
The bulge luminosity, however, is hardly affected by the bar component.

{\it PG 0923+201} (Fig. 29) --- \ \
The host is fit with a single bulge component ($n=4$).   There are three 
nearby companions that were fit simultaneously.

{\it PG 0953+414} (Fig. 30) --- \ \
The host is fit with a single bulge component ($n=4$), which appears 
slightly disturbed based on the amplitude of the Fourier mode ($a_1=0.12$).

{\it PG 1004+130} (Fig. 31) --- \ \
The host is fit with a single bulge component ($n=4$).  The residual image 
shows significant PSF mismatch, which might affect the fitting result. 

{\it PG 1012+008} (Fig. 32) --- \ \
The image of the host shows clear evidence of interaction with a spiral galaxy 
and a smaller, compact neighbor. The primary host can be fit with a single 
bulge component ($n=4$), but the residual image shows significant structure.

{\it PG 1116+215} (Fig. 33) --- \ \
The residual image indicates that the central core is slightly affected by PSF 
mismatch.  The host galaxy, however, is well-represented by a single-component 
bulge with $n$ fixed to 4.  If we allow $n$ to be free, the best-fitting 
value of $n=1.84$ yields a bulge luminosity that is 0.3 mag fainter.

{\it PG 1202+281} (Fig. 34) --- \ \
This source has a number of nearby galaxies, the brightest and nearest of 
which we fit simultaneously.  The primary host is well described by a 
single-component bulge ($n=4$).

{\it PG 1211+143} (Fig. 35) --- \ \
Like PG 0804+761, there appears to be a faint central feature that resembles 
a bar or highly inclined disk-like structure.  It is unclear if this is 
an artifact due to PSF mismatch. Depending on whether this extra component is
included, the bulge luminosity ranges 
between 16.7 and 17.2 mag, with a best-fit value of 16.9 mag.

{\it PG 1226+023} (Fig. 36) --- \ \
Because of the strong bleeding regions from the saturated core, we did not 
perform nonparametric aperture photometry. 
The host is reasonably well fit with a single bulge component ($n=4$). The 
residual image shows the well-known jet of this source (3C 273), as well
as some diffuse, extended emission.  The box-like imprint in the residual 
image results from the PSF image being smaller than the science image.

{\it PG 1302$-$102} (Fig. 37) --- \ \
This object was observed in two different filters (F606W and F702W), but both
images are saturated in the core.  The image contains two compact sources 
superposed on the main host, which we fit simultaneously.  The host can 
be fit with a single bulge component, although significant structure on large
scales remain in the residual image.  The fits from both filters 
are in good agreement to within the uncertainty.   

{\it PG 1307+085} (Fig. 38) --- \ \
Although the core of the image is saturated, the host is well-described by a 
single bulge component ($n=4$).

{\it PG 1309+355} (Fig. 39) --- \ \
The core of the image is saturated.  We fit the host with a single 
bulge component ($n=4$).  The residual image shows evidence of spiral-like 
substructure, but we did not attempt to model it.

{\it PG 1351+640} (Fig. 40) --- \ \
This is an almost face-on system with spiral arms. The fit is done with a 
classical bulge ($n=3.8$) and an exponential disk.

{\it PG 1411+442} (Fig. 41) --- \ \
This is an extremely disturbed object that appears to have a nearby companion.
We fit this object with a classical bulge ($n=4$) and a spiral disk 
with Fourier modes.

{\it PG 1416$-$129} (Fig. 42) --- \ \
The host is well-described by a single bulge component ($n=4$).

{\it PG 1426+015} (Fig. 43) --- \ \
This object has a tidal tail and a small companion. The best-fitting
model for the host consists of a pseudobulge ($n=2.1$) and a disk with Fourier 
modes. Modeling the host with only a single component yields a much
brighter bulge (14.2 vs. 16.08 mag), but the residuals of the fit are 
significantly worse than those of the two-component fit.

{\it PG 1444+407} (Fig. 44) --- \ \
The host is fit with a bulge ($n=4$) and a somewhat disturbed disk 
component ($n=0.31$, $a_1 = 0.10$).
The residuals suggest that a ring-like component might be present.

{\it PG 1613+658} (Fig. 45) --- \ \
This is a highly disturbed object in a merging system. The fit is ambiguous.
The best-fitting model for the host consists of a classical bulge ($n=4$) and a
disturbed disk ($n=1$) with Fourier modes. The bulge is $\sim$0.7 mag fainter
than the best-fitting case if the fit is done with a single bulge component 
($n=4$), but the residuals of the fit are 
significantly worse than those of the two-component fit.

{\it PG 1617+175} (Fig. 46) --- \ \
Although the best-fitting model for the host contains a bulge and a disk, 
a single-component model also works reasonably well.
In the two-component fits, the bulge luminosities range from 17.7 mag 
($n=4$) to 17.9 mag ($n=1$).

{\it PG 1700+518} (Fig. 47) --- \ \
The host is well-represented by a single bulge component ($n=4$) plus a tidal
tail, which is fit with Fourier modes.

{\it PG 2130+099} (Fig. 48) --- \ \
This object is very similar to PG 0052+251; both have ring-like spiral disk.
Since the central few pixels are saturated, the fit is slightly uncertain.
The model for the host consists of a pseudobulge ($n=0.44$) 
and a disk ($n=0.33$) with Fourier modes.

{\it PHL 909} (Fig. 49) --- \ \
The host is fit with a classical bulge ($n=4$), but the residual image shows
a faint central feature that resembles a tiny bar or highly inclined disk-like 
structure.  The host galaxy is slightly disturbed.

{\it PHL 1093} (Fig. 50) --- \ \
The host is well fit with a single bulge component ($n=4$).  Three small 
nearby companions are included in the fit simultaneously.

{\it PKS 0736+01} (Fig. 51) --- \ \
The host is well fit with a single bulge component ($n=4$).  There are several
faint blobs nearby, but it is unclear whether these are associated with the 
primary host.

{\it PKS 1020$-$103} (Fig. 52) --- \ \
A single bulge component ($n=4$) is enough to describe the host. 

{\it PKS 1217+02} (Fig. 53) --- \ \
The host is well fit with a single bulge component ($n=4$).

{\it PKS 2135$-$14} (Fig. 54) --- \ \
We fit the host with a single bulge component ($n=4$).  Two nearby objects 
are included simultaneously in the fit.

{\it PKS 2349$-$01} (Fig. 55) --- \ \
There are two curved tidal tails. The best-fit model for the host consists of 
a classical bulge ($n=4$) and a highly distorted disk with $n\approx 0.88$.

{\it PKS 2355$-$082} (Fig. 56) --- \ \
The host is well fit with a single bulge component ($n=4$).

\clearpage
\begin{figure}
\figurenum{12}
\epsscale{0.9}
\plotone{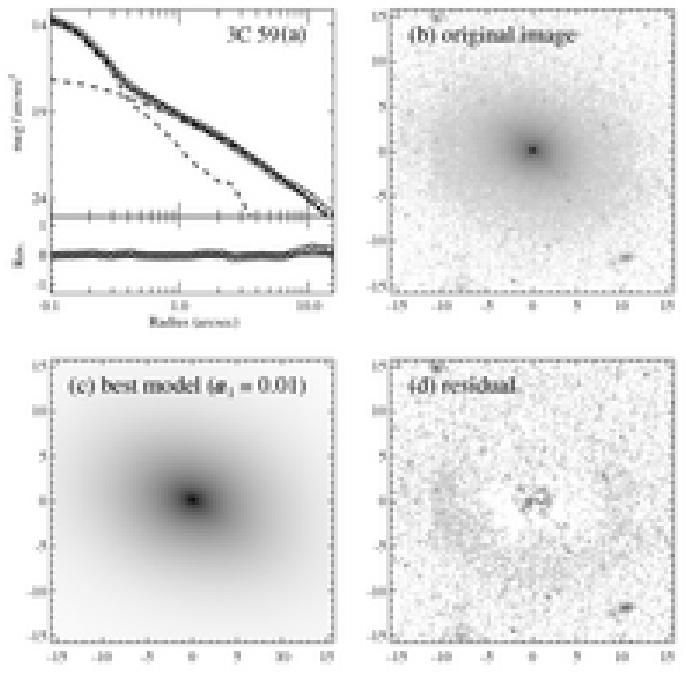}
\caption{
GALFIT decomposition for 3C 59.
({\it a}) Azimuthally averaged profile, showing the original data
({\it open circles}), the best fit ({\it solid line}), and the sub-components
(PSF and bulge; {\it dashed lines}). The residuals are plotted on the
{\it bottom}.  We present the 2-D image of the original data ({\it b}), the
best-fit model for the host (the AGN is excluded to better highlight the
host), with the amplitude of the first Fourier mode ($a_1$) labeled ({\it c}),
and the residuals ({\it d}).  The units of the images are in arcseconds. All
images are on an asinh stretch.
}
\end{figure}

\clearpage
\begin{figure}
\figurenum{13}
\epsscale{0.9}
\plotone{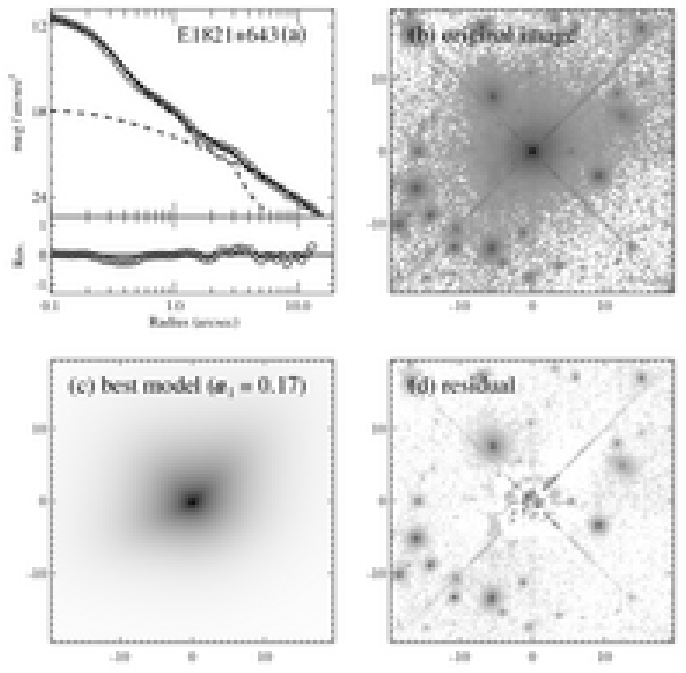}
\caption{
GALFIT decomposition for E 1821+643; symbols and conventions as in Figure 12.
}
\end{figure}

\clearpage
\begin{figure}
\figurenum{14}
\epsscale{0.9}
\plotone{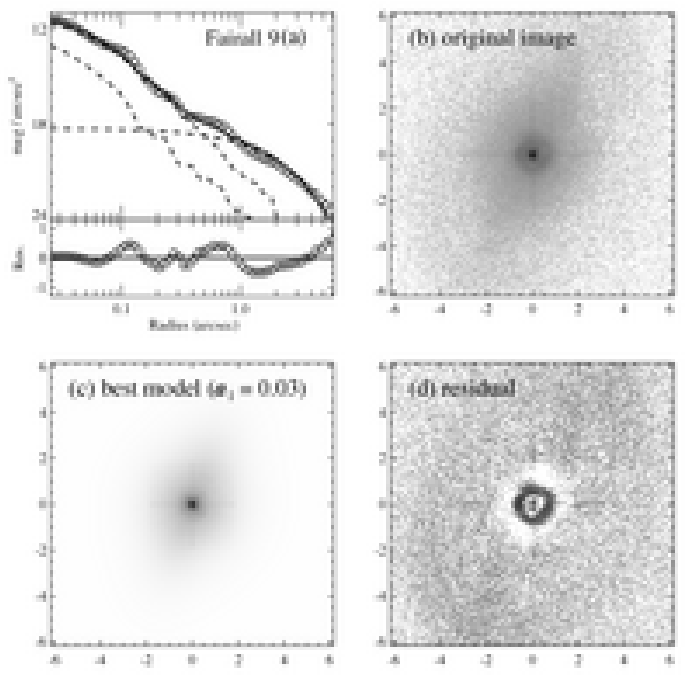}
\caption{
GALFIT decomposition for Fairall 9; symbols and conventions as in Figure 12.
}
\end{figure}

\clearpage
\begin{figure}
\figurenum{15}
\epsscale{0.9}
\plotone{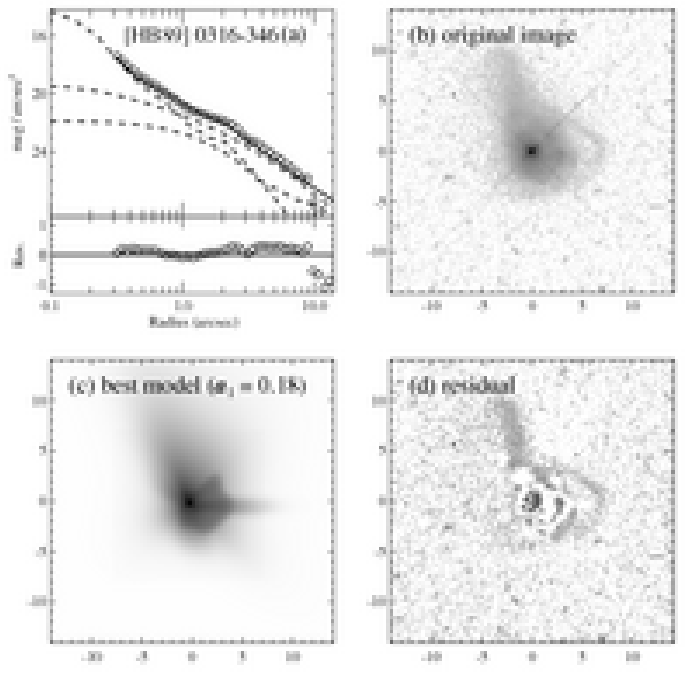}
\caption{
GALFIT decomposition for [HB89] 0316$-$346; symbols and conventions as in Figure 12.
}
\end{figure}

\clearpage
\begin{figure}
\figurenum{16}
\epsscale{0.9}
\plotone{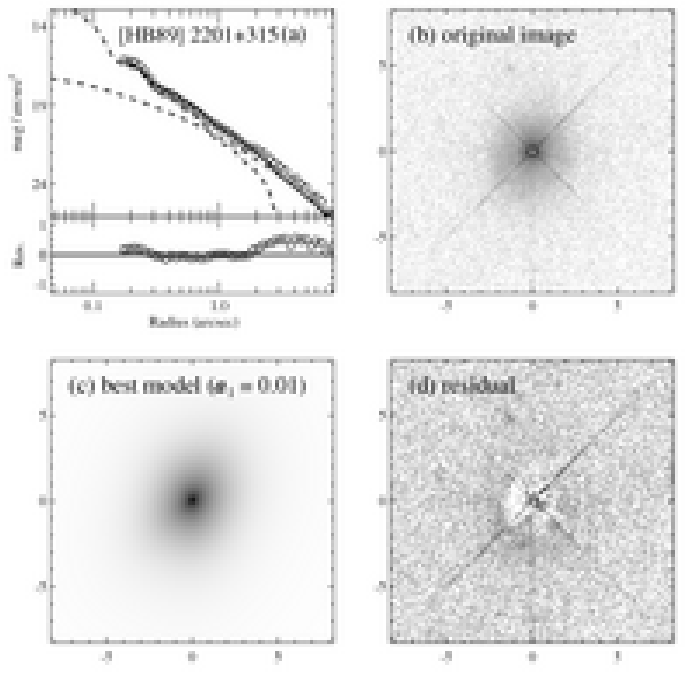}
\caption{
GALFIT decomposition for [HB89] 2201+315 (PC/F555W); symbols and conventions as in Figure 12.
}
\end{figure}

\clearpage
\begin{figure}
\figurenum{17}
\epsscale{0.9}
\plotone{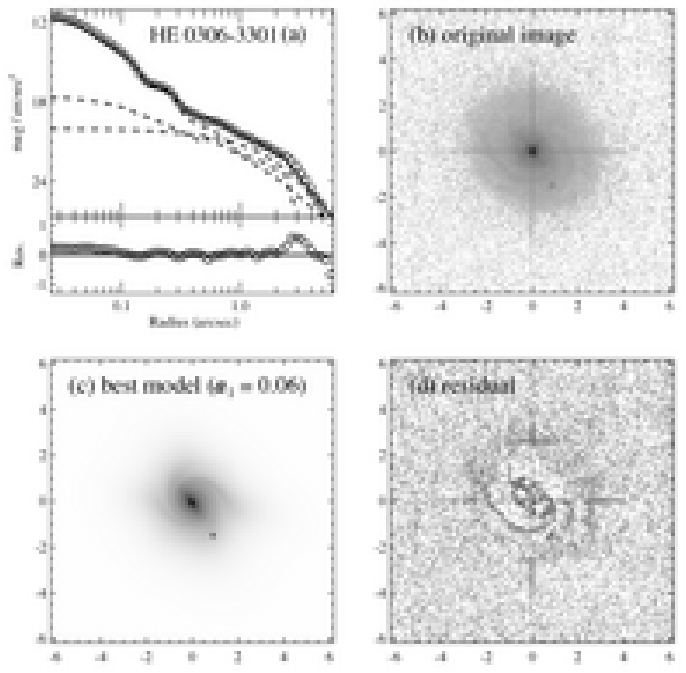}
\caption{
GALFIT decomposition for HE 0306$-$3301; symbols and conventions as in 
Figure 12.
}
\end{figure}

\clearpage
\begin{figure}
\figurenum{18}
\epsscale{0.9}
\plotone{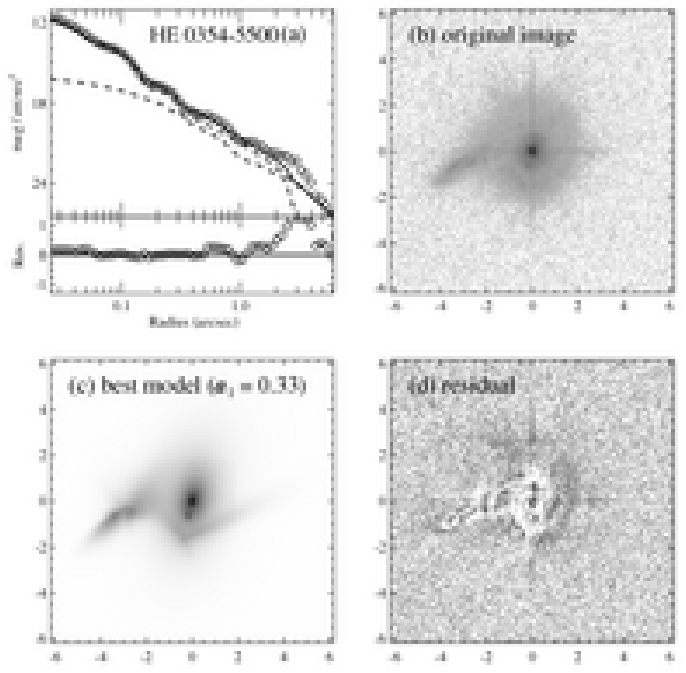}
\caption{
GALFIT decomposition for HE 0354$-$5500; symbols and conventions as in Figure 12.
}
\end{figure}

\clearpage
\begin{figure}
\figurenum{19}
\epsscale{0.9}
\plotone{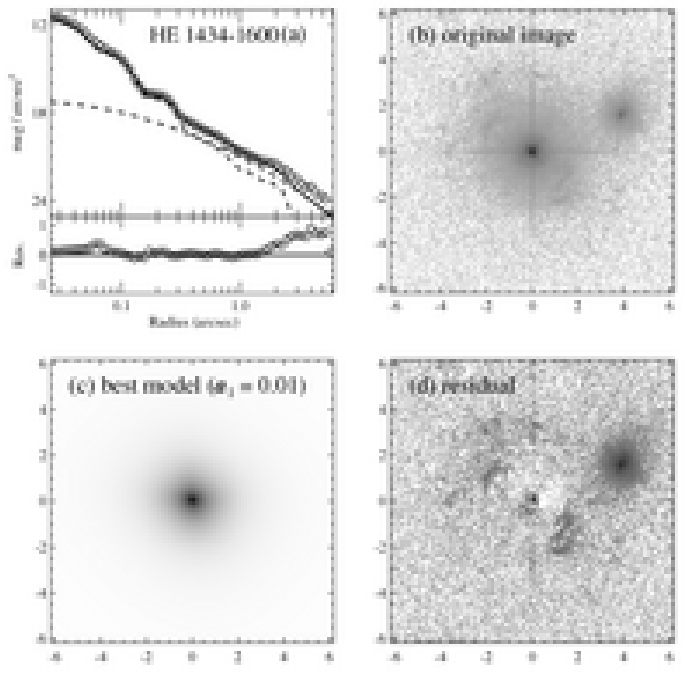}
\caption{
GALFIT decomposition for HE 1434$-$1600; symbols and conventions as in Figure 12.
}
\end{figure}

\clearpage
\begin{figure}
\figurenum{20}
\epsscale{0.9}
\plotone{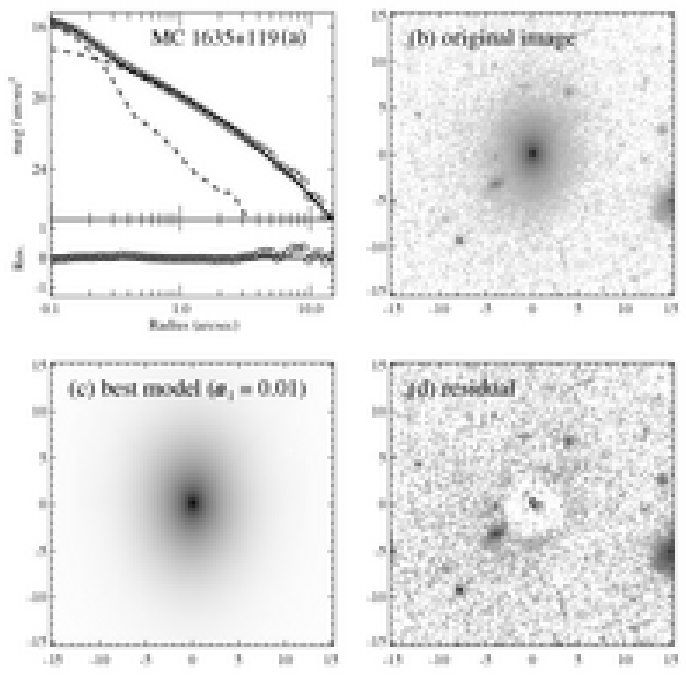}
\caption{
GALFIT decomposition for MC 1635+119; symbols and conventions as in Figure 12.
}
\end{figure}

\clearpage
\begin{figure}
\figurenum{21}
\epsscale{0.9}
\plotone{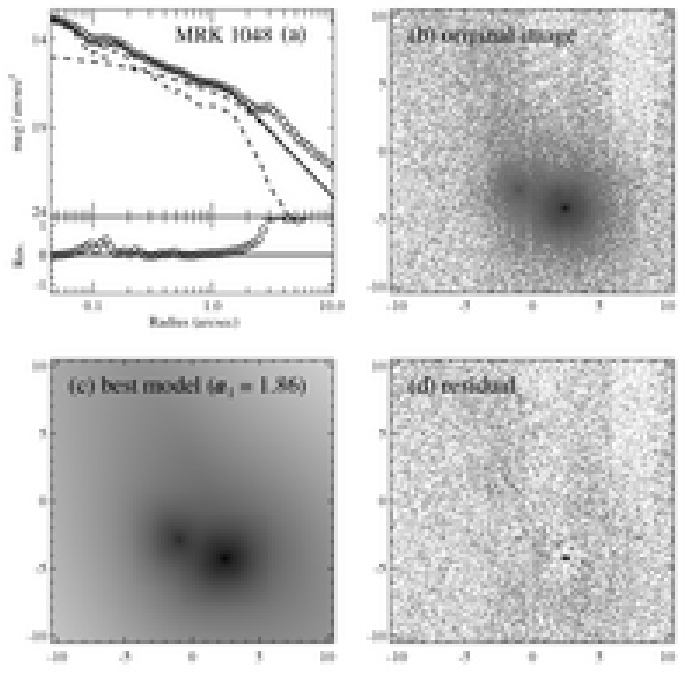}
\caption{
GALFIT decomposition for MRK 1048; symbols and conventions as in Figure 12.
}
\end{figure}

\clearpage
\begin{figure}
\figurenum{22}
\epsscale{0.9}
\plotone{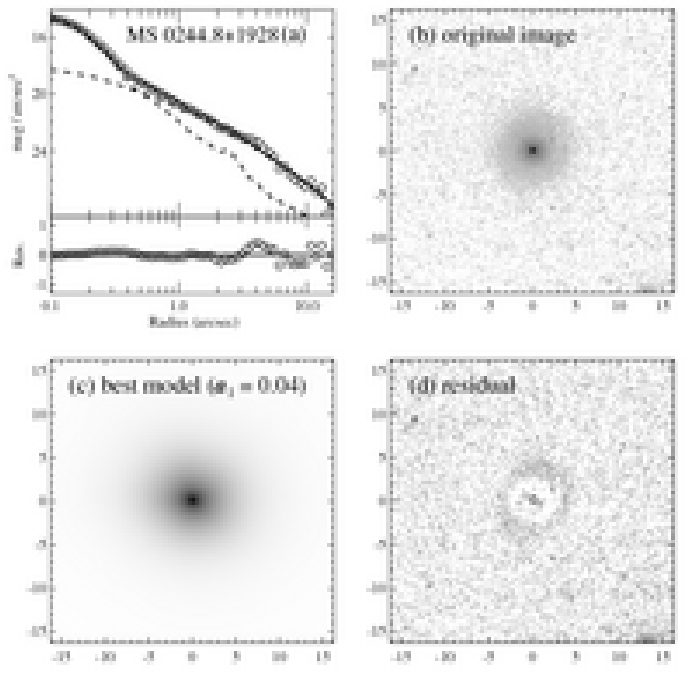}
\caption{
GALFIT decomposition for MS 0244.8+1928; symbols and conventions as in Figure 12.
}
\end{figure}

\clearpage
\begin{figure}
\figurenum{23}
\epsscale{0.9}
\plotone{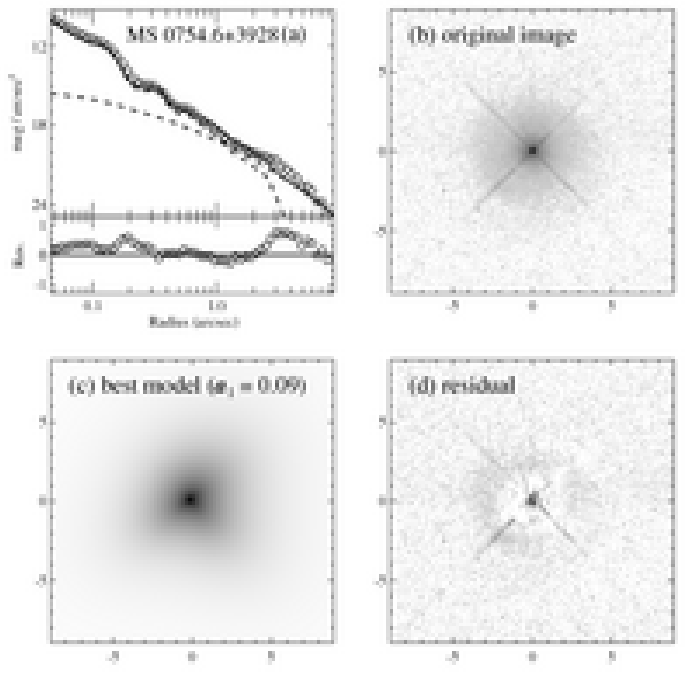}
\caption{GALFIT decomposition for MS 0754.6+3928; symbols and conventions as in Figure 12.
}
\end{figure}

\clearpage
\begin{figure}
\figurenum{24}
\epsscale{0.9}
\plotone{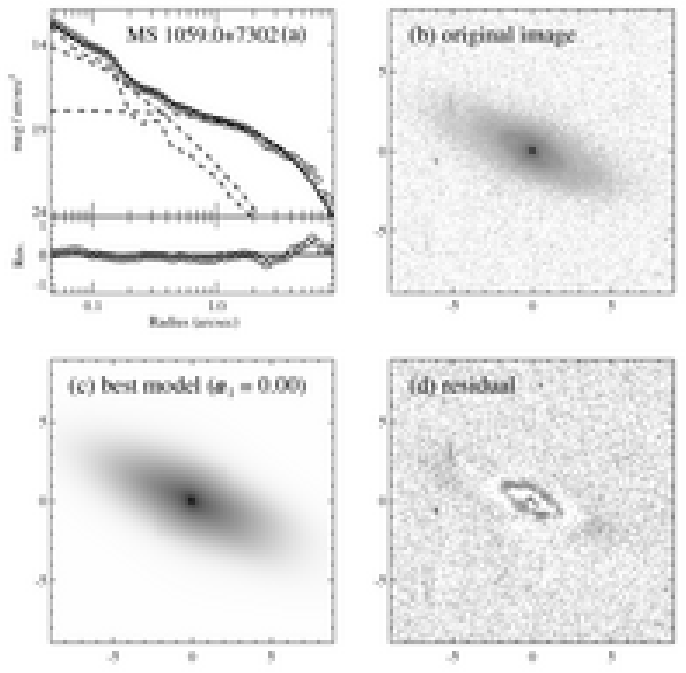}
\caption{GALFIT decomposition for MS 1059.0+7302; symbols and conventions as in Figure 12.
}
\end{figure}

\clearpage
\begin{figure}
\figurenum{25}
\epsscale{0.9}
\plotone{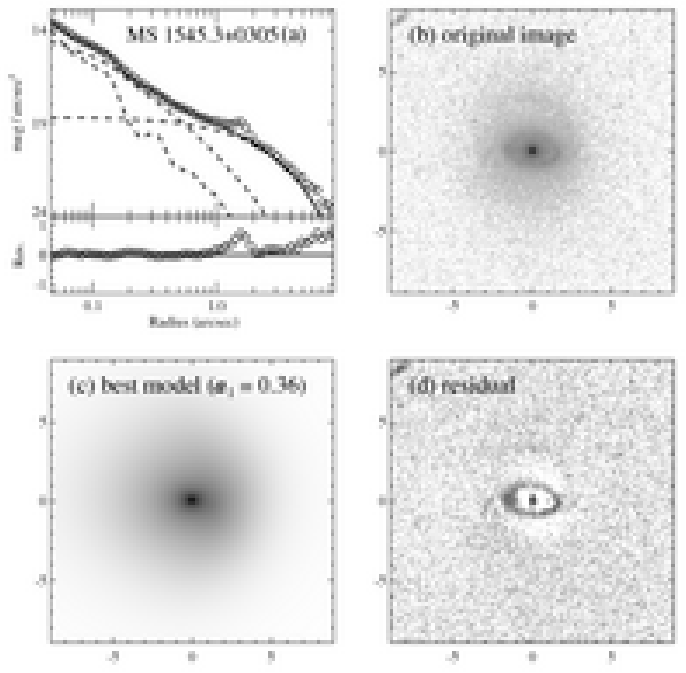}
\caption{GALFIT decomposition for MS 1545.3+0305; symbols and conventions as in Figure 12.
}
\end{figure}

\clearpage
\begin{figure}
\figurenum{26}
\epsscale{0.9}
\plotone{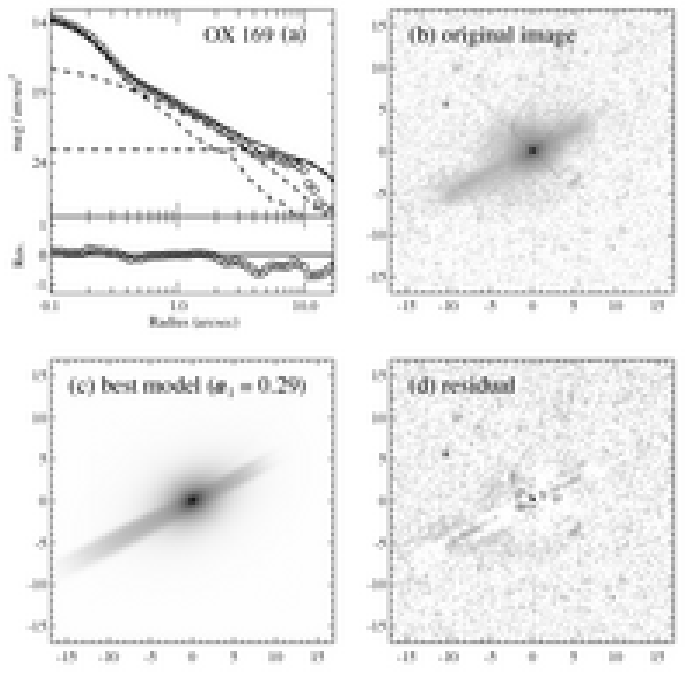}
\caption{GALFIT decomposition for OX 169; symbols and conventions as in Figure 12.
}
\end{figure}

\clearpage
\begin{figure}
\figurenum{27}
\epsscale{0.9}
\plotone{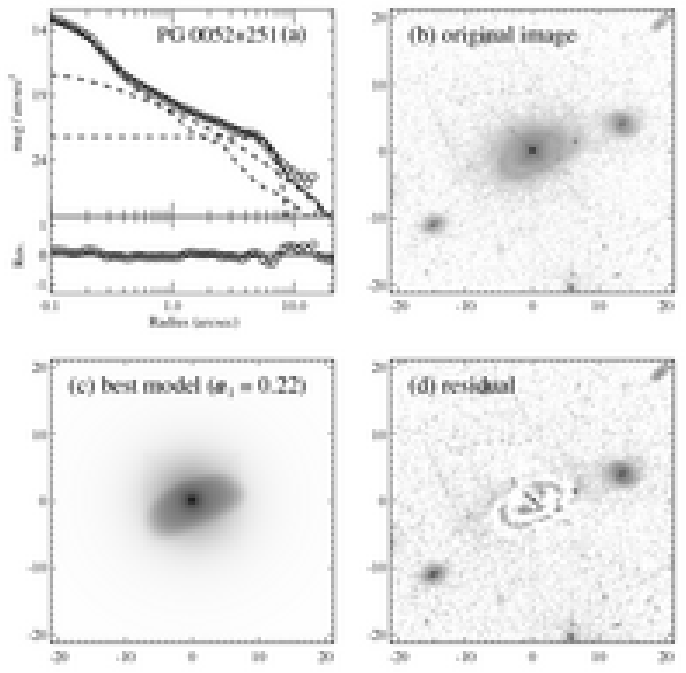}
\caption{GALFIT decomposition for PG 0052+251; symbols and conventions as in Figure 12.
}
\end{figure}

\clearpage
\begin{figure}
\figurenum{28}
\epsscale{0.9}
\plotone{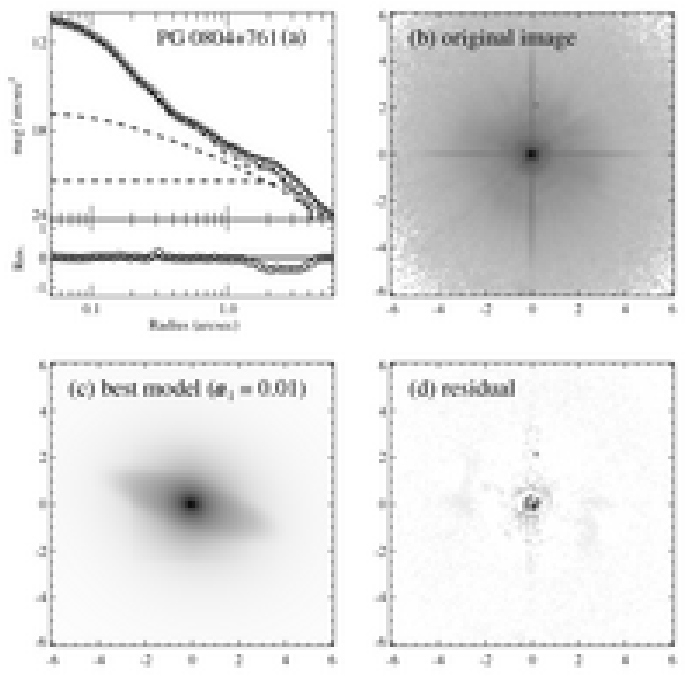}
\caption{GALFIT decomposition for PG 0804+761; symbols and conventions as in Figure 12.
}
\end{figure}

\clearpage
\begin{figure}
\figurenum{29}
\epsscale{0.9}
\plotone{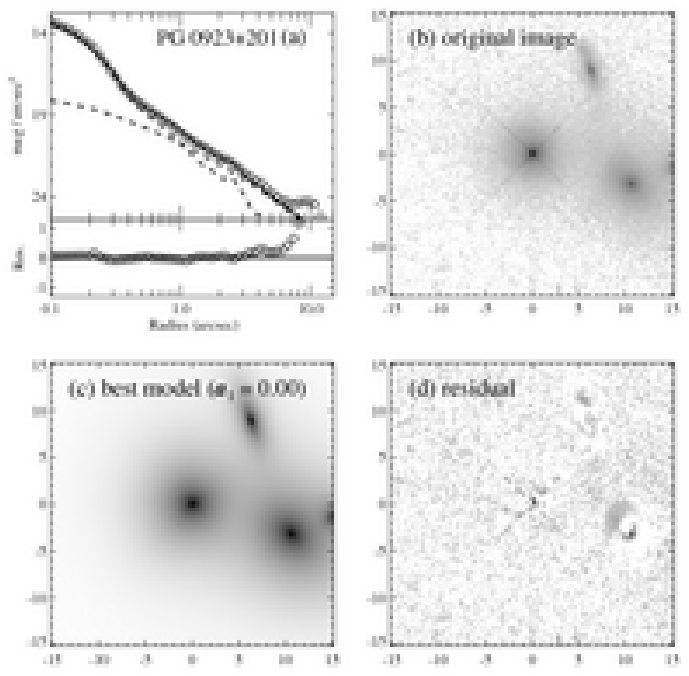}
\caption{GALFIT decomposition for PG 0923+201; symbols and conventions as in Figure 12.
}
\end{figure}

\clearpage
\begin{figure}
\figurenum{30}
\epsscale{0.9}
\plotone{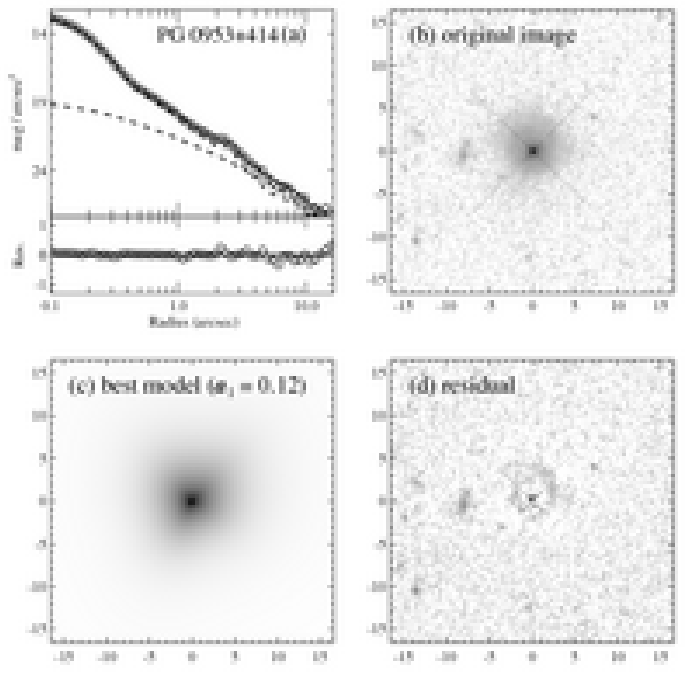}
\caption{GALFIT decomposition for PG 0953+414; symbols and conventions as in Figure 12.
}
\end{figure}

\clearpage
\begin{figure}
\figurenum{31}
\epsscale{0.9}
\plotone{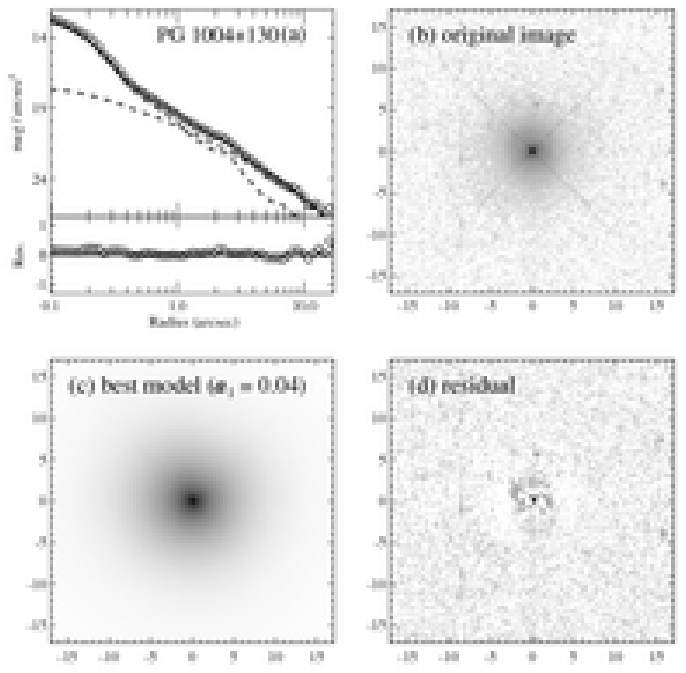}
\caption{
GALFIT decomposition for PG 1004+130; symbols and conventions as in Figure 12.
}
\end{figure}

\clearpage
\begin{figure}
\figurenum{32}
\epsscale{0.9}
\plotone{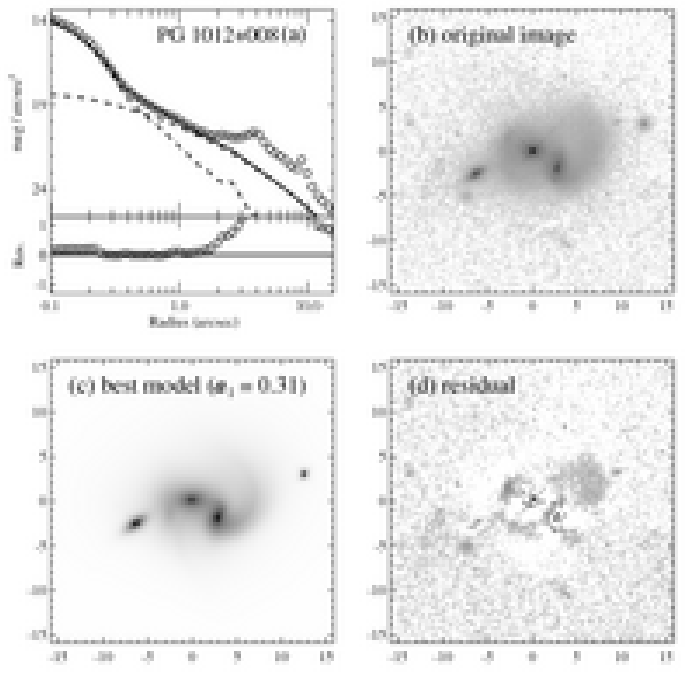}
\caption{
GALFIT decomposition for PG 1012+008; symbols and conventions as in Figure 12.
}
\end{figure}

\clearpage
\begin{figure}
\figurenum{33}
\epsscale{0.9}
\plotone{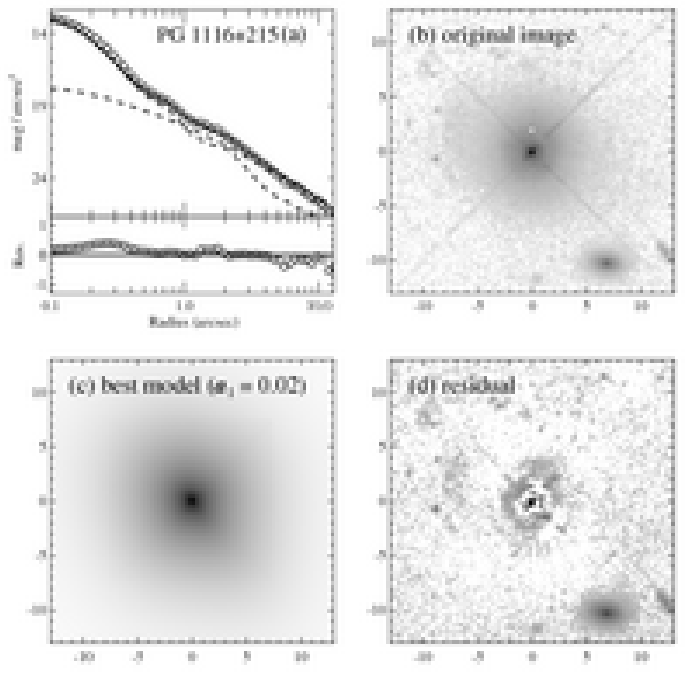}
\caption{GALFIT decomposition for PG 1116+215; symbols and conventions as in Figure 12.
}
\end{figure}

\clearpage
\begin{figure}
\figurenum{34}
\epsscale{0.9}
\plotone{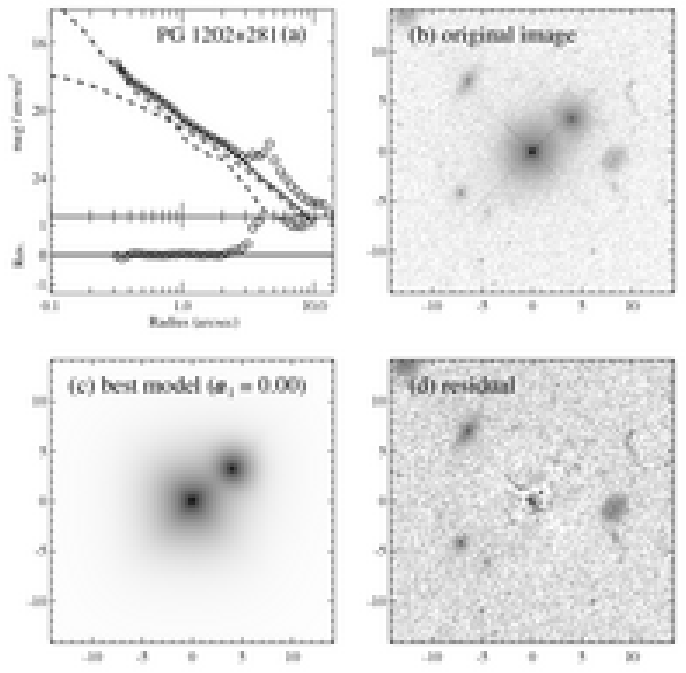}
\caption{
GALFIT decomposition for PG 1202+281; symbols and conventions as in Figure 12.
}
\end{figure}

\clearpage
\begin{figure}
\figurenum{35}
\epsscale{0.9}
\plotone{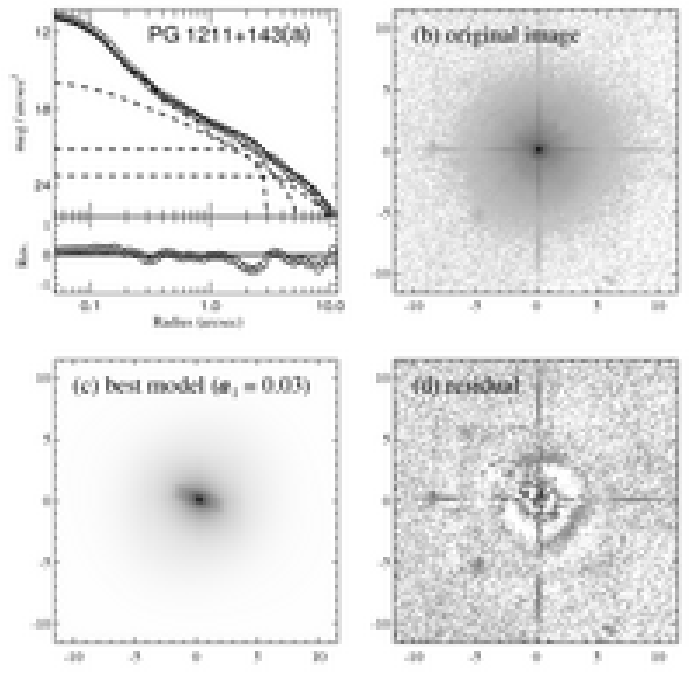}
\caption{
GALFIT decomposition for PG 1211+143; symbols and conventions as in Figure 12.
}
\end{figure}

\clearpage
\begin{figure}
\figurenum{36}
\epsscale{0.9}
\plotone{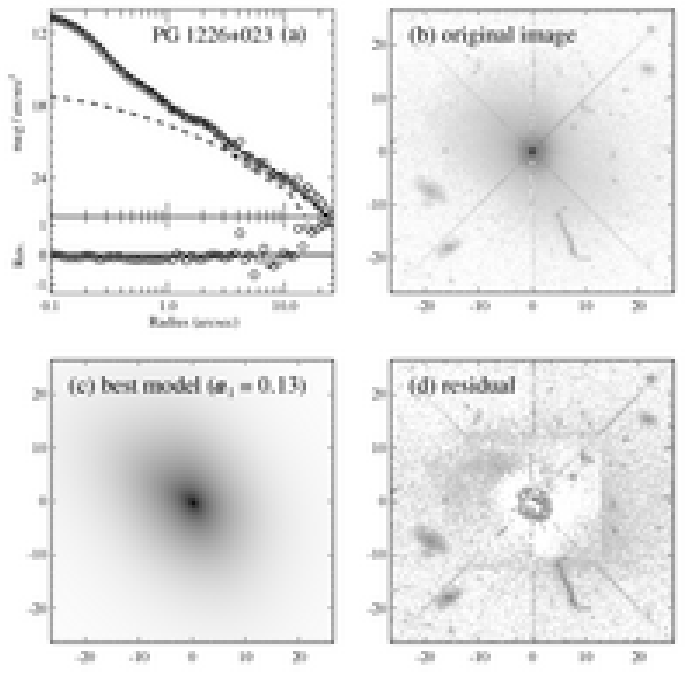}
\caption{
GALFIT decomposition for PG 1226+023; symbols and conventions as in Figure 12.
}
\end{figure}

\clearpage
\begin{figure}
\figurenum{37}
\epsscale{0.9}
\plotone{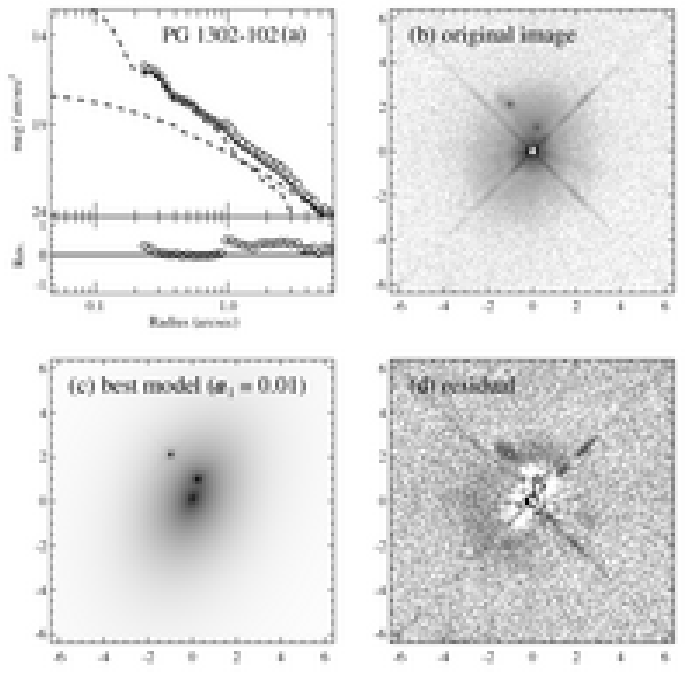}
\caption{
GALFIT decomposition for PG 1302$-$102 (PC/F702W); symbols and conventions as in Figure 12.
}
\end{figure}

\clearpage
\begin{figure}
\figurenum{38}
\epsscale{0.9}
\plotone{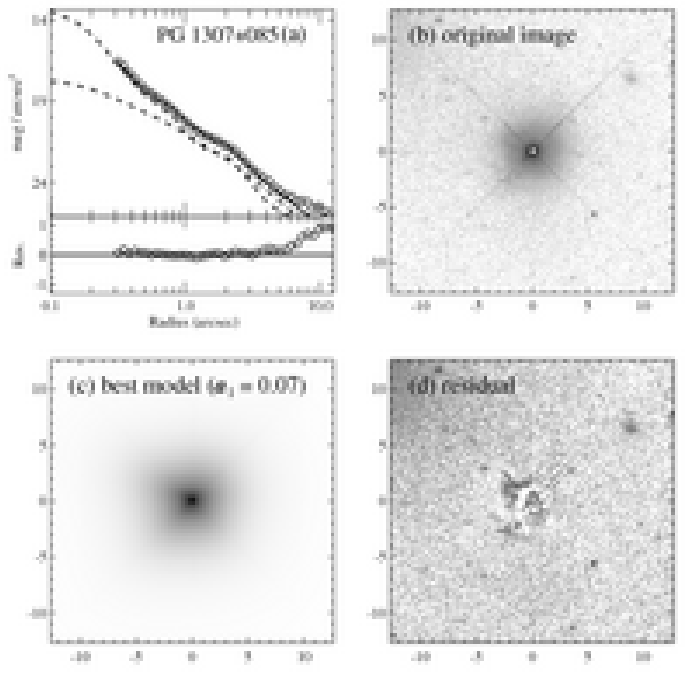}
\caption{
GALFIT decomposition for PG 1307+085; symbols and conventions as in Figure 12.
}
\end{figure}

\clearpage
\begin{figure}
\figurenum{39}
\epsscale{0.9}
\plotone{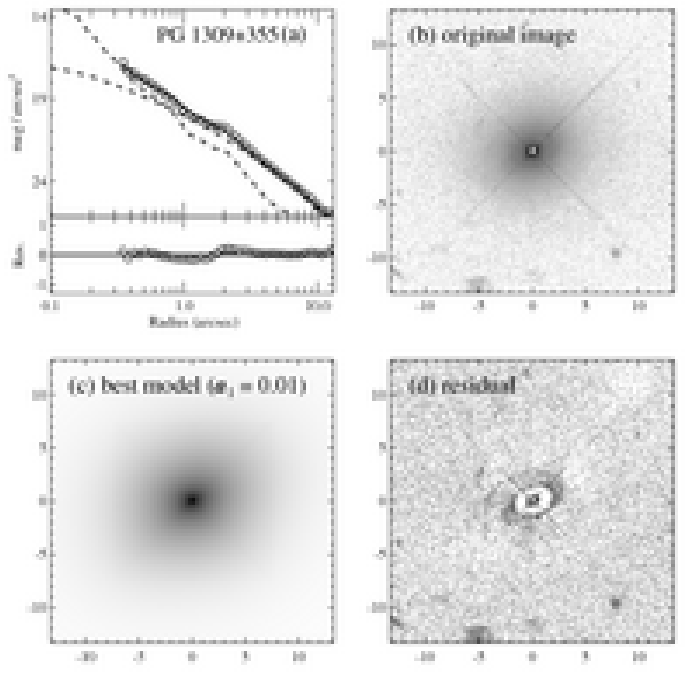}
\caption{
GALFIT decomposition for PG 1309+355; symbols and conventions as in Figure 12.
}
\end{figure}

\clearpage
\begin{figure}
\figurenum{40}
\epsscale{0.9}
\plotone{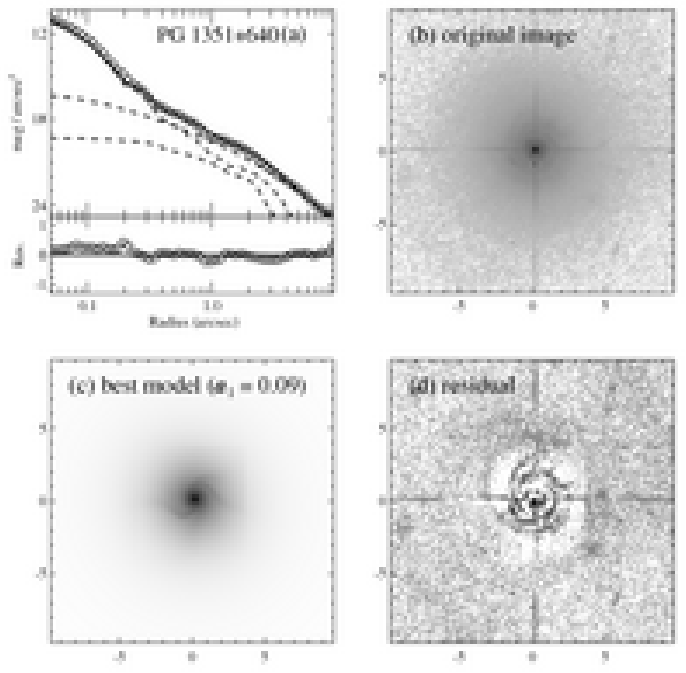}
\caption{
GALFIT decomposition for PG 1351+640; symbols and conventions as in Figure 12.
}
\end{figure}

\clearpage
\begin{figure}
\figurenum{41}
\epsscale{0.9}
\plotone{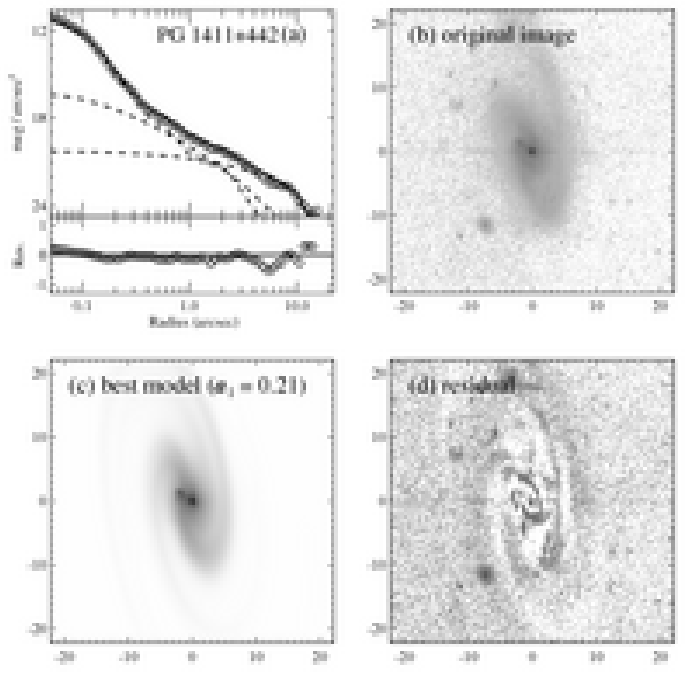}
\caption{
GALFIT decomposition for PG 1411+442; symbols and conventions as in Figure 12.
}
\end{figure}

\clearpage
\begin{figure}
\figurenum{42}
\epsscale{0.9}
\plotone{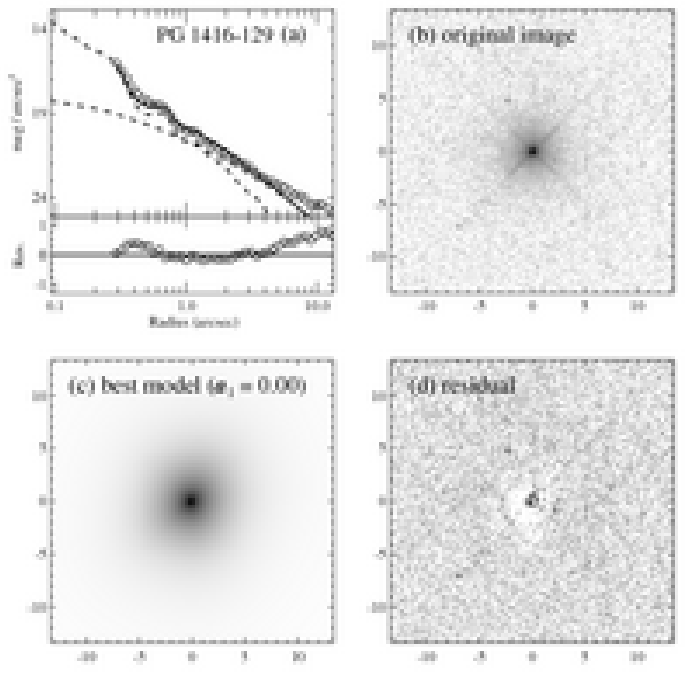}
\caption{
GALFIT decomposition for PG 1416$-$129; symbols and conventions as in Figure 12.
}
\end{figure}

\clearpage
\begin{figure}
\figurenum{43}
\epsscale{0.9}
\plotone{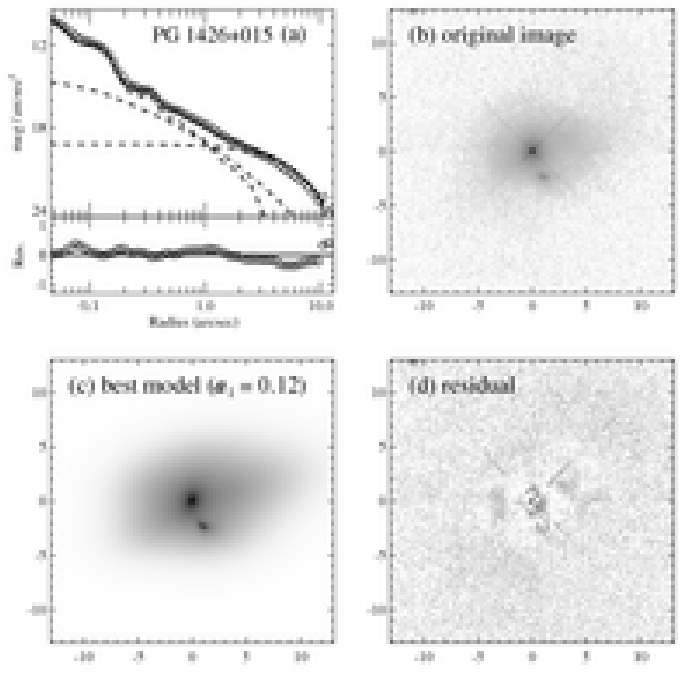}
\caption{
GALFIT decomposition for PG 1426+015; symbols and conventions as in Figure 12.
}
\end{figure}

\clearpage
\begin{figure}
\figurenum{44}
\epsscale{0.9}
\plotone{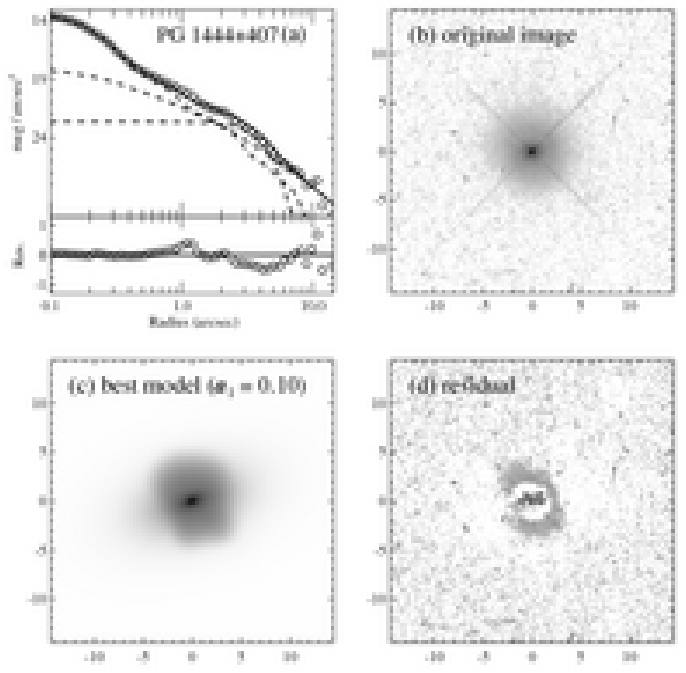}
\caption{GALFIT decomposition for PG 1444+407; symbols and conventions as in Figure 12.
}
\end{figure}

\clearpage
\begin{figure}
\figurenum{45}
\epsscale{0.9}
\plotone{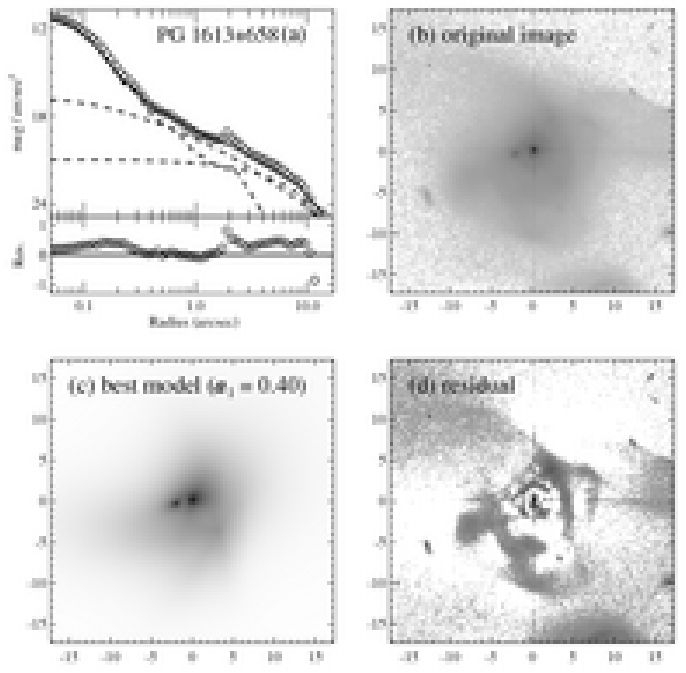}
\caption{
GALFIT decomposition for PG 1613+658; symbols and conventions as in Figure 12.
}
\end{figure}

\clearpage
\begin{figure}
\figurenum{46}
\epsscale{0.9}
\plotone{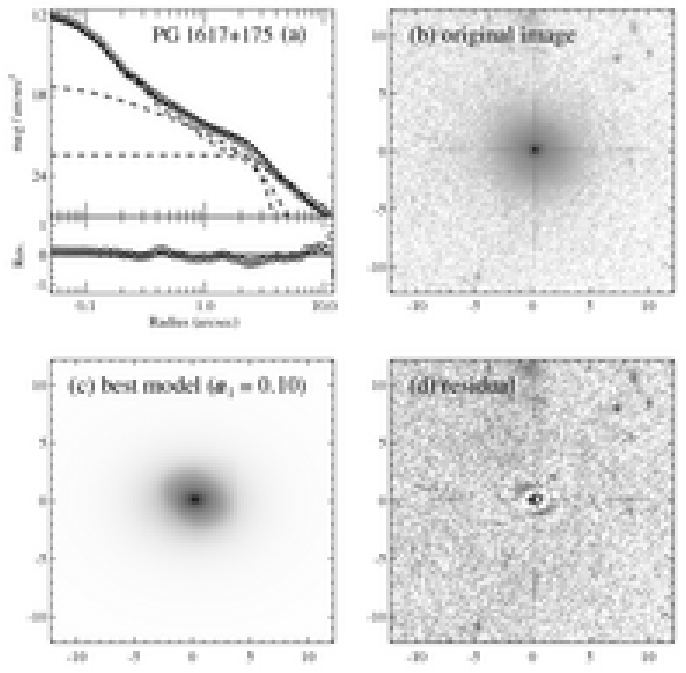}
\caption{
GALFIT decomposition for PG 1617+175; symbols and conventions as in Figure 12.
}
\end{figure}

\clearpage
\begin{figure}
\figurenum{47}
\epsscale{0.9}
\plotone{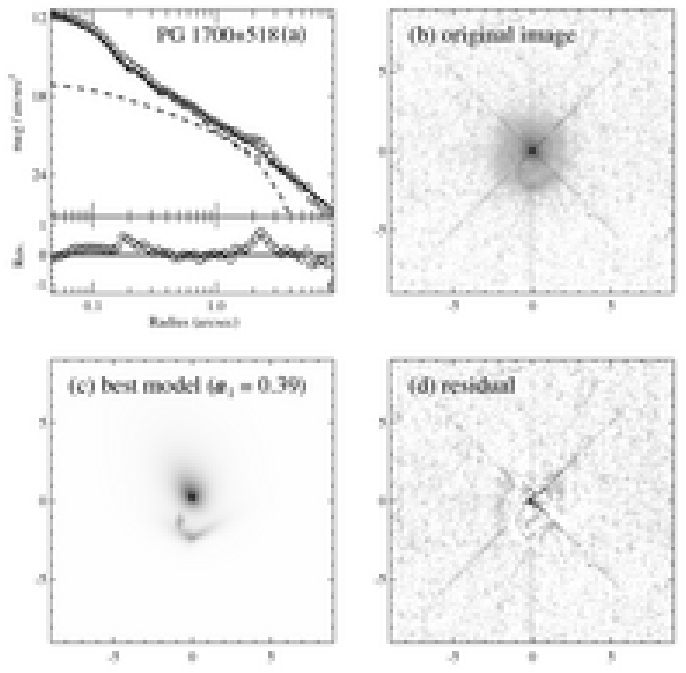}
\caption{
GALFIT decomposition for PG 1700+518; symbols and conventions as in Figure 12.
}
\end{figure}

\clearpage
\begin{figure}
\figurenum{48}
\epsscale{0.9}
\plotone{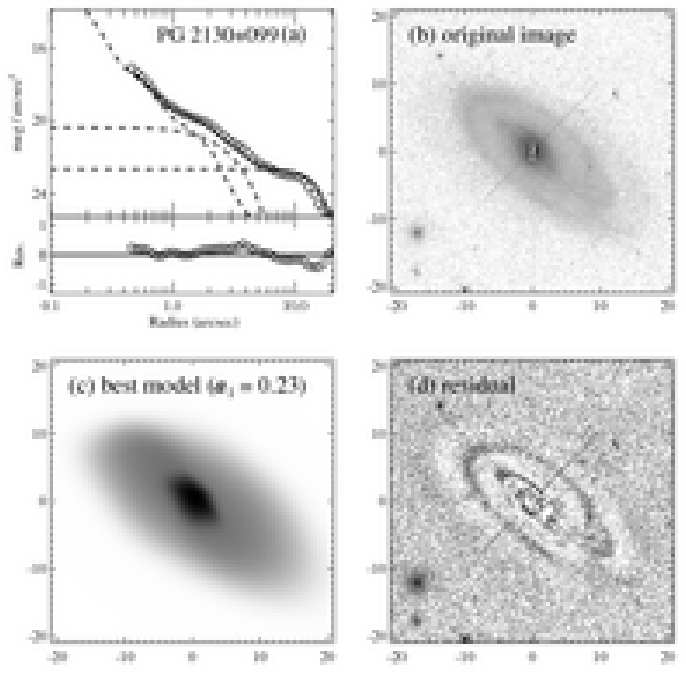}
\caption{
Example of GALFIT decomposition for PG 2130+099; symbols and conventions as in Figure 12.
}
\end{figure}

\clearpage
\begin{figure}
\figurenum{49}
\epsscale{0.9}
\plotone{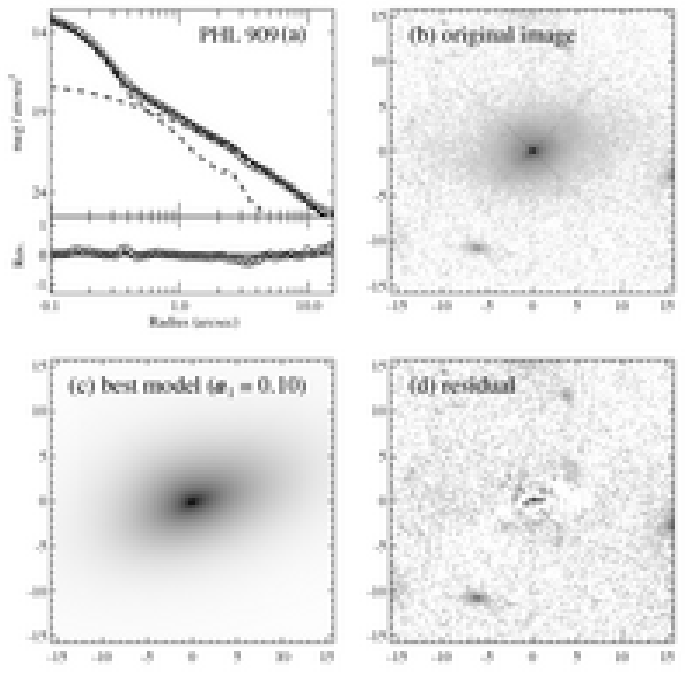}
\caption{GALFIT decomposition for PHL 909; symbols and conventions as in Figure 12.
}
\end{figure}

\clearpage
\begin{figure}
\figurenum{50}
\epsscale{0.9}
\plotone{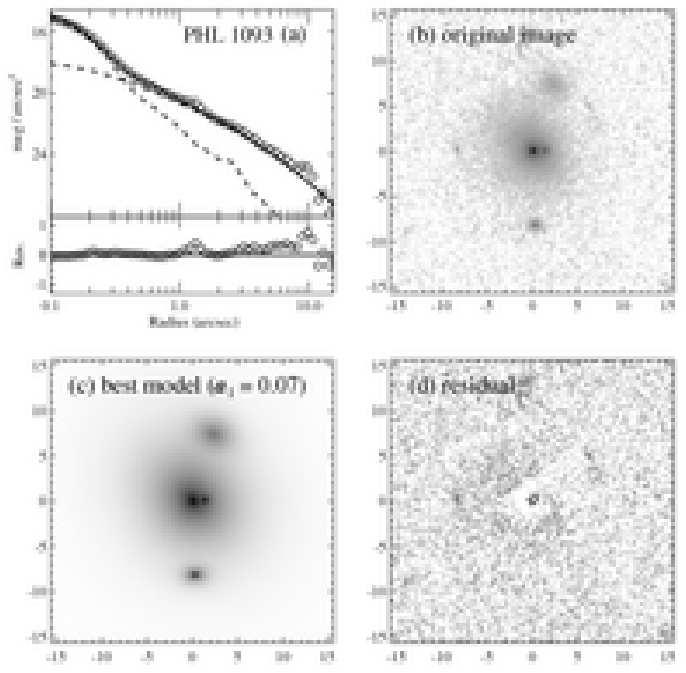}
\caption{
GALFIT decomposition for PHL 1093; symbols and conventions as in Figure 12.
}
\end{figure}

\clearpage
\begin{figure}
\figurenum{51}
\epsscale{0.9}
\plotone{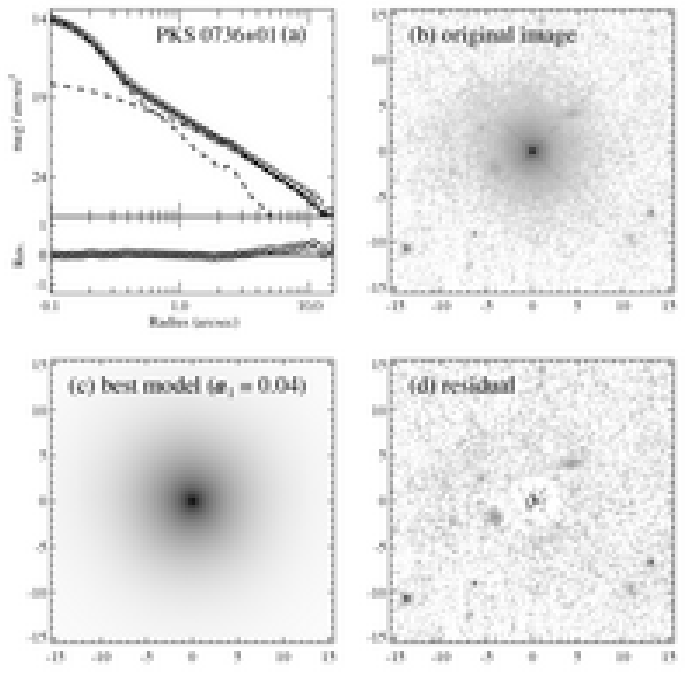}
\caption{
GALFIT decomposition for PKS 0736+01; symbols and conventions as in Figure 12.
}
\end{figure}

\clearpage
\begin{figure}
\figurenum{52}
\epsscale{0.9}
\plotone{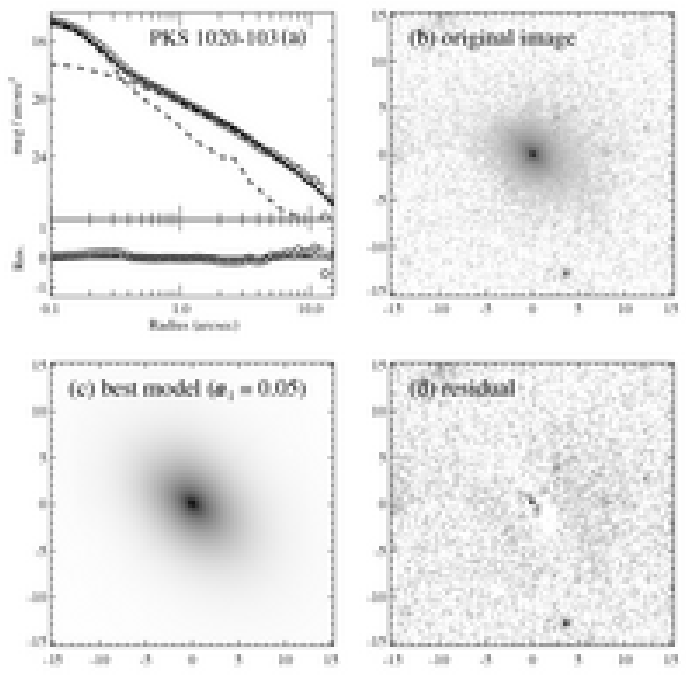}
\caption{
GALFIT decomposition for PKS 1020$-$103; symbols and conventions as in Figure 12.
}
\end{figure}

\clearpage
\begin{figure}
\figurenum{53}
\epsscale{0.9}
\plotone{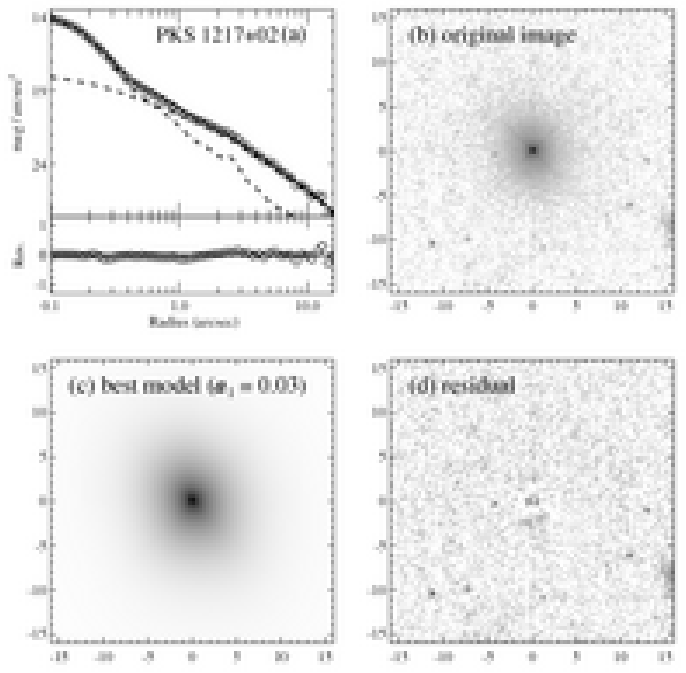}
\caption{
GALFIT decomposition for PKS 1217+02; symbols and conventions as in Figure 12.
}
\end{figure}

\clearpage
\begin{figure}
\figurenum{54}
\epsscale{0.9}
\plotone{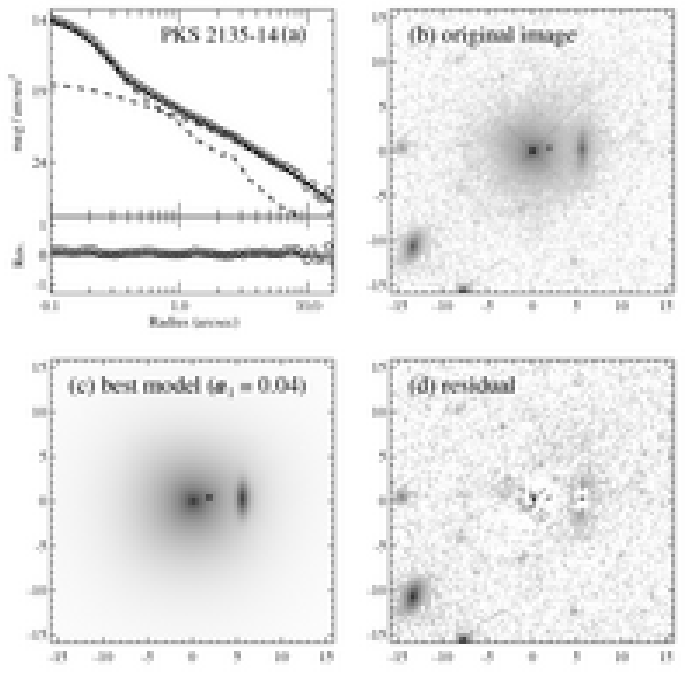}
\caption{
GALFIT decomposition for PKS 2135$-$14; symbols and conventions as in Figure 12.
}
\end{figure}

\clearpage
\begin{figure}
\figurenum{55}
\epsscale{0.9}
\plotone{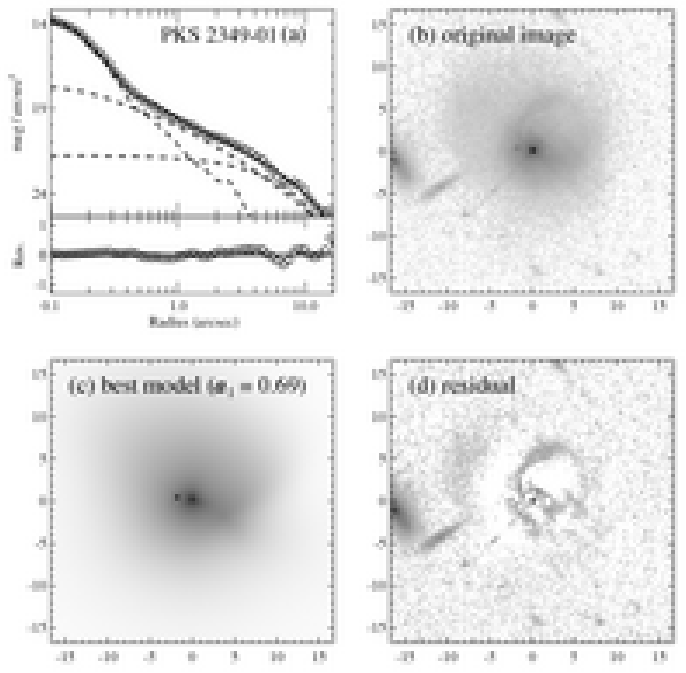}
\caption{
GALFIT decomposition for PKS 2349$-$01; symbols and conventions as in Figure 12.
}
\end{figure}

\clearpage
\begin{figure}
\figurenum{56}
\epsscale{0.9}
\plotone{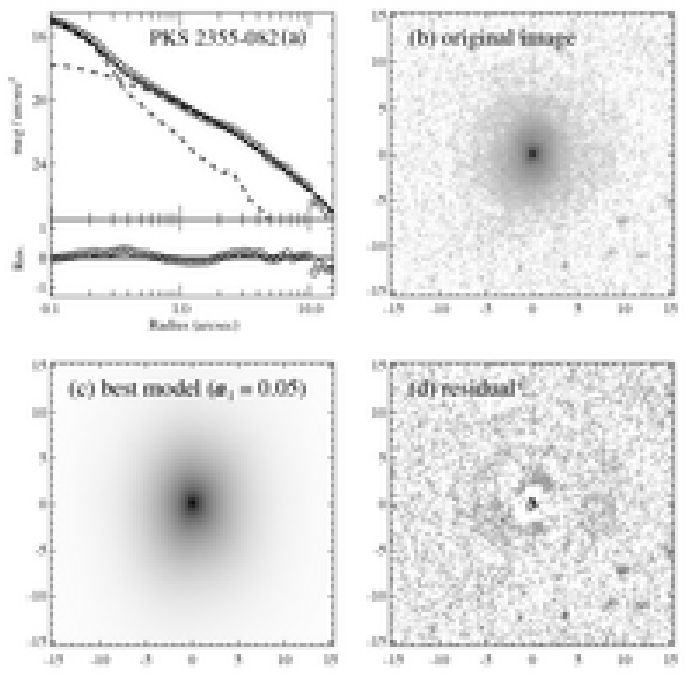}
\caption{
GALFIT decomposition for PKS 2355$-$082; symbols and conventions as in Figure 12.
}
\end{figure}

\end{document}